\pdfoutput=1
\documentclass{biophys-new}
\usepackage[utf8]{inputenc}
\usepackage{graphicx}
\usepackage[colorlinks,allcolors=cyan!70!black]{hyperref}
\usepackage{siunitx}
\usepackage{mhchem}
\usepackage[labelfont=bf]{caption}
 \usepackage{setspace}
 \usepackage{xr}
\usepackage[sort&compress,numbers]{natbib}
\DeclareSIUnit{\molar}{M}
\DeclareSIUnit{\plt}{plt}

\usepackage{lipsum}

\title{Modeling Platelet P2Y$_1$/$_{12}$ Pathway to Integrin Activation}
\runningtitle{Modeling Platelet P2Y$_1$/$_{12}$ Pathway} 

\author[1]{Keshav B. Patel}
\author[2,3]{Wolfgang Bergmeier}
\author[1,4*]{Aaron L. Fogelson}
\runningauthor{Patel, Bergmeier, and Fogelson} 

\affil[1]{Department of Mathematics, University of Utah, Salt Lake City, UT, 84112}
\affil[2]{Department of Biochemistry and Biophysics, University of North Carolina at Chapel Hill, Chapel Hill, NC, 27599}
\affil[3]{Blood Research Center, University of North Carolina at Chapel Hill, Chapel Hill, NC, 27599}
\affil[4]{Department of Bioengineering, University of Utah, Salt Lake City, UT, 84112}

\corrauthor[*]{fogelson@math.utah.edu}

\papertype{Article}

\begin{document}

\begin{frontmatter}

\begin{abstract}
Through experimental studies, many details of the pathway of integrin $\alpha_{\rm IIb}\beta_3$ activation by ADP during the platelet aggregation process have been mapped out.
ADP binds to two separate G protein coupled receptors on platelet surfaces, leading to alterations in the regulation of the small GTPase RAP1.
We seek to (1) gain insights into the relative contributions of both pathways to RAP1-mediated integrin activation and to (2) predict wildtype and mutated cell behavior in response to a continuous range of external agonist concentrations. 
To this end, we develop a dynamical systems model detailing the action of each protein in the two pathways up to the regulation of RAP1.
We perform a parameter estimation using flow cytometry data to determine a number of unknown rate constants.
We then validate with already published data; in particular, the model confirmed the effect of impaired P2Y$_1$ receptor desensitization or reduced RASA3 expression on RAP1 activation.
We then predict the effect of protein expression levels on integrin activation and show that components of the P2Y$_{12}$ pathway are critical to the regulation of integrin.
This model aids in our understanding of interindividual variability in platelet response to ADP and therapeutic P2Y$_{12}$ inhibition.
It also provides a more detailed view of platelet activation in the ongoing mathematical study of platelet aggregation.
\end{abstract}

\begin{sigstatement}
A detailed dynamical systems model of integrin $\alpha_{\rm IIb}\beta_3$ activation mediated by ADP is presented.
This model takes in years of experimental literature on the relevant pathways into simulations that are efficient to run and simple to manipulate parameters to match a platelet mutation.
Our model can finely adjust any parameter and analyze how a platelet responds to any number of experimental settings.
We present the model's utility through parameter study on the expression levels of proteins key to integrin activation and show that integrin response changes more significantly in response to changes in the P2Y$_{12}$ pathway over the P2Y$_1$ pathway.
\textcolor{black}{This insight can be used to understand how certain antithrombotic therapies, like clopidogrel, can affect a platelet's intracellular signaling pathways.}
\end{sigstatement}
\end{frontmatter}


\section*{Introduction}

Platelets are the primary cellular component in arterial blood clots, supporting the coagulation response and recruitment of other cells/proteins to a growing thrombus. 
They perform a variety of processes that aid in clot growth and stabilization, such as releasing agonists into the environment to recruit other platelets, changing shape through cytoskeletal remodeling to increase surface area for reactions, and activating integrins on their surfaces for binding crosslinking proteins.
For the purposes of this work, we will use the words ``integrin activation'' to refer to the conversion of integrin $\alpha_{\rm IIb}\beta_3$ from a low fibrinogen affinity state to a high-affinity state. \par

Several molecules serve as receptor ligands that lead to integrin activation \cite{Lee_ch18}.
Most ligands, like thrombin or collagen, initiate pathways that lead to irreversible activation of the integrin.
Signaling pathways for these receptor-ligand bindings have been studied by both experimentalists and modelers. 
Dunster et al. modeled the activation signals associated with collagen binding the GPVI receptor \cite{dunster2015regulation}. Lenoci et al. \cite{lenoci2011mathematical} and Sveshnikova et al. \cite{sveshnikova2015compartmentalized} modeled platelet activation through the action of thrombin on the PAR1 receptor.

\textcolor{black}{We focused our modeling efforts on platelet inside-out integrin activation induced by ADP, an autocrine/paracrine small molecule agonist released by activated platelets. 
Our main reasons for choosing ADP as the agonist include 
(i) its physiological relevance as a key recruiter of free-flowing platelets to a growing aggregate \textit{in vivo} and is the target of certain antithrombotic therapies such as clopidogrel \cite{moshfegh2000antiplatelet}, 
(ii) the well-defined signaling pathways connecting the ADP receptors, P2Y$_1$ and P2Y$_{12}$, to integrin activation mediated by the adapter protein, Talin1, and 
(iii) a relatively weak and reversible integrin activation response that provides a perfect system to model both positive and negative modulation of these signaling pathways.
Furthermore, we focus specifically on the effect of inside-out signaling in our first version of this model.
We experimentally isolate the effect of inside-out signaling by using the fluorescent molecule JON/A(PE) \cite{bergmeier2002flow}, which selectively binds to the active conformation of the integrin and prevents outside-in signaling.}

\textcolor{black}{ADP binds to two platelet GPCRs, the G$_{\rm q}$-coupled P2Y$_1$ and G$_{\rm i}$-coupled P2Y$_{12}$ receptors. 
Activation of P2Y$_1$ induces activation of phospholipase C$\beta$ (PLC$\beta$) and formation of Ca$^{2+}$ and diacylglycerol (DAG), while activation of P2Y$_{12}$ induces PI3 kinase-dependent formation phosphatidylinositol-3,4,5-trisphosphate (PIP$_3$). 
Ca$^{2+}$ activates the RAP-GEF, CalDAG-GEFI (CDGI), while PIP$_3$ inhibits the RAP-GAP, RASA3 \cite{Stefanini2015}. 
Activation of both pathways, but not P2Y$_1$ or P2Y$_{12}$ signaling alone, leads to robust and prolonged activation of small GTPase RAP1, a key regulatory component of the integrin activation complex (IAC) \cite{stefanini2018negative,andre2003p2y,Paul_Blatt_Schug_Clark_etAl}. 
The IAC consists of RAP1 and adapter proteins such as Talin1 \cite{ma2007platelet}. 
Following RAP1 activation, these proteins assemble at the intracellular tail of the integrin to induce a conformational change required for ligand binding. 
Once RAP1-GTP is converted to its GDP bound form, however, these proteins quickly unbind, and the integrin returns to its inactive conformation. 
Thus, the dynamics of RAP1 activation can be used as a proxy for integrin activation. 
It is also important to note that ADP-dependent mobilization of intracellular calcium stores is also linked to the influx of calcium from the extracellular space. 
However, this so-called store-operated calcium entry (SOCE) does not impact ADP-induced integrin inside-out activation \cite{Braun_etAl_2009}. 
Furthermore, P2Y$_{12}$ signaling leads to decreased levels of cyclic AMP, a second messenger that leads to the inhibition of various platelet activation responses \cite{smolenski2012novel}. 
We have not included these pathways in our model as our previous work demonstrated that signals directly affecting CalDAG-GEFI and RASA3 function are most relevant for RAP1-dependent integrin inside-out activation.}

Other models \cite{filkova2019quantitative} have been developed to study integrin activation and aggregation mediated by ADP, but to the authors' knowledge, no detailed models of signaling pathways for ADP-dependent activation have been developed. 
In this work, we convert the collected information on both pathways to a system of ordinary differential equations (ODEs) that can be solved numerically to simulate a single platelet's response to a given amount of P2Y$_1$ and/or P2Y$_{12}$ agonist.
We use experimental data to estimate unknown parameters and validate the model against recently published data.
We then explore how a simulated platelet responds to agonists under a change to the expression level of P2Y$_1$, P2Y$_{12}$, RASA3, and CalDAG-GEFI.
Unlike in experimental settings, we are able to tune expression with much finer precision to gain a more quantitative measure of sensitivity.
We begin by describing the chemical signaling pathways triggered upon platelet GPCR binding to ADP.

\subsubsection*{Biological Background}

\begin{figure*}[htb]
\centering
     \includegraphics{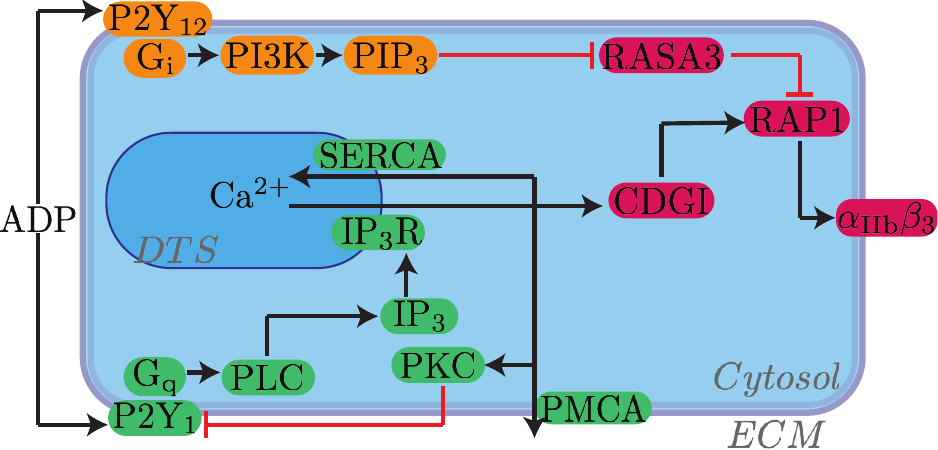}
\caption{\textbf{Schematic of intracellular pathways involved in integrin activation.} ADP binds to G protein Coupled Receptors P2Y$_1$ and P2Y$_{12}$ on platelet surfaces, each initiating its own signaling pathway and both contributing to a change in RAP1-GTP concentration.}
\label{fig:schematic}
\end{figure*}

Figure~\ref{fig:schematic} shows schematically how the P2Y$_1$ and P2Y$_{12}$ receptors are activated by ADP and, through different pathways, cause changes to both the activator and inhibitor of RAP1.

The extracellular domains of the platelet P2Y$_1$ and P2Y$_{12}$ receptors bind to ADP in the blood plasma.  
Upon binding to ADP, these receptors act as GEF enzymes on the membrane-associated G proteins G$_{\rm q}$ and G$_{\rm i}$, respectively.
The G proteins then unbind from the receptor and bind to membrane-associated enzymes PLC and PI3K, respectively.
Both PLC and PI3K use the lipid PIP$_2$ as substrate: PLC converts PIP$_{2}$ to the membrane-associated molecule DAG and the cytosolic molecule IP$_3$, and PI3K converts PIP$_2$ to membrane-associated PIP$_{3}$.  
These inositol lipids are part of a large cycle of formation and degradation.
IP$_3$ binds to IP$_3$ receptors (IP$_3$R) embedded in the dense tubular system (DTS) membrane and triggers the release of calcium ions from the DTS into the cytosolic space. 

The chemically-gated IP$_3$R channel is a four-subunit protein; each subunit contains an activating binding site for IP$_3$, an activating binding site for calcium, and a slower inactivating binding site for calcium \cite{SneydDufour2002}.  
Thus, the initial release of calcium from the DTS leads to a stage of positive feedback followed by negative feedback.
\textcolor{black}{In resting platelets, the cytosolic Ca$^{2+}$ concentration is $\sim$20-100 nM, a level that is maintained by active sequestration of Ca$^{2+}$ in the DTS and the extracellular space.}
In addition to the IP$_3$R channel, the model incorporates the active PMCA pump and passive leak current on the plasma membrane, and the active SERCA pump across the DTS membrane.

Calcium's effect on integrin activation is mediated through its binding to the GEF enzyme CalDAG-GEFI.  
CalDAG-GEFI contains two EF domains, which each binds calcium with a dissociation constant $K_{\rm GEF,M} = 80$ \si{\nano\molar} \cite{Iwig2013}.  
Calcium-bound CalDAG-GEFI converts RAP1 from its inactive GDP-bound form to its active GTP-bound form.  
The active form of RAP1 complexes with a variety of proteins on the cytosolic tail of $\alpha_{\rm IIb}\beta_3$, for example, TALIN, to trigger integrin activation.
RASA3 is a membrane-associated protein constitutively active in converting active RAP1 to its inactive GDP-bound form.  
RASA3 can be inactivated by interacting with PIP$_{3}$, although the exact mechanism has not been determined \cite{Battram2017}.  
CalDAG-GEFI and RASA3 have nearly equal copy numbers in mouse platelets (approximately 30,000 \si{\per\plt}), whereas RAP1 in its two major isoforms has a total copy number of approximately 200,000 \si{\per\plt} \cite{Zeiler2014}.

\section*{Materials and Methods}

We used Mass Action or Michalis-Menten kinetics to describe each reaction in the P2Y$_1$ and P2Y$_{12}$ pathways.  
The concentration, denoted by $[\cdot]$, or surface density, denoted by $[\cdot]_S$, of each type of molecule in its various states was tracked using an ordinary differential equation (ODE).  
The overall model, therefore, comprised an extensive coupled system of ODEs that we must solve simultaneously, and we accomplished this using LLNL's DLSODE solver package \cite{Hindmarsh1983}.

Here, we discuss the setup of the differential equation system. 
We used the model of Purvis et al. \cite{Purvis2008} as a starting point for our modeling; below, we discuss explicitly where we deviated and extended their model.
The system was described by 82 differential equations; see the supplemental material for a complete list of the model's equations.
We then describe the statistical methods used to inform unknown model parameters.
Finally, we describe the experimental methods used to generate the data against which we validate the model.

\subsubsection*{Model components and numerical setup}

\paragraph{ADP Receptors:}
Following prior modeling work on G protein coupled receptors \cite{Kinzer-Ursem2007}, we explicitly tracked the state changes of the P2Y$_1$ and P2Y$_{12}$ receptors as they bind to ADP and their specific G protein.
Each binding, unbinding, and catalytic reaction is described using the Law of Mass Action.
Receptors bound to an inactive G protein can activate it, regardless of whether ADP is also bound to the receptor. However, the presence of ADP significantly increases the receptor's efficiency. \par

We made two notable changes to the model as implemented in Purvis et al. \cite{Purvis2008}. 
First, this and other previous models include an "inactive" GPCR state converted into an "active" state through a reversible, unimolecular reaction.  
We noted through numerical simulations (not shown) that the concentration for each inactive state quickly reached an equilibrium value proportional to the concentration for the corresponding active state.  
Therefore, we performed a quasi-steady-state (QSS) reduction to reduce the number of explicit states. 

Second, we included a PKC-dependent inactivated state for P2Y$_1$.  
We assumed that ADP-bound P2Y$_1$ can be inactivated via phosphorylation of an intracellular site and reactivated via dephosphorylation. 
PKC mediates the phosphorylation process.
The kinetic parameters for this reaction were estimated in this study.

\paragraph{Second Messengers:}
The model explicitly tracked the second messenger proteins PLC and PI3K in their inactive state and active (i.e., bound to G protein) state \cite{BallaIyengar1999}.
Binding and unbinding of G proteins were tracked through Mass Action kinetics, with an additional irreversible hydrolysis term, in which the second messenger both converts the $\alpha$ subunit to a GDP bound form and releases it into the cytosol.
Once activated, PLC and PI3K act upon the phosphoinositol species PIP$_2$, converting it to IP$_3$ and PIP$_3$, respectively.
These inositol species are recycled through a variety of states by various other phosphatases and kinases \cite{KANOH1990120,URUMOW1990152,MATZARIS19943397,ATACK1993305,MOYER19871018,MITCHELL19898873,KELLEY198714563,VARGAS1984123}.

\paragraph{Calcium Dynamics:}
The flux of calcium ions across passive elements (namely, the IP$_3$R and PM leak) are governed by the Nernst Equation.
For example, the rate at which calcium moves across the DTS membrane through conducting IP$_3$R channels is given by:

\begin{equation}
R_{\rm DTS} = \frac{N_{\rm IP_3R}}{4}P_0\gamma_{\rm IP_3R} \frac{RT}{(zF)^2}\log\bigg(\frac{[{\rm Ca}^{2+}]_{\rm dts}}{[{\rm Ca}^{2+}]_{\rm cyt}}\bigg),
\label{eqn:Nernst}
\end{equation}

\noindent where $N_{\rm IP_3R}$ is the copy number of IP$_3$R subunits, $P_0$ is the probability a single channel is in a conducting state, $\gamma$ is the conductance of the channel, $R$ is the universal gas constant, $T$ is the temperature, $z$ is the charge of a calcium ion, $F$ is Faraday's constant, and $[{\rm Ca}^{2+}]_{\rm dts}$ and $[{\rm Ca}^{2+}]_{\rm cyt}$ are the free calcium concentrations in the DTS and cytosol, respectively. 
We model the IP$_3$R channel using the six-state model given by Sneyd and Dufour \cite{SneydDufour2002} and use their formula for computing $P_0$, which is dependent on the fraction of receptors in the two conducting states. \par

Once calcium is released into the cytosol, it can bind to various pumps, enzymes, and buffers.  
We modeled both the SERCA and PMCA pumps using 6-state transport models \cite{Dode2002}, where two calcium ions at a time are transported across their respective membranes. 
We explicitly modeled PKC binding of calcium and DAG using Mass Action kinetics.  
As described above, PKC molecules bound to both calcium and DAG can inactivate the P2Y$_1$ receptor.  
To match the experimental setup, we included the binding of calcium to a fluorescent experimental probe with a concentration of 5 \si{\micro\molar} and a dissociation constant of 100 \si{\nano\molar}.
This probe was only included during the parameter estimation that relied on calcium data.
In all other experiments, we excluded the probe.

\paragraph{GEF/GAP Enzymes:}
We modeled both the binding of calcium with CalDAG-GEFI \cite{Iwig2013} and the binding of PIP$_3$ with RASA3 using Mass Action kinetics.
We then modeled the activation and inactivation of RAP1 using Michalis-Menten reaction terms, such that the differential equation for the surface density of RAP1-GTP is given by:

\begin{align}
\begin{split}
S_{\rm PM}\frac{d}{dt}[{\rm RAP1-GTP}]_S = V_{\rm cyt}\frac{k_{\rm GEF}[{\rm CalDAG-GEFI}-{\rm 2Ca}][{\rm RAP1-GDP}]_S}{(\frac{V_{\rm cyt}}{S_{\rm PM}}\big)K_{\rm GEF,M} + [{\rm RAP1-GDP}]_S}  - S_{\rm PM}\frac{k_{\rm GAP}[{\rm RASA3}]_S[{\rm RAP1-GTP}]_S}{(\frac{V_{\rm cyt}}{S_{\rm PM}}\big)K_{\rm GAP,M} + [{\rm RAP1-GTP}]_S},
\end{split}
\label{eqn:RAP1}
\end{align}
\noindent where $S_{\rm PM}$ is the surface area of the cell membrane, $V_{\rm cyt}$ is the volume of the cytosol, and the catalytic rates and Michalis-Menten constants are estimated in this work.

\paragraph{Simulation Protocol:}
The copy number of each protein within mouse platelets has been documented using proteomics \cite{Zeiler2014}. 
From the data, we computed the concentration or surface density of every species in the model.
The model is run in two stages: an equilibrium stage and an experimental stage.
In the equilibrium stage, we started by assuming that every molecule is in a form that is not bound to any other molecule, and we assumed there is no external ADP.
We then ran the system to steady-state, which we determined by computing the relative change in protein states at each iteration and stopping when changes decreased below a threshold.
The concentration and surface density values at steady state were then used as the initial conditions in the experimental stage, where at $t=30$ \si{\second}, the external ADP concentration was instantaneously increased to a set value, and the simulation was run for 600 s.

\paragraph{Sensitivity Analysis:}
We briefly describe the Method of Morris, a one-at-a-time global sensitivity analysis algorithm.
The algorithm gives a way to estimate the derivative of a model output with respect to model parameters when no closed form of the output exists and when dealing with many model parameters.
In this study, we are interested in describing the derivative of the maximum RAP1-GTP value with respect to protein copy numbers.

The method begins by defining a hypercube in parameter space; for each dimension, the center is given by the average value of a specified protein's copy number, and the half-width is given by the standard deviation of that protein's copy number.
Beginning at the center, the algorithm defines several paths to the boundary of the hypercube such that only one parameter changes with each step along a path.
The paths were computed using Python's SALib package \cite{Iwanaga2022,Herman2017}.
Each vertex on a path specifies a parameter set used for a single simulation.
In each simulation, we recorded the peak number of RAP1-GTP and then used finite differences to approximate the derivative of the maximum RAP1-GTP concentration with respect to the single parameter changed in that step.
This process generated a sampling of the derivative of peak RAP1-GTP as a function of each parameter within the hypercube; for each parameter, we reported the mean $\mu$ and standard deviation $\sigma$ of the approximation to the derivative. 
A large value of $\mu$ suggests a high sensitivity with respect to the given parameter, and a large value of $\sigma$ suggests a high correlation between the given parameter and other parameters in the model.
We again relied on SALib to compute $\mu$ and $\sigma$ for each parameter.

\paragraph{Total integrin response metric:}
JON/A is a fluorescent probe that binds tightly to the integrin $\alpha_{\rm IIb}\beta_3$ in its high-affinity state.
JON/A is assumed to bind irreversibly to the integrin, and enough JON/A is added to ensure rapid binding between the two species.
To compare model outputs to previously published JON/A binding assays, we reported the integral with respect to time of RAP1-GTP over the 600 s after agonist application. 
See the supplemental for further details.

\subsubsection*{Experimental procedures and data extraction}

\paragraph{Flow cytometry experiments:}
Washed platelets in Tyrode's buffer were diluted to $10^7$ platelets/mL and loaded with 5 $\mu$M Fluo-4 (Thermo Fisher Scientific) for 30 minutes at 37$^{\circ}$C in the dark. 
Afterward, the samples were diluted to $10^6$ platelets/mL in Tyrode's buffer and activated with 10 $\mu$M ADP in the presence of 1 mM Ca$^{2+}$ and Alexa647-labeled fibrinogen (100 $\mu$g/ml, Sigma) while being continuously sampled on a BD C6 Plus flow cytometer. 
Kinetic calcium mobilization and fibrinogen binding were analyzed in Flow Jo (Version 10) as mean fluorescence intensities over time.

\begin{figure*}[htb]
\centering
     \includegraphics{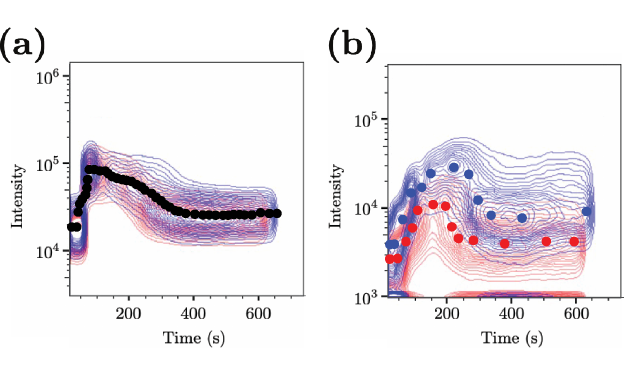}
        \caption{\textbf{Experimental studies show transient activation of platelets in response to ADP}. Flow cytometry experiments for WT (red) and RASA3$^{+/-}$ (blue) platelets when 10 $\mu$M of ADP was applied at 30 s. 
        Scatter points indicate data that were collected for parameter estimation.
        Points were selected at approximately the peak fluorescence intensity for each specified time. 
        (a) Fluorescent intensity of calcium-bound probe over time. (b) Fluorescent intensity of fibrinogen over time. }
        \label{fig:data}
\end{figure*}

\paragraph{Parameter estimation:} 
Previous models \cite{Purvis2008} of calcium signaling in cells and experimental work \cite{Zeiler2014} with mice platelets have provided many of the rate constants and concentrations needed for this study.
However, the rates governing the GEF and GAP action on RAP1 have yet to be entirely determined, and parameters estimated to fit prior calcium data were inconsistent with ours.
Therefore, we performed a parameter estimation study to compare model outputs to our experimental data given in Figure~\ref{fig:data}.
We first describe the method of converting fluorescence intensity to a measure of concentration and then the parameter estimation method.

For each dataset provided, we recorded the maximum intensity at select times to generate a single representative curve for each experiment.
To convert calcium probe fluorescence intensity to concentration, we assumed that a relative increase in fluorescent intensity corresponds to an equivalent relative increase in the protein of interest. 
For calcium, we assumed a resting concentration of 40 \si{\nano\molar} and a peak concentration of 200 \si{\nano\molar}, in line with the literature \cite{Lee_ch18}.
This data was used to estimate kinetic parameters for P2Y$_1$, $G_{\rm q}$, SERCA, and PMCA. 
To reduce the number of parameters to estimate, we make the simplifying assumption that the kinetic parameters for P2Y$_{12}$, G$_{\rm i}$, and PI3K are the same as those for P2Y$_1$, G$_{\rm q}$, and PLC.

For fibrinogen data, we performed a similar rescaling, then assumed 1) that there are no fibrinogen bound to resting platelets and 2) a WT platelet activated by a saturating level of ADP allows an \textit{a priori} known quantity of integrins to activate and 3) fibrinogen binds instantly to such an integrin.
Finally, we used a 1:1 stochiometric relationship between RAP1-GTP and integrin to arrive at the amount of RAP1-GTP that is active intracellularly.
We assume that either a maximum of 1\% or 10\% of integrins become activated, which equated to a maximum of 500 or 5000 RAP1-GTP, respectively; for either assumption, we ran a separate parameter estimation for rates governing CalDAG-GEFI, RASA3, and RAP1.

For any choice of parameters, we ran the model according to the simulation protocol described above. 
From the list of model outputs, we take the concentration of free cytosolic calcium and the number of activated RAP1 proteins, then compute the least squared error between the model and the datasets at specified times.
The goal of the parameter estimation algorithm is to minimize this error, and for this, we relied on MATLAB's \texttt{fmincon} function.

\section*{Results and Discussion}

\begin{figure*}[htbp!]
\centering
     \includegraphics{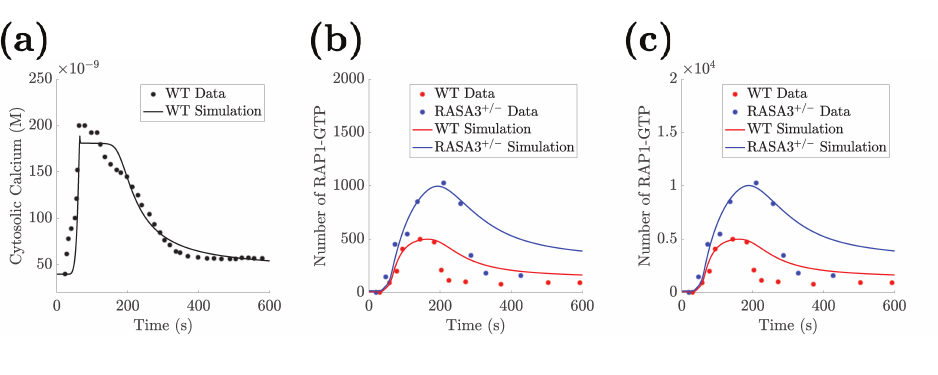}
        \caption{\textbf{Parameter estimation allows for agreement between data and computational model.} Solution to ODE system when using best-fit parameters. 
        Fluorescence intensity readings were rescaled and presented as scatter points using selected time points from the flow cytometry data.
        Solid curves depict results from the computational model.
        Black traces represent data/simulated calcium data, red traces represent data/simulated wildtype platelet RAP1-GTP, and blue traces represent data/simulated RASA3$^{+/-}$ platelet RAP1-GTP.
        In simulations and experiments, 10 $\mu$M of ADP was assumed to be added instantaneously at 30 s. 
        (a) Calcium data was rescaled, assuming a resting concentration of 40 nM and peak concentration of 200 nM, and then it was overlayed with the computed free calcium concentration.
        (b) Fibrinogen binding data was assumed to correspond directly to RAP1-GTP levels at any given time and rescaled assuming no activated RAP1 at rest and a peak number of activated RAP1 of 500.
        (c) Using the same fibrinogen binding data, we assume a peak number of activated RAP1 of 5000.}
        \label{fig:wholeCell}
\end{figure*}

\paragraph{Model outputs estimated to match experimental data:}
Figure~\ref{fig:wholeCell}a compares a calcium spike dataset to the model's calcium spike.
Because the data and simulated calcium timecourse in the wildtype and RASA3 heterozygote are nearly identical, only the wildtype timecourse is given.
The simulated calcium curves approximated the data well but smoothed over some of the dynamics near the peak.

Figures~\ref{fig:wholeCell}b-c show a comparison between the fibrinogen binding data and the simulated RAP1-GTP spike using the lower bound approximation (see Methods) of the data in Figure~\ref{fig:wholeCell}b and the upper bound approximation of the data in Figure~\ref{fig:wholeCell}c.
For each of these assumptions about the number of integrins activated by ADP, our estimation procedure sought to minimize the sum of the mean squared differences between data and simulation for \textit{both} the WT and RASA3$^{\rm +/-}$ cases. 
Our model approximated the data well when ADP was applied to the peak RAP1-GTP response.
However, simulations showed a slower return to baseline than the experimental data.
For estimated parameter values, see Supplemental Tables~S1 and~S2.

\paragraph{Model sensitivity to protein copy number is unaffected by our data approximation:}
Many of the simulations conducted in this work involve adjusting protein copy numbers. 
Therefore, understanding how our assumptions about the copy number data affected model outputs was useful.
Thus, the global sensitivity analysis algorithm, the Method of Morris, was used to determine how changes in protein copy numbers change the maximum amount of RAP1-GTP seen in simulations.
All protein copy numbers in the model were selected from the experimental literature \cite{Zeiler2014}, and each copy number was increased or decreased by one standard deviation from the mean.

\begin{figure*}[htbp!]
\centering
     \includegraphics{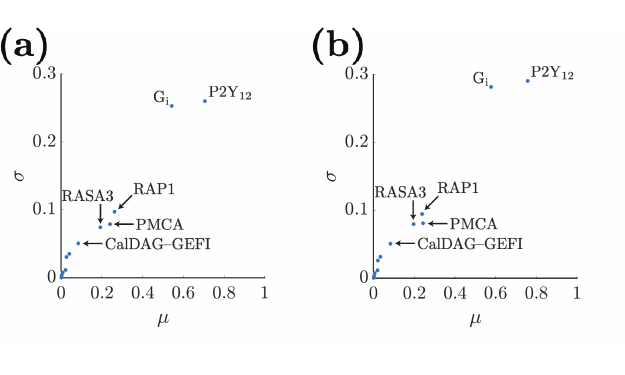}
        \caption{\textbf{Sensitivity analysis reveals agreement between model outputs using either parameter set.} 
        Sensitivity of the maximum RAP1 response to protein copy numbers. 
        Protein copy numbers are adjusted one at a time within a standard deviation of the literature value. 
        For each adjustment made, the ratio of the relative change in maximum RAP1-GTP over the relative change in copy number is computed. The process is repeated for each protein at many points in parameter space. The ratios' average ($\mu$) and standard deviation ($\sigma$) are presented. (a) Sensitivity of the model when using the lower bound estimated parameters as shown in Figure~\ref{fig:wholeCell}b. (b) Sensitivity of the model when using the upper bound estimated parameters as shown in Figure~\ref{fig:wholeCell}c. While thirteen proteins were examined for this study, only the proteins with a significant sensitivity were explicitly labeled.}
        \label{fig:concSensitivity}
\end{figure*}

Figure~\ref{fig:concSensitivity} shows the sensitivity of the maximum RAP1 response to the protein copy numbers.
Comparing the two subfigures, we see little difference in the sensitivity of the model to copy numbers.
We also note that in both figures, the same proteins have the most sensitivity, namely G$_{\rm i}$, P2Y$_{12}$, PMCA, CalDAG-GEFI, and RASA3, meaning that a change in copy number of these proteins should yield similar relative responses.
This served as evidence that under the appropriate normalization, the results generated by one estimated parameter set are similar to those generated by the other. 
We therefore present results using only our upper bound estimate parameter set for the remainder of this work.

Moreover, Figure~\ref{fig:concSensitivity} shows that the peak level of RAP1-GTP was most sensitive to changes in the P2Y$_{12}$ receptor and G$_{\rm i}$ protein. 
This implies that the variability in expression of P2Y$_{12}$ or G$_{\rm i}$ had the most significant effect on RAP1 activation. 
In particular, equivalent variability in the expression of RASA3, the immediate effector of RAP1, did not produce the same variation in the RAP1-GTP signal.

\begin{figure*}[htbp!]
\centering
     \includegraphics{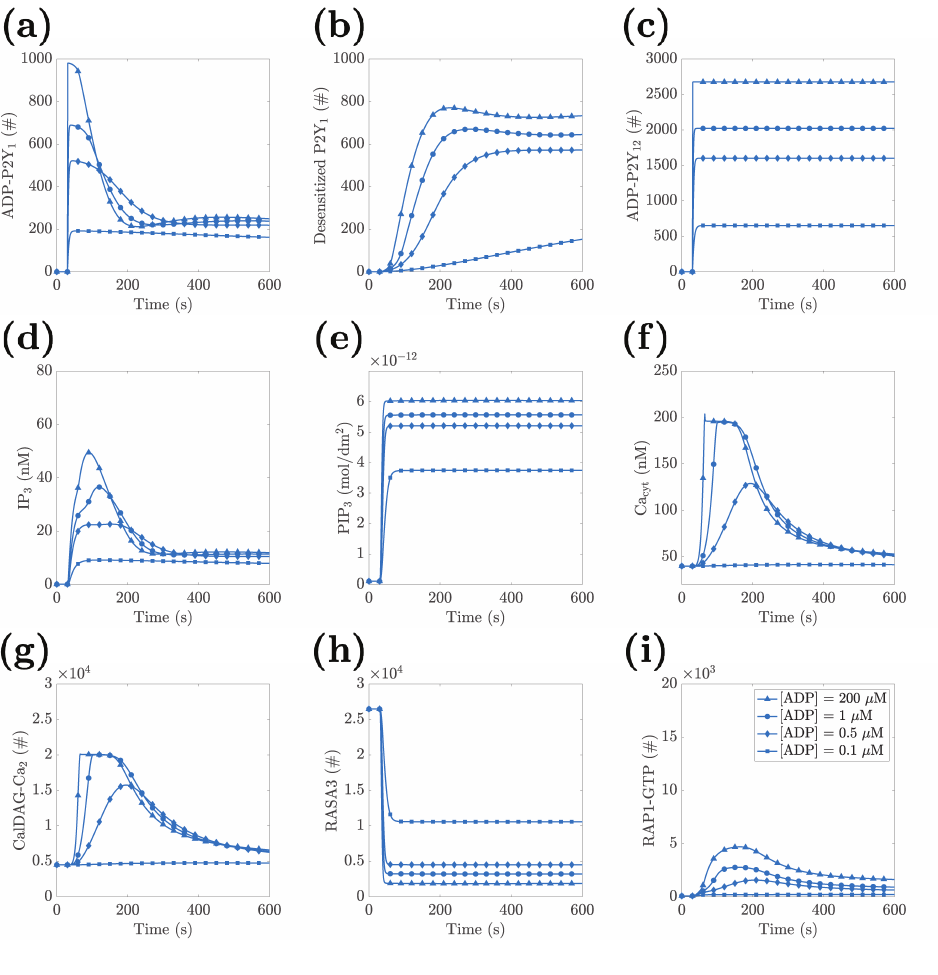}
        \caption{\textbf{Model predicts transient activation of RAP1-GTP across ADP concentrations.}
        At $t=30$ s, a specified concentration of ADP was instantaneously added to the extracellular space. 
        The curves shown are for ADP at 0.1 $\mu$M (squares), 0.5 $\mu$M (diamonds), 1 $\mu$M (circles), and 200 $\mu$M (triangles). 
        (a) the total number of P2Y$_1$ receptors bound to ADP and not desensitized, (b) the total number of desensitized P2Y$_1$ receptors, (c) the total number of P2Y$_{12}$ receptors bound to ADP, (d) the total concentration of IP$_3$, (e) the total surface density of PIP$_3$, (f) the concentration of free cytosolic calcium, (g) the number of CalDAG-GEFI molecules bound with two calcium ions, (h) the number of RASA3 not bound to PIP$_3$, and (i) the number of RAP1 bound to GTP. See Supplemental Section S2 for details on which model equations were used in calculating the total number or concentration.}
        \label{fig:ADPLoop}
\end{figure*}

\paragraph{Model predicts saturating integrin response to ADP:}
We then examined how our simulated platelet responds to a range of applied agonist concentrations.
Figure~\ref{fig:ADPLoop} shows how the numbers or concentrations of various chemicals changed over time for a given application of ADP at $t=30$ s.
Figures~\ref{fig:ADPLoop}a and~\ref{fig:ADPLoop}c show that immediately upon application of ADP into the system, there was a rapid rise in the number of P2Y$_1$ and P2Y$_{12}$ receptors, respectively, bound with ADP. 
When 200 $\mu$M of ADP was applied, essentially all P2Y$_1$ and P2Y$_{12}$ receptors were bound to ADP at the peak. 
Due to PKC's desensitization of the P2Y$_1$ receptor, the number of active P2Y$_1$ receptors decreased from its peak value as the simulation progressed.
Interestingly, the number of active P2Y$_1$ receptors decreased most rapidly for the highest concentration of ADP. 
This decrease was matched by a fast rise in the number of desensitized P2Y$_1$ receptors, as shown in Figure~\ref{fig:ADPLoop}b.
Therefore, as the concentration of ADP applied increased, the maximum response increased, but the time interval on which a near-peak level was maintained decreased.

Following receptor activation, second messengers PLC and PI3K became active by binding the appropriate G protein, leading to a rise in the production of the inositol species IP$_3$ and PIP$_3$, respectively. 
Figure~\ref{fig:ADPLoop}d shows that the IP$_3$ concentration rose with a slight delay compared to the receptor concentrations.
Later in the simulations, as the P2Y$_1$ receptors become desensitized, the rate of formation of IP$_3$ decreased, leading to a decline in its concentration.

IP$_3$ then activated IP$_3$R, which led to an influx of calcium into the cytosol from the DTS.  
As seen in Figure~\ref{fig:ADPLoop}f, the level of free cytosolic calcium increased, plateaued, and then decreased throughout the simulation.
The rise in cytosolic calcium levels led to the activation of proteins PKC and CalDAG-GEFI, the latter of which is seen in Figure~\ref{fig:ADPLoop}g following a similar pattern of rise, plateau, and decrease as the corresponding calcium trace.

Concurrently with IP$_3$, calcium, and CalDAG-GEFI spiking, the P2Y$_{12}$ pathway led to the formation of the PIP$_3$ inositol species as seen in Figure~\ref{fig:ADPLoop}e, followed by the inactivation of RASA3, as seen in Figure~\ref{fig:ADPLoop}h.
Unlike in the P2Y$_1$ pathway, the lack of a desensitization mechanism for P2Y$_{12}$ meant that PIP$_3$ was constantly created by PI3K during the simulation, leading to a permanent decrease in the number of active RASA3 molecules.

The activation of CalDAG-GEFI and inactivation of RASA3 caused an increase in the amount of RAP1 bound to GTP, shown in Figure~\ref{fig:ADPLoop}i.
When the CalDAG-GEFI levels decreased, RAP1-GTP levels also decreased and approached their baseline levels.
As shown more explicitly in Figure~\ref{fig:ADPvsMRS}, the model's response to ADP saturated around 10 $\mu$M, so the 200 $\mu$M ADP simulated experiment represents the maximum ADP response.

\paragraph{RAP1 activation response increases with applied agonist:} 
We next attempted to validate our model's output by comparison to data previously published \cite{Paul_Blatt_Schug_Clark_etAl}.
To do this, we simulated platelets' exposure to ADP, which binds to both P2Y$_1$ and P2Y$_{12}$ receptors, and platelets' exposure to the synthetic agonist MRS2365, which binds only to the P2Y$_1$ receptor.
To accommodate the synthetic agonist, the model was changed in two ways: 1) by setting the binding rate between agonist and P2Y$_{12}$ to zero, i.e., $k_{\rm ADP}^{\rm P2Y_{12}}=0$, and 2) by lowering the dissociation constant for the agonist binding to P2Y$_1$ from 0.6 to 0.01 $\mu$M, i.e. $K_{\rm ADP}^{\rm P2Y_1} = 0.01\ \mu$M.

\begin{figure*}[htbp!]
\centering
     \includegraphics{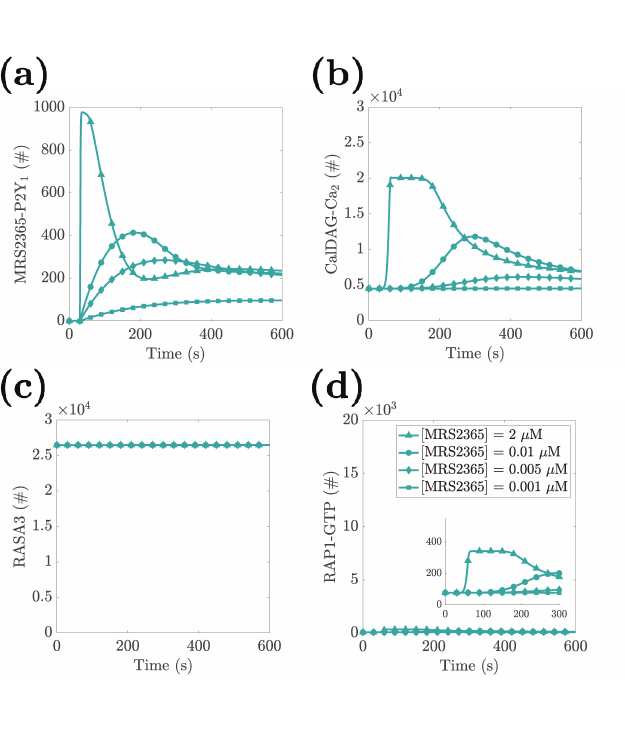}
        \caption{\textbf{Simulated platelets exposed to MRS2365 show significantly reduced RAP1-GTP response.} 
        At $t=30$ s, a specified concentration of MRS2365 was instantaneously added to the extracellular space. 
        The curves shown are for MRS2365 at 0.001 $\mu$M (squares), 0.005 $\mu$M (diamonds), 0.01 $\mu$M (circles), and 2 $\mu$M (triangles). 
        (a) the total number of P2Y$_1$ receptors bound to MRS2365 and not desensitized, (b) the number of CalDAG-GEFI molecules bound with two calcium ions, (c) the number of RASA3 not bound to PIP$_3$, and (d) the number of RAP1 bound to GTP. See Supplemental Section S2 for details on which model equations were used in calculating the total number or concentration.}
        \label{fig:MRSLoop}
\end{figure*}

Figure~\ref{fig:MRSLoop} shows how WT platelets respond to the synthetic agonist.
We found that the application of a saturating amount of MRS2365 yielded similar levels of P2Y$_1$ and CalDAG-GEFI activation compared to those evoked by the application of a saturating amount of ADP (compare Figure~\ref{fig:MRSLoop}a with Figure~\ref{fig:ADPLoop}a and Figure~\ref{fig:MRSLoop}b with Figure~\ref{fig:ADPLoop}g).
However, because the concentration of RASA3 remained high throughout the MRS2365 simulations, RAP1 activation was greatly limited.

\begin{figure*}[htbp!]
\centering
     \includegraphics{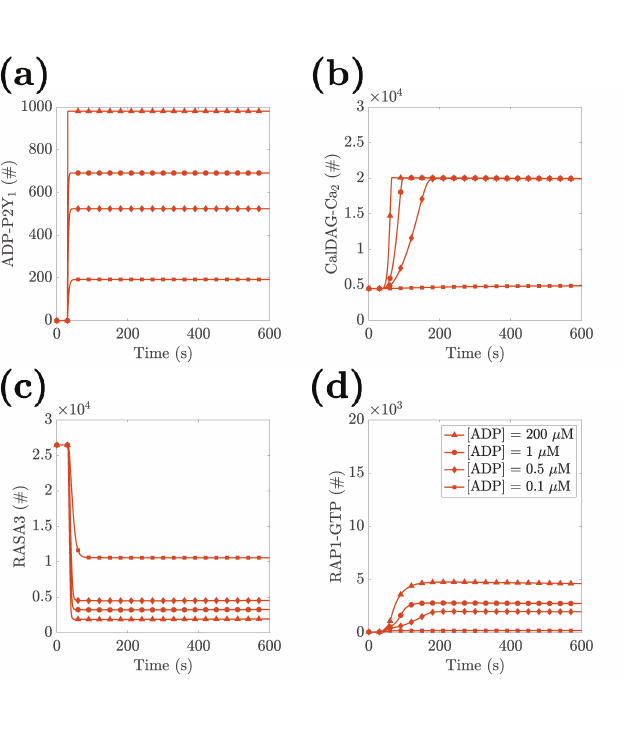}
     \caption{\textbf{Simulated platelets lacking PKC-dependent P2Y$_1$ desensitization show a significantly prolonged RAP1-GTP signal.}
     At $t=30$ s, a specified concentration of ADP was instantaneously added to the extracellular space of P2Y$_1^{\rm 340-0P/340-0P}$ platelets. 
     The curves shown are for ADP at 0.1 $\mu$M (squares), 0.5 $\mu$M (diamonds), 1 $\mu$M (circles), and 200 $\mu$M (triangles). 
        (a) the total number of P2Y$_1$ receptors bound to ADP, (b) the number of CalDAG-GEFI molecules bound with two calcium ions, (c) the number of RASA3 not bound to PIP$_3$, and (d) the number of RAP1 bound to GTP. See Supplemental Section S2 for details on which model equations were used in calculating the total number or concentration.}
        \label{fig:ADPLoop_340-0P}
\end{figure*}

\begin{figure*}[htbp!]
\centering
     \includegraphics{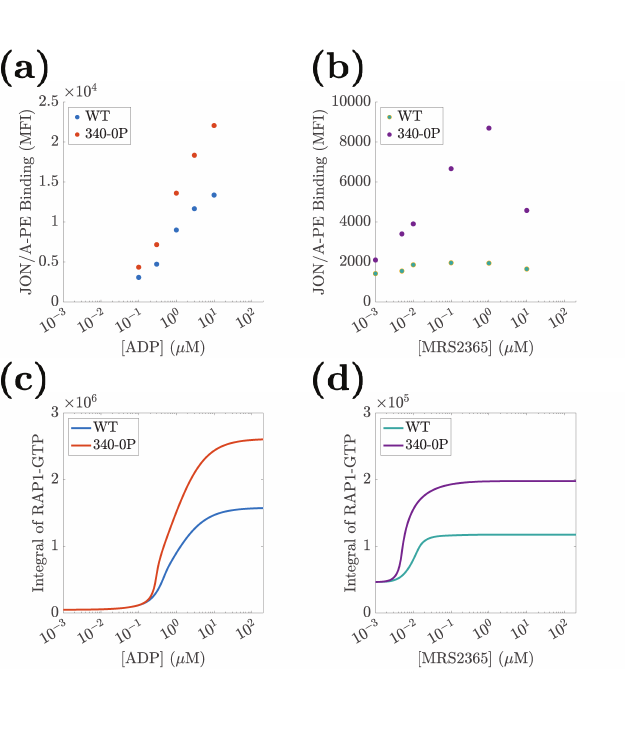}
     \caption{\textbf{Total integrin response displays a saturating response to two types of P2Y agonists.}
        Wildtype and P2Y$_1^{\rm 340-0P/340-0P}$ platelets are stimulated with a specified concentration of agonists for 10 minutes, and the time-integral of the RAP1-GTP curve is recorded.
        Subfigures (a) and (b) are recreated from \cite{Paul_Blatt_Schug_Clark_etAl} Figures 2c and 4c.
        (a) Mean Fluorescence Intensity of JON/A-PE binding to active $\alpha_{\rm IIb}\beta_3$ integrins after a specified concentration of ADP is applied. 
        (b) Mean Fluorescence Intensity of JON/A-PE binding to active $\alpha_{\rm IIb}\beta_3$ integrins after a specified concentration of MRS2365 is applied. 
        (c) Platelets are exposed to ADP, and (d) platelets are exposed to P2Y$_1$ agonist MRS2365.
        Note the differences in vertical scales.}
        \label{fig:ADPvsMRS}
\end{figure*}

We next considered platelets with a mutation to the intracellular tail of their P2Y$_1$ receptor, which makes it unable to be phosphorylated and therefore the receptor is prevented from being desensitized to agonists and being taken up into the cell \cite{Paul_Blatt_Schug_Clark_etAl}.  
These platelets are known as P2Y$_1^{\rm 340-0P/340-0P}$ platelets or simply 340-0P. 
Figure~\ref{fig:ADPLoop_340-0P} shows how a 340-0P platelet responds to ADP.
While the peak value of various proteins was similar to the wildtype case seen in Figure~\ref{fig:ADPLoop}, the simulated mutant platelet's P2Y$_1$ signal never declined, and consequently the downstream calcium, CDGI, and RAP1 signals also did not decline.

To gauge variations in platelet responses due to variations in agonist concentration, we performed computational experiments for a wide range of applied ADP and MRS2365 concentrations. 
In the corresponding physical experiments, the readout was the amount of the integrin probe JON/A bound irreversibly to activated integrin. 
As a proxy for the amount of JON/A bound at time $t$, we used the time-integral of the RAP1-GTP curve up to that time. 
Figure~\ref{fig:ADPvsMRS} shows the results of these experiments. 
As expected, for both agonists, the RAP1-GTP response increased as the concentration of the agonist was increased.  
This response was consistently greater for 340-0P platelets than for WT ones. 
This is attributed to the fact that for the P2Y$_1^{\rm 340-0P/340-0P}$ platelets, the P2Y$_1$ receptors remained active and continued to perform their GEF action on the G$_{\rm q}$ protein for the duration of the simulation. 
Consequently, the IP$_3$, calcium, and RAP1-GTP responses are prolonged, and therefore, the area under the RAP1-GTP curve increased.

Two differences were evident in how both wildtype and 340-0P platelets responded to the two stimuli. 
First, the response to MRS2365 began at lower concentrations than that to ADP, consistent with the higher affinity of MRS2365 for the P2Y$_1$ receptor. 
Second, the response to a saturating concentration of MRS2365 was lower than for a saturating concentration of ADP, consistent with unabated RASA3 GAP activity in the MRS2365-stimulated platelets.

We compared the results of Figures 2c and 4c in \cite{Paul_Blatt_Schug_Clark_etAl}, recreated in Figure~\ref{fig:ADPvsMRS}a and~\ref{fig:ADPvsMRS}b, to the results of Figures~\ref{fig:ADPvsMRS}c and~\ref{fig:ADPvsMRS}d.
First, comparing the experiments in which platelets are stimulated with ADP in Figure~\ref{fig:ADPvsMRS}a and~\ref{fig:ADPvsMRS}c, we saw an increase in the integrin response over the same range of applied ADP concentration for both wildtype and P2Y$_1^{\rm 340-0P/340-0P}$ platelets.
Additionally, we saw that for a given applied ADP concentration, the 340-0P platelet signal was consistently larger than the wildtype platelet.
We note that in Figure~\ref{fig:ADPvsMRS}b and~\ref{fig:ADPvsMRS}d, we saw a similar relationship between mutant and wildtype platelets.
Deviations between the model and experiment are evident when comparing simulated versus experimental mutant platelets stimulated with MRS2365, where the experiments show a more gradual increase in integrin activation as a function of applied agonist concentration compared to simulations.
Comparing our model to results from \cite{Paul_Blatt_Schug_Clark_etAl} showed good agreement, even when a single experimental modification is considered.
As more modifications are combined into simulated platelets, the results may deviate from experiments.

\paragraph{Changes to protein copy number in the P2Y$_{12}$ pathway affect RAP1-GTP response more than in the P2Y$_1$ pathway:} 
The model allowed for the continuous variation of parameters, specifically protein copy numbers.
We leveraged this to assess the effects of protein expression levels on the integrin activation response \textcolor{black}{over a range of ADP stimuli.}  
In each set of simulations, we set the copy number of one of the proteins RASA3, CDGI, P2Y$_1$, or P2Y$_{12}$ to a value between 5\% and 150\% of its literature value (while holding other copy numbers at their literature values). 

\begin{figure*}[htbp!]
\centering
     \includegraphics{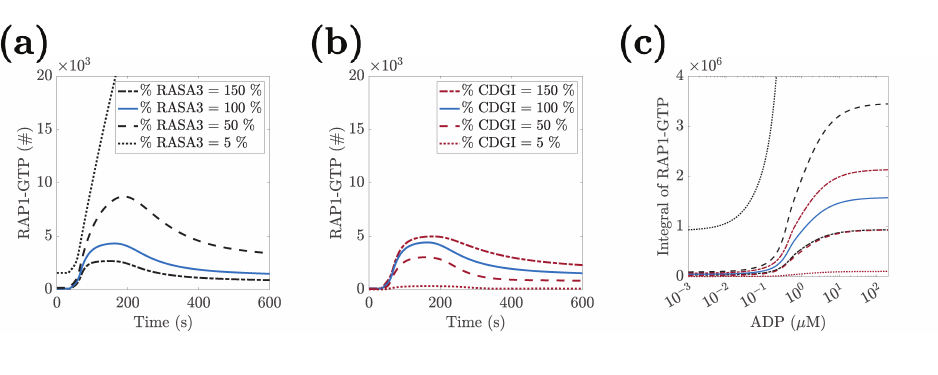}
        \caption{\textbf{Total integrin response is more sensitive to RASA3 levels than CalDAG-GEFI levels.} In each simulation, the starting copy number of one protein is varied between 5\% and 150\% of the literature value before applying 10 $\mu$M ADP. 
        \textcolor{black}{In all figures, the solid blue curve indicates the WT case.} 
        (a) Timecourse of RAP1-GTP levels in platelets expressing a specified percentage of the base RASA3 copy number. (b) Timecourse of RAP1-GTP levels in platelets expressing a certain percentage of the base CalDAG-GEFI copy number. (c) Integral of the RAP1-GTP curve \textcolor{black}{as a function of ADP concentration applied to platelets with a specified change to a protein's expression level. Legends in (a) and (b) apply to (c).}}
        \label{fig:proteinvsprotein}
\end{figure*}

Figure~\ref{fig:proteinvsprotein} shows the results of varying the RASA3 or CDGI copy numbers.  
We saw that a decrease in the number of RASA3 molecules or an increase in the number of CDGI molecules elicits an increase in the RAP1-GTP response.
Figure~\ref{fig:proteinvsprotein}a shows that when the copy number of RASA3 is decreased by 50\%, integrin activation was stronger and more prolonged than the same simulations with 100\% RASA3 expression.  
In contrast, Figure~\ref{fig:proteinvsprotein}b shows that increasing the amount of CDGI by 50\% did not significantly change the peak level but did increase the length of the RAP1-GTP response. 

\textcolor{black}{Figure~\ref{fig:proteinvsprotein}c plots the RAP1-GTP response of platelets with specified expression level changes as a function of ADP concentration.
We found that as we vary the expression of RASA3, the dose-response curve deviates further from the WT curve than for equivalent variations in the expression of CDGI. 
Furthermore, the variation in dose-response occurs across applied ADP concentrations.
For example, at 10 nM ADP, there are modest increases in the integrin response as we vary expression levels within 50\% of baseline and much more dramatic increases as we decrease RASA3 expression down to 5\%.}
\textcolor{black}{This result mirrors behaviors seen experimentally in Stefanini et al. \cite{Stefanini2015}, where a specific mouse platelet mutation resulting in 5\% RASA3 expression led to an increase in both the baseline and stimulated integrin response.}
\textcolor{black}{When we examined the integrin response as a function of expression level, we found that the dose-response curve scales approximately linearly with changes in CDGI expression (Supplemental Figure~S1b). 
In contrast, platelet response as a function of RASA3 expression is highly nonlinear, with a dramatic dose-response curve as the RASA3 copy number decreases to and below 50\% of its baseline value.}

\begin{figure*}[htbp!]
\centering
     \includegraphics{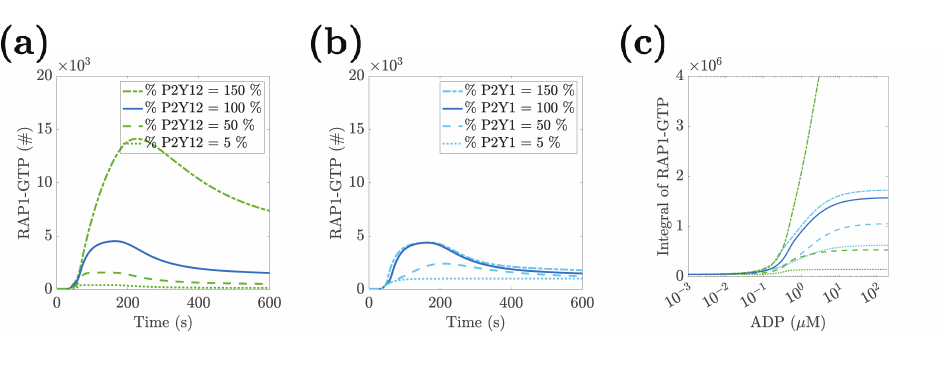}
        \caption{\textbf{Total integrin response is more sensitive to P2Y$_{12}$ levels than to P2Y$_1$ levels.} In each simulation, the starting copy number of one protein is varied between 5\% and 150\% of the literature-derived copy number before the application of 10 $\mu$M ADP. 
        \textcolor{black}{In all figures, the solid blue curve indicates the WT case.}
         (a) Timecourse of RAP1-GTP levels in platelets expressing a specified percentage of the base P2Y$_{12}$ copy number. (b) Timecourse of RAP1-GTP levels in platelets expressing a certain percentage of the base P2Y$_1$ copy number. (c) Integral of the RAP1-GTP curve \textcolor{black}{as a function of ADP concentration applied to platelets with a specified change to a protein's expression level. Legends in (a) and (b) apply to (c).}}
        \label{fig:receptorvsreceptor}
\end{figure*}

\textcolor{black}{We next were interested in determining how platelets would respond to ADP when their receptor expression levels varied. 
In Figure~\ref{fig:receptorvsreceptor}, we vary the expression of P2Y$_{1}$ and P2Y$_{12}$ individually to see the platelet's response to ADP over time and the total dose response.
We expected that the integrin response should be positively correlated with the expression levels of either receptor, and Figures~\ref{fig:receptorvsreceptor}a-b agree.
Examining model outputs further verifies that the calcium timecourse is affected by variation in P2Y$_1$ expression levels but not by variation in P2Y$_{12}$ expression levels (data not shown).
Figure~\ref{fig:receptorvsreceptor}c shows the dose-response curves for different receptor expression levels.
Similar to RASA3 in Figure~S1b, we found that variation in P2Y$_{12}$ yielded larger changes to the dose-response curve than in P2Y$_1$.}

\textcolor{black}{Interestingly, Figures~\ref{fig:receptorvsreceptor}b-c show that an increase in the P2Y$_1$ receptor to 150\% leads to an almost negligible change in integrin response.
Further analysis of model outputs shows that although increasing P2Y$_{1}$ expression led to an increase in peak IP$_3$ concentrations, the peak concentration of calcium did not increase significantly beyond 200 nM (data not shown).
We show in Supplemental Figure~S2a that the relative integrin response between 100\% and 150\% P2Y$_1$ expression is more significant when 0.5 $\mu$M ADP is used due to a difference in peak cytosolic calcium levels.
Similar results were seen experimentally in Heckler et al. \cite{hechler2003lineage}, in which they overexpress P2Y$_1$ by 84\% and notice an increase in aggregation response to low doses of ADP.}

\section*{Conclusion}

In this report, we developed a mathematical model of inside-out integrin activation on a platelet surface and fitted it to experimental data. 
Thanks to years of experimental studies \cite{Lee_ch18}, a chemical pathway has been defined sufficiently to allow such a model to exist.
Previous platelet activation models rely on other major indicators, such as calcium signals or aggregometry readings \cite{LeeDiamond2015, AleksandraEtAl2019}.
While effective, inferring integrin activation information from only one of the two ADP-dependent pathways does not fully encapsulate the regulatory pathway's complexity, making it difficult to extend to particular mutant platelets.
Our model can be used in testing and making predictions without the need for a large number of experimental trials.
It is quite economical to run; for reference, a single simulation takes approximately one minute to run ten simulated minutes on a standard personal computer.
Thus, data generation is significantly faster than generation of an equivalent amount of data in the lab.

\textcolor{black}{The results in this work rely on a combination of kinetic rates found in the literature or estimated from data.
Copy numbers of each protein were based on proteomics analysis of mouse platelets \cite{Zeiler2014}.
In some cases, due to a lack of available data, rate constants associated with a protein isoform found in human platelets were used \cite{KANOH1990120,MATZARIS19943397,ATACK1993305,MITCHELL19898873,VARGAS1984123}.
This choice allows us to estimate fewer parameters under the assumption that the action of the protein isoform found in mouse platelets acts similarly on its substrate compared to these human platelet isoforms.}

While our estimated parameters yielded a model that matched our experimental dataset from the point of ADP addition up to the peak, our model tended to lengthen the duration of the RAP1-GTP signal.
Under the assumption that our estimated parameters are the ideal parameters for the model, a mechanism for desensitization of some signaling protein in the P2Y$_{12}$ pathway, leading to a recovery of RASA3 levels, could shorten the RAP1-GTP signal duration.
Without sufficient evidence for a biological mechanism, we save modeling this portion of the pathway for future work.

In experiments where we varied the applied ADP concentration, we found that a platelet's response saturates beyond 10 $\mu$M, consistent with experiments.
Our highly detailed model allowed us to examine the individual responses of each protein in time.
We showed the usefulness of our model beyond directly comparing to experimental data by varying parameters that would be difficult to vary continuously in the lab, like protein expression levels, and making predictions on integrin activation.
In particular, we showed that the total integrin response varies more significantly with changes to proteins in the P2Y$_{12}$ pathway compared to the P2Y$_1$ pathway.
\textcolor{black}{By decreasing the expression level of RASA3 to well below 50\%, we found that both a resting and stimulated platelet presented a marked increase in integrin response, similar to results seen in Stefanini et al. \cite{Stefanini2015}.}
\textcolor{black}{By changing expression levels of the receptors, we show that the integrin response can be modulated only above 0.1 $\mu$M ADP.
In Supplemental Figure~S2a, we show that when using 50 nM ADP, we start to see a change in the integrin response between 100\% and 150\% expression of P2Y$_1$, akin to the results of Hechler et al \cite{hechler2003lineage} in which they overexpress P2Y$_1$.}

\textcolor{black}{The results from this study can be used to understand how certain pharmacological agents affect the integrin activation process. 
In particular, the P2Y$_{12}$ antagonist clopidogrel is taken long-term as an antithrombotic treatment, and its action is analogous to a reduction in the overall expression of P2Y$_{12}$. 
More detailed modeling is needed to show this connection explicitly, and will be the subject of future work.}

This model can be used to conduct initial tests on the behavior of platelets under any form of experimental condition before running experiments.
In addition, the outputs of this model describe integrin activation over time in far more detail than is typical in mathematical models for hemostasis and thrombosis \cite{DuFogelson2017, XuEtAl2008}, where the tracking of platelet aggregation and fibrin formation is of interest.
This may lead to improvements in these larger-scale models.

\section*{Author Contributions}

Patel and Fogelson constructed the mathematical model and wrote the article. Patel carried out all simulations. Bergmeier conducted all experiments.

\section*{Declaration of Interests}

The authors declare no competing interests.

\section*{Acknowledgments}

This work was supported by the National Institutes of Health, National Heart, Lung, and Blood Institute (grant R35 HL144976 to WB and R01 HL151984 to AF), as well as the National Science Foundation's Graduate Research Fellowship to KP.

\bibliography{main}

\pagebreak

\section*{Figure Legends}

\end{document}


\beginsupplement

\tableofcontents

\pagebreak

\section{Comparing integral of RAP1-GTP to JON/A binding assay}

Given that we already assume a one-to-one stoichiometric relationship between RAP1-GTP and integrin, we can write down the following ODE for the rate at which integrin binds to JON/A:
\begin{equation}
\frac{d}{dt}[{\rm integrin-probe}]_S = k[{\rm probe}][{\rm RAP1-GTP}]_S.
\end{equation}
Assuming that 1) ADP is instantaneously added to the extracellular space at $t=30$ s and 2) the total amount of probe is large compared to the number of integrins, we can explicitly integrate the equation over 600 s to find that:
\begin{equation}
[{\rm integrin-probe}]_S(630) = k[{\rm probe}]\int_{30}^{630}[{\rm RAP1-GTP}]_S(t)dt.
\end{equation}
Thus, to find the amount of integrin bound to probe, we simply integrate the RAP1-GTP curve.

\section{Grouped quantities reported in main text}
\label{groupedQuantities}

In some cases in our model, a chemical species of interest (say, the $\rm \pTWOyONE$ receptor), is further split into multiple states.
If we would like to know the total number of receptors with an ADP bound to it at time $t$, we sum up the amounts from each state and report the summed quantity. 
In particular, we make use of the following sums:
\begin{itemize}
\item $\rm ADP-\pTWOyONE\ =\ \surf{ADP-\pTWOyONE}+\surf{ADP-\pTWOyONE-\gq GDP}+\surf{ADP-\pTWOyONE-\gq GTP}+\surf{ADP-\pTWOyONE-\gq}$ 
\item $\rm ADP-\pTWOyTWELVE\ =\ \surf{ADP-\pTWOyTWELVE}+\surf{ADP-\pTWOyTWELVE-\gi GDP}+\surf{ADP-\pTWOyTWELVE-\gi GTP}+\surf{ADP-\pTWOyTWELVE-\gi}$ 
\item $\rm Desensitized\ \pTWOyONE\ = \surf{ADP-p\pTWOyONE}+\surf{ADP-p\pTWOyONE-\gq GDP}$ 
\item $\rm \ipTHREE\ =\ \conc{\ipTHREE}+\frac{S_{\rm IM}}{V_{\rm cyt}}\Big(\surf{\ipTHREEr_o}+\surf{\ipTHREEr_a}+\surf{\ipTHREEr_{i2}}+\surf{\ipTHREEr_s}\Big)$
\end{itemize}

\section{Supplemental Figures}
\label{supplementalFigures}

\begin{figure*}[htbp!]
     \centering
     \begin{subfigure}[b]{.3\textwidth}
         \centering
         \caption{}
         \includegraphics[trim={3cm 6.5cm 3.5cm 6.5cm},clip,width=\textwidth]{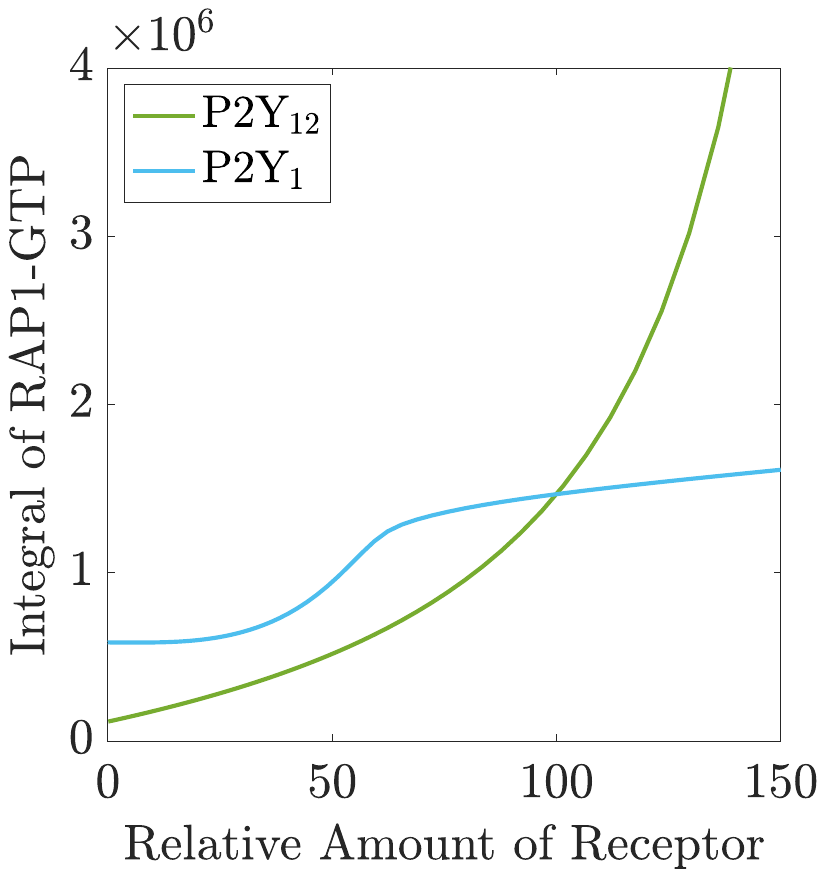}
         \label{fig:p2y1vsp2y12}
     \end{subfigure}
     \hspace{48pt}
     \begin{subfigure}[b]{.3\textwidth}
         \centering
         \caption{}
         \includegraphics[trim={3cm 6.5cm 3.5cm 6.5cm},clip,width=\textwidth]{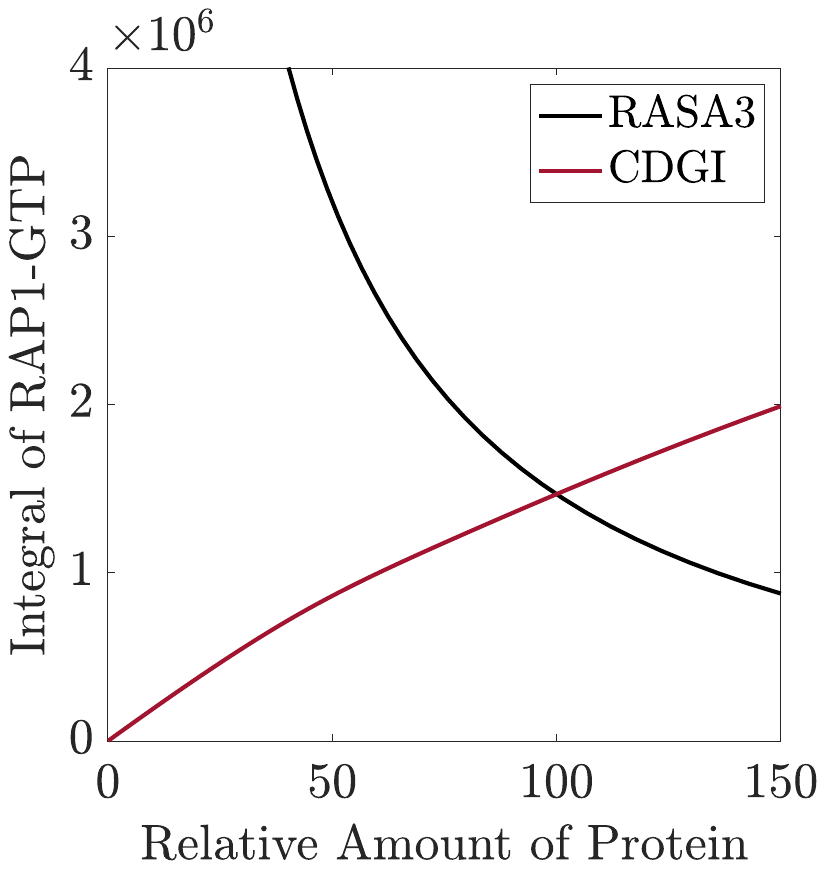}
         \label{fig:rasa3vscdgi}
     \end{subfigure}
        \caption{\textbf{Integrin response as a function of protein expression.} For each curve, the expression level of a particular protein is varied then the platelet is simulated with the application of 10 $\mu$M ADP.
        (a) Receptor expression levels are varied. (b) GAP/GEF protein expression levels are varied.}
        \label{fig:expressionLevels}
\end{figure*}

\begin{figure*}[htbp!]
     \centering
     \begin{subfigure}[b]{.3\textwidth}
         \centering
         \caption{}
         \includegraphics[trim={3cm 6.5cm 3.5cm 6.5cm},clip,width=\textwidth]{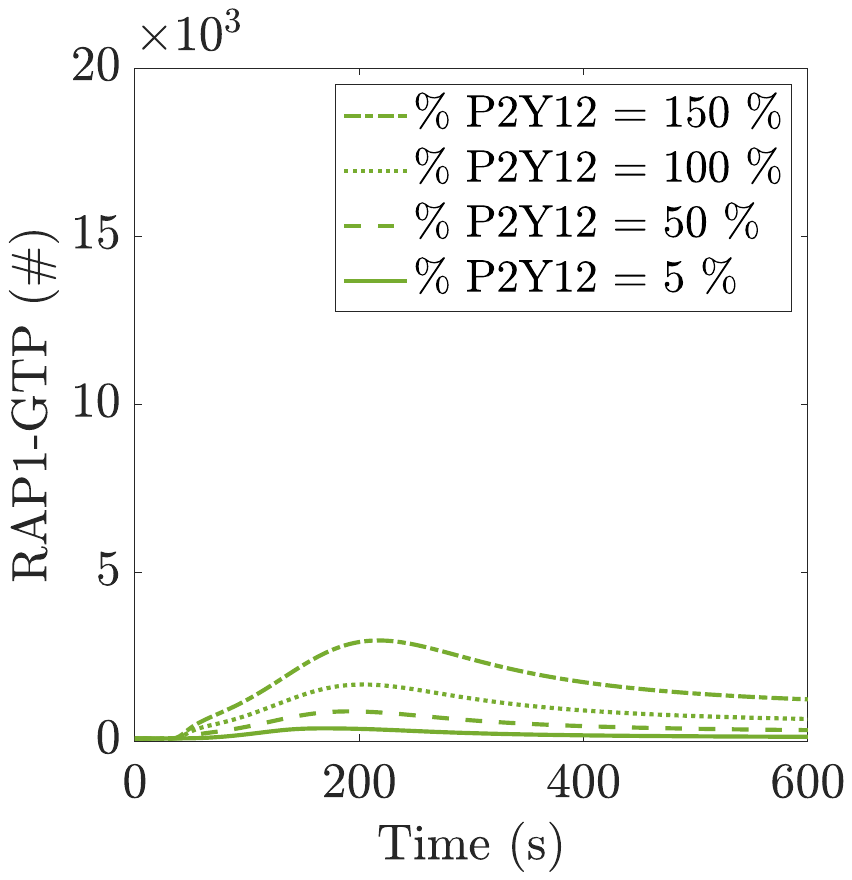}
         \label{fig:p2y1_adp_pt5}
     \end{subfigure}
     \hspace{48pt}
     \begin{subfigure}[b]{.3\textwidth}
         \centering
         \caption{}
         \includegraphics[trim={3cm 6.5cm 3.5cm 6.5cm},clip,width=\textwidth]{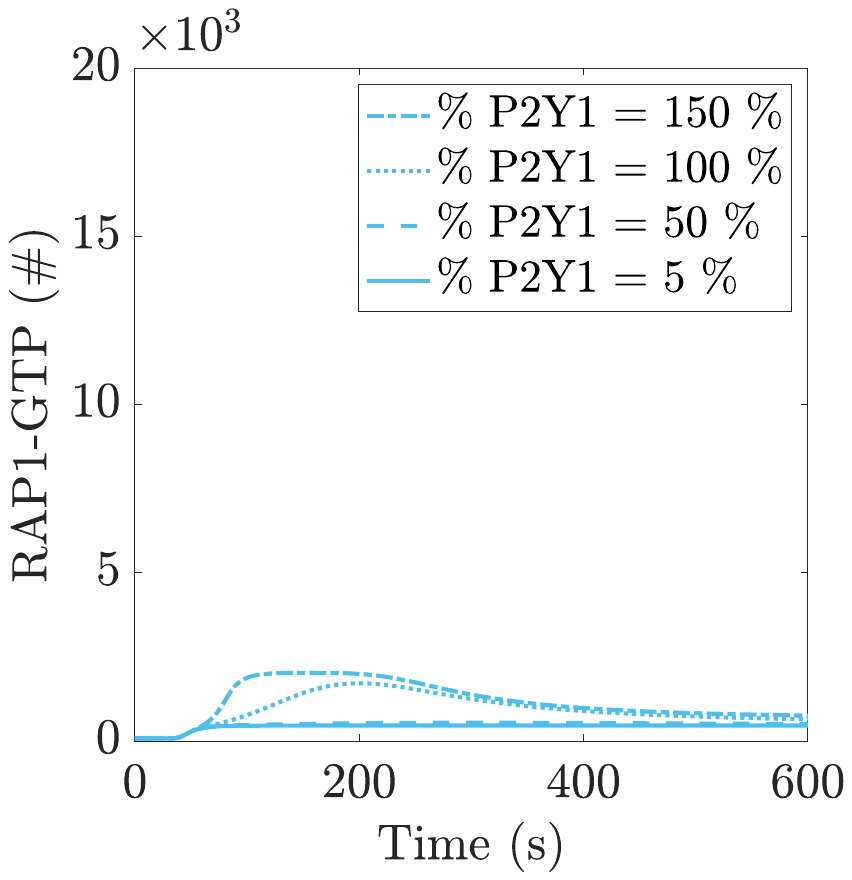}
         \label{fig:p2y12_adp_pt5}
     \end{subfigure}
        \caption{\textbf{Total integrin response is more sensitive to P2Y$_{12}$ levels than to P2Y$_1$ levels.} In each simulation, the starting copy number of one protein is varied between 5\% and 150\% of the literature-derived copy number before the application of 0.5 $\mu$M ADP. 
         In all figures, the solid blue curve indicates the WT case.
         (a) Timecourse of RAP1-GTP levels in platelets expressing a specified percentage of the base P2Y$_{12}$ copy number. (b) Timecourse of RAP1-GTP levels in platelets expressing a certain percentage of the base P2Y$_1$ copy number.}
        \label{fig:receptorvsreceptor_adp_pt5}
\end{figure*}

\section{Estimated Parameters}
\begin{longtable}{| c  c  c  c |}
   \hline
      Parameter & Value & Value & Units \\ 
      & (Max Calcium 100 nM) & (Max Calcium 200 nM) & \\ \hline
$\bigk{act}{\pTWOyONE}$ \& $\bigk{act}{\pTWOyTWELVE}$		& $4.906\cdot10^{-3}$	& $4.806\cdot10^{-3}$ & n.d. \\ \hline
$\alphaS{\pTWOyONE}$ \& $\alphaS{\pTWOyTWELVE}$			& $1.782$                  & $1.782$ & n.d. \\ \hline
$\betaS{\pTWOyONE}$ \& $\betaS{\pTWOyTWELVE}$			& $8.573$                  & $8.508$ &  n.d. \\ \hline
$\deltaS{\pTWOyONE}$ \& $\deltaS{\pTWOyTWELVE}$			& $1.46\cdot10^{1}$       & $1.466\cdot10^{1}$ & n.d. \\ \hline
$\gammaS{\pTWOyONE}$ \& $\gammaS{\pTWOyTWELVE}$		& $2.324\cdot10^{1}$      & $2.361\cdot10^{1}$ & n.d.\\ \hline
$\smallk{\gq GDP}{\pTWOyONE}$ \& $\smallk{\gi GDP}{\pTWOyTWELVE}$		& $6.947\cdot10^{13}$     & $6.947\cdot10^{13}$ & dm$^2$/mol-s \\ \hline
$\smallk{-\gq GDP}{\pTWOyONE}$ \& $\smallk{-\gi GDP}{\pTWOyTWELVE}$		& $5.858\cdot10^{3}$     & $5.858\cdot10^{3}$ & 1/s \\ \hline
$\smallk{ADP}{\pTWOyONE}$ \& $\smallk{ADP}{\pTWOyTWELVE}$			& $5.716\cdot10^{5}$      & $5.983\cdot10^{5}$ & 1/M-s \\ \hline
$\smallk{-ADP}{\pTWOyONE}$ \& $\smallk{-ADP}{\pTWOyTWELVE}$			& $0.362$      & $0.362$ & 1/s \\ \hline
$\smallk{-GDP}{\pTWOyONE}$ \& $\smallk{-GDP}{\pTWOyTWELVE}$		& $2.625\cdot10^{1}$      & $2.601\cdot10^{1}$ & 1/s \\ \hline
$\smallk{GDP}{\pTWOyONE}$ \& $\smallk{GDP}{\pTWOyTWELVE}$		& $1.248\cdot10^{6}$      & $1.30\cdot10^{6}$  & 1/M-s \\ \hline
$\smallk{-GTP}{\pTWOyONE}$ \& $\smallk{-GTP}{\pTWOyTWELVE}$                            & $8.057$                  & $8.077$ & 1/s \\ \hline
$\smallk{GTP}{\pTWOyONE}$ \& $\smallk{GTP}{\pTWOyTWELVE}$                              & $2.185\cdot10^{4}$      & $2.189\cdot10^{4}$ & 1/M-s \\ \hline
$\smallk{-\gq GTP}{\pTWOyONE}$ \& $\smallk{-\gi GTP}{\pTWOyTWELVE}$                      & $3.358\cdot10^{-1}$     & $3.315\cdot10^{-1}$ & 1/s \\ \hline
$\smallk{\gq GTP}{\pTWOyONE}$ \& $\smallk{\gi GTP}{\pTWOyTWELVE}$                         &  $2.65\cdot10^{15}$     & $2.65\cdot10^{15}$ & dm$^2$/mol-s \\ \hline
$\smallk{-\gqa GDP}{\plc}$ \& $\smallk{-\gia GDP}{\piTHREEk}$ 	& 	$1.247\cdot10^{1}$ 	&	$1.247\cdot10^{1}$		&	1/s \\ \hline
$\smallk{-\gqa GTP}{\plc}$ \& $\smallk{-\gia GTP}{\piTHREEk}$ 	& 	$1.207\cdot10^{-3}$ 	&	$1.207\cdot10^{-3}$		&	1/s \\ \hline
$\smallk{\gqa GDP}{\plc}$ \& $\smallk{\gia GDP}{\piTHREEk}$ 	& 	$6.098\cdot10^{11}$ &	$6.098\cdot10^{11}$		&	dm$^2$/mol-s \\ \hline
$\smallk{\gqa GTP}{\plc}$ \& $\smallk{\gia GTP}{\piTHREEk}$ 	& 	$6.044\cdot10^{16}$ &	$6.044\cdot10^{16}$		&	dm$^2$/mol-s \\ \hline
$\smallk{hydrolyze}{\plc}$ \& $\smallk{hydrolyze}{\piTHREEk}$                                & $1.293$                  & $1.189$ & 1/s \\ \hline
$\smallk{GTP}{}$                                                                      & $1.776\cdot10^{1}$      & $1.851\cdot10^{1}$ & 1/s \\ \hline
$\smallk{phos}{\pTWOyONE}$                                                          & $2.78\cdot10^{8}$       & $1.485\cdot10^{8}$ & 1/M-s \\ \hline
$\smallk{dephos}{\pTWOyONE}$                                                        & $1.313\cdot10^{-3}$     & $4.152\cdot10^{-3}$ & 1/s \\ \hline
$\gammaS{\ipTHREEr}$                                                                & $4.286\cdot10^{-10}$     & $4.171\cdot10^{-10}$ & S \\ \hline
$\gammaS{leak}$                                                                & $7\cdot10^{-12}$         & $7.026\cdot10^{-12}$ & S/dm$^2$ \\ \hline
$\smallk{2f}{\serca}$                                                           & $1.033\cdot10^{12}$      & $8.877\cdot10^{11}$ & 1/M$^2$-s \\ \hline
$\smallk{2r}{\serca}$                                                           & $1.375\cdot10^{2}$      & $1.368\cdot10^{2}$ & 1/s \\ \hline
$\smallk{3f}{\serca}$                                                           & $5.532\cdot10^{2}$      & $5.59 \cdot10^{2}$ & 1/s \\ \hline
$\smallk{3r}{\serca}$                                                           & $1.003\cdot10^{1}$      & $9.887$ & 1/s \\ \hline
$\smallk{4f}{\serca}$                                                           & $3.563\cdot10^{2}$      & $3.829\cdot10^{2}$ & 1/s \\ \hline
$\smallk{4r}{\serca}$                                                           & $8.559\cdot10^{1}$      & $8.419\cdot10^{1}$ & 1/s \\ \hline
$\smallk{5f}{\serca}$                                                           & $5.724\cdot10^{2}$      & $2.193\cdot10^{1}$ & 1/s \\ \hline
$\smallk{5r}{\serca}$                                                           & $3.662\cdot10^{9}$      & $3.619\cdot10^{9}$ & 1/M$^2$-s \\ \hline
$\smallk{2f}{\pmca}$                                                            & $4.484\cdot10^{10}$      & $4.704\cdot10^{10}$ & 1/M$^2$-s \\ \hline
$\smallk{2r}{\pmca}$                                                            & $4.742\cdot10^{2}$      & $4.423\cdot10^{2}$ & 1/s \\ \hline
$\smallk{3f}{\pmca}$                                                            & $2.071\cdot10^{9}$      & $2.23 \cdot10^{9}$ & 1/s \\ \hline
$\smallk{3r}{\pmca}$                                                            & $2.773\cdot10^{2}$      & $2.641\cdot10^{2}$ & 1/s \\ \hline
$\smallk{4f}{\pmca}$                                                            & $1.456$                  & $1.337$ & 1/s \\ \hline
$\smallk{4r}{\pmca}$                                                            & $1.355\cdot10^{3}$      & $1.338\cdot10^{3}$ & 1/s \\ \hline
$\smallk{5f}{\pmca}$                                                            & $7.614\cdot10^{-1}$     & $1.925$ & 1/s \\ \hline
$\smallk{5r}{\pmca}$                                                            & $1.843\cdot10^{5}$      & $1.105\cdot10^{4}$ & 1/M$^2$-s \\ \hline
   \caption{Parameter values estimated against calcium data. The parameter estimation was run twice; once where it was assumed the peak calcium concentration reached 100 nM, and the second where it was assumed the peak calcium concentration reached 200 nM. The estimated values from the 200 nM assumption were used in the results of the main paper.}
   \label{tab:CaEstimatedParams}
\end{longtable}

\begin{table}
	\footnotesize
   \centering
   \begin{tabular}{| c  c  c  c  c  c |} 
   \hline
      Parameter & max RAP1-GTP = 500 & max RAP1-GTP = 5000 & max RAP1-GTP = 500 & max RAP1-GTP = 5000 & Units \\
      & (Max Ca 100 nM) & (Max Ca 100 nM) & (Max Ca 200 nM) & (Max Ca 200 nM) & \\ \hline
 	$\smallk{\pipTWO}{\piTHREEk}$      & $1\cdot10^8$ & $1\cdot10^8$ & $1\cdot10^8$ & $1\cdot10^8$ & 1/M-s \\ \hline
 	$\smallk{-\pipTWO}{\piTHREEk}$     & 6870 & 6870 & 6870 & 6870       & 1/s   \\ \hline
	$\smallk{\pipTWO,cat}{\piTHREEk}$  & 33.15 & 33.15 & 33.15 & 33.15      & 1/s   \\ \hline
       $\smallk{-\pipTHREE,cat}{\pipTHREE Phosphatase}$ &  20 &  20 &  20 &  20      & 1/s  \\ \hline
       $\bigk{-\pipTHREE,M}{\pipTHREE Phosphatase}$   &  $2.5\cdot10^{-7}$ &  $2.5\cdot10^{-7}$ &  $2.5\cdot10^{-7}$ &  $2.5\cdot10^{-7}$ & M  \\ \hline
      $\smallk{\pipTHREE}{\rasaTHREE}$   & $4.8\cdot10^{10}$          & $4.8\cdot10^{10}$ & $4.8\cdot10^{10}$ & $4.8\cdot10^{10}$ & 1/M-s \\ \hline
      $\smallk{-\pipTHREE}{\rasaTHREE}$  & $1.6\cdot10^3$             & $1.6\cdot10^3$ & $1.6\cdot10^{3}$ & $1.6\cdot10^{3}$ & 1/s \\ \hline
      $\smallk{cat}{GAP}$      & $1.6\cdot10^{-2}$           & $1.011\cdot10^{-1}$ & $1.585\cdot10^{1}$ & $1.239\cdot10^{1}$ & 1/s \\ \hline
      $\bigk{M}{GAP}$        & $4.922\cdot10^{-8}$  & $8.078\cdot10^{-7}$ & $2.441\cdot10^{-4}$ & $1.666\cdot10^{-4}$ & M \\ \hline
      $\smallk{cat}{GEF}$      & $2.305\cdot10^{-3}$  & $1.5\cdot10^{-2}$ & $6.652\cdot10^{-3}$ & $4.0\cdot10^{-2}$ & 1/s \\ \hline
      $\bigk{M}{GEF}$        & $4.896\cdot10^{-6}$  & $2.412\cdot10^{-5}$ & $3.848\cdot10^{-4}$ & $1.915\cdot10^{-4}$ & M \\ \hline  
   \end{tabular}
   \caption{Parameter values estimated against integrin binding data. The parameter estimation was run four times, or twice for each choice of peak calcium concentration assumption. For each choice, the parameter estimation was conducted under the assumption that the peak RAP1-GTP number was 500, and again under the assumption that the peak RAP1-GTP number was 5000. The estimated values assuming a peak calcium concentration of 200 nM and a peak RAP1-GTP number of 5000 were used in the results of the main paper unless otherwise stated.}
   \label{tab:RAP1EstimatedParams}
\end{table}

\newpage

\section{Full List of Model Equations}

All volume concentrations are denoted by $[\cdot]$ and all surface densities are denoted by $[\cdot]_S$.
The volumes of this system are the platelet rich plasma (PRP), cytosol, and dense tubular structure (DTS), the volumes of which are denoted by $V_{\rm PRP}$, $V_{\rm cyt}$, and $V_{\rm DTS}$, respectively.
The surfaces of this system are the plasma membrane and DTS membrane, the surface areas of which are denoted by $S_{\rm PM}$ and $S_{\rm IM}$, respectively.

\begin{table}[htbp]
   \centering
   \begin{tabular}{| c  c  c |} 
   \hline
      Parameter & Value & Units \\ \hline 
  $V_{\rm PRP}$ & $1\cdot10^{-3}$ & dm$^3$ \\
  $S_{\rm PM}$  & $7.35\cdot10^{-9}$ & dm$^2$ \\
  $V_{\rm cyt}$ & $6\cdot10^{-15}$ & dm$^3$ \\
  $S_{\rm IM}$  & $7.35\cdot10^{-8}$ & dm$^2$ \\
  $V_{\rm DTS}$ & $3.7\cdot10^{-16}$ & dm$^3$ \\ \hline
   \end{tabular}
   \caption{Parameter values associated with the volumes of each container.}
   \label{tab:CDGIParams}
\end{table}

\pagebreak
\subsection{CalDAG-GEFI, RASA3, and RAP1}
\subsubsection{CalDAG-GEFI}

\begin{table}[h]
   \centering
   \begin{tabular}{| c  c  c  c |} 
   \hline
      Parameter & Value & Units & Reference \\ \hline 
  $\smallk{\ca}{\cdgi}$ & $6\cdot10^6$ & 1/M-s & \cite{Cook2018} \\ 
  $\smallk{-\ca}{\cdgi}$ & 0.48 & 1/s & \cite{Cook2018} \\
  $\NS{\cdgi}$ & 31595 & \# & \cite{Zeiler2014} \\ \hline
   \end{tabular}
   \caption{Parameter values associated with CalDAG-GEFI. At time $t=0$ s in the steady-state simulation, $[{\rm CalDAG}] = \NS{\cdgi}/V_{\rm cyt}$ and all other states are initialized to 0.}
   \label{tab:CDGIParams}
\end{table}

\begin{align}
\begin{split}
V_{\rm cyt}\frac{d}{dt}\conc{\cdgi}         ={}& -V_{\rm cyt}\smallk{\ca}{\cdgi}\conc{\cacyt}\conc{\cdgi} \\
& \underbrace{ \qquad +V_{\rm cyt}\smallk{-Ca}{\cdgi}\conc{\cdgi-\ca}}_{\text{CalDAG-GEFI binding calcium}} 
\end{split} \\[1em]
\begin{split}
V_{\rm cyt}\frac{d}{dt}\conc{\cdgi-\ca}         ={}& V_{\rm cyt}\smallk{\ca}{\cdgi}\conc{\cacyt}\conc{\cdgi} \\
& \underbrace{ \qquad  - V_{\rm cyt}\smallk{-Ca}{\cdgi}\conc{\cdgi-\ca}}_{\text{CalDAG-GEFI binding calcium}} 
\\ {}& -V_{\rm cyt}\smallk{\ca}{\cdgi}\conc{\cacyt}\conc{\cdgi-Ca} \\
& \underbrace{\qquad + V_{\rm cyt}\smallk{-Ca}{\cdgi}\conc{\cdgi-2Ca}}_{\text{CalDAG-GEFI-Ca binding calcium}}
\end{split} \\[1em]
\begin{split}
V_{\rm cyt}\frac{d}{dt}\conc{\cdgi-2\ca}         ={}& V_{\rm cyt}\smallk{\ca}{\cdgi}\conc{\cacyt}\conc{\cdgi-\ca} \\
& \underbrace{\qquad - V_{\rm cyt}\smallk{-Ca}{\cdgi}\conc{\cdgi-2\ca}}_{\text{CalDAG-GEFI-Ca binding calcium}}
\end{split}
\end{align}

\pagebreak
\subsubsection{RASA3}
RASA3 inactivation is not well known experimentally, so for now we will treat it as a mass-action binding between RASA3 and PIP$_3$.

\begin{table}[htbp]
   \centering
   \begin{tabular}{| c  c  c  c |} 
   \hline
      Parameter & Value & Units & Reference \\ \hline 
  $\smallk{\pipTHREE}{\rasaTHREE}$ & See Table \ref{tab:RAP1EstimatedParams} & 1/M-s & \\ 
  $\smallk{-\pipTHREE}{\rasaTHREE}$ & See Table \ref{tab:RAP1EstimatedParams} & 1/s & \\
  $\NS{\rasaTHREE}$ & 26959 & \# & \cite{Zeiler2014} \\ \hline
   \end{tabular}
   \caption{Parameter values associated with RASA3. At time $t=0$ s in the steady-state simulation, $[{\rm RASA3}]_S = \NS{\rasaTHREE}/S_{\rm PM}$ and all other states are initialized to 0.}
   \label{tab:RASA3Params}
\end{table}

\begin{align}
\begin{split}
S_{\rm PM}\frac{d}{dt}[{\rm RASA3}]_S ={}& \underbrace{S_{\rm PM}\smallk{-\pipTHREE}{\rasaTHREE}[{\rm RASA3-PIP_3}]_S-S_{\rm PM}\big(\frac{S_{\rm PM}}{V_{\rm cyt}}\smallk{\pipTHREE}{\rasaTHREE}\big)[{\rm RASA3}]_S[{\rm PIP_3}]_S}_{\text{PIP$_3$ inactivation of RASA3}}
\end{split} \\[1em]
\begin{split}
S_{\rm PM}\frac{d}{dt}[{\rm RASA3-PIP_3}]_S ={}& \underbrace{-S_{\rm PM}\smallk{-\pipTHREE}{\rasaTHREE}[{\rm RASA3-PIP_3}]_S+S_{\rm PM}\big(\frac{S_{\rm PM}}{V_{\rm cyt}}\smallk{\pipTHREE}{\rasaTHREE}\big)[{\rm RASA3}]_S[{\rm PIP_3}]_S}_{\text{PIP$_3$ inactivation of RASA3}}
\end{split}
\end{align}

\pagebreak
\subsubsection{RAP1}

\begin{table}[htbp]
   \centering
   \begin{tabular}{| c  c  c  c |} 
   \hline
      Parameter & Value & Units & Reference \\ \hline 
  $\smallk{cat}{GAP}$ & See Table \ref{tab:RAP1EstimatedParams} & 1/s & \\
  $\bigk{M}{GAP}$ & See Table \ref{tab:RAP1EstimatedParams} & M & \\
  $\smallk{cat}{GEF}$ & See Table \ref{tab:RAP1EstimatedParams} & 1/s & \\
  $\bigk{M}{GAP}$ & See Table \ref{tab:RAP1EstimatedParams} & M & \\
  $\NS{\rapONE}$ & 209601 & \# & \cite{Zeiler2014} \\ \hline
   \end{tabular}
   \caption{Parameter values associated with RAP1. At time $t=0$ s in the steady-state simulation, $[{\rm RAP1-GDP}]_S = \NS{\rapONE}/S_{\rm PM}$ and all other states are initialized to 0.}
   \label{tab:RAP1Params}
\end{table}

\begin{align}
\begin{split}
S_{\rm PM}\frac{d}{dt}\surf{\rapONE-GTP} ={}& \underbrace{-S_{\rm PM}\frac{\smallk{cat}{GAP}\surf{RASA3}}{\big(\frac{V_{\rm cyt}}{S_{\rm PM}}\bigk{M}{GAP}\big)+\surf{\rapONE-GTP}}\surf{\rapONE-GTP}}_{\text{RASA3 inactivation of RAP1}} \\[.5em]
& \underbrace{+V_{\rm cyt}\frac{\smallk{cat}{GEF}\conc{\cdgi-2\ca}}{\big(\frac{V_{\rm cyt}}{S_{\rm PM}}\bigk{M}{GEF}\big)+\surf{\rapONE-GDP}}\surf{\rapONE-GDP}}_{\text{CalDAG activation of RAP1}}
\end{split}
\end{align}
\begin{align}
\begin{split}
S_{\rm PM}\frac{d}{dt}\surf{\rapONE-GDP} ={}& \underbrace{S_{\rm PM}\frac{\smallk{cat}{GAP}\surf{RASA3}}{\big(\frac{V_{\rm cyt}}{S_{\rm PM}}\bigk{M}{GAP}\big)+\surf{\rapONE-GTP}}\surf{\rapONE-GTP}}_{\text{RASA3 inactivation of RAP1}} \\[.25em]
& \underbrace{-V_{\rm cyt}\frac{\smallk{cat}{GEF}\conc{\cdgi-2\ca}}{\big(\frac{V_{\rm cyt}}{S_{\rm PM}}\bigk{M}{GEF}\big)+\surf{\rapONE-GDP}}\surf{\rapONE-GDP}}_{\text{CalDAG activation of RAP1}}
\end{split}
\end{align}

\pagebreak
\subsection{ADP}

\begin{align}
\begin{split}
V_{\rm PRP}\frac{d}{dt}[{\rm ADP}] ={}& \underbrace{-S_{\rm PM}\bigg(\smallk{ADP}{\pTWOyTWELVE}\frac{\alphaS{\pTWOyTWELVE} \bigk{act}{\pTWOyTWELVE}+1}{\bigk{act}{\pTWOyTWELVE}+1}\bigg)\conc{ADP}\surf{\pTWOyTWELVE} + S_{\rm PM}\smallk{-ADP}{\pTWOyTWELVE}\surf{ADP-\pTWOyTWELVE}}_{\text{ADP Binding to P2Y$_{12}$}} \\[.5em]
& -S_{\rm PM}\bigg(\smallk{ADP}{\pTWOyTWELVE}\frac{\alphaS{\pTWOyTWELVE}\betaS{\pTWOyTWELVE} \bigk{act}{\pTWOyTWELVE}+1}{\betaS{\pTWOyTWELVE} \bigk{act}{\pTWOyTWELVE}+1}\bigg)\conc{ADP}\surf{\pTWOyTWELVE-\gi GDP} \\
& \underbrace{\qquad +S_{\rm PM}\bigg(\frac{\smallk{-ADP}{\pTWOyTWELVE}}{\gammaS{\pTWOyTWELVE}}\frac{\alphaS{\pTWOyTWELVE}\betaS{\pTWOyTWELVE} \bigk{act}{\pTWOyTWELVE}+1}{\alphaS{\pTWOyTWELVE}\betaS{\pTWOyTWELVE}\deltaS{\pTWOyTWELVE} \bigk{act}{\pTWOyTWELVE}+1}\bigg)\surf{ADP-\pTWOyTWELVE-\gi GDP}}_{\text{ADP Binding to P2Y$_{12}$-G$_i$GDP}} \\[.5em]
& \underbrace{-S_{\rm PM}\bigg(\smallk{ADP}{\pTWOyONE}\frac{\alphaS{\pTWOyONE} \bigk{act}{\pTWOyONE}+1}{\bigk{act}{\pTWOyONE}+1}\bigg)\conc{ADP}\surf{\pTWOyONE} + S_{\rm PM}\smallk{-ADP}{\pTWOyONE}\surf{ADP-P2Y_1}}_{\text{ADP Binding to P2Y$_1$}} \\[.5em]
& -S_{\rm PM}\bigg(\smallk{ADP}{\pTWOyONE}\frac{\alphaS{\pTWOyONE}\betaS{\pTWOyONE} \bigk{act}{\pTWOyONE}+1}{\betaS{\pTWOyONE} \bigk{act}{\pTWOyONE}+1}\bigg)\conc{ADP}\surf{\pTWOyONE-\gq GDP} \\
& \underbrace{\qquad +S_{\rm PM}\bigg(\frac{\smallk{-ADP}{\pTWOyONE}}{\gammaS{\pTWOyONE}}\frac{\alphaS{\pTWOyONE}\betaS{\pTWOyONE} \bigk{act}{\pTWOyONE}+1}{\alphaS{\pTWOyONE}\betaS{\pTWOyONE}\deltaS{\pTWOyONE} \bigk{act}{\pTWOyONE}+1}\bigg)\surf{ADP-P2Y_1-\gq GDP}}_{\text{ADP Binding to P2Y$_1$-G$_q$GDP}} \\[.5em]
\end{split}
\end{align}

\pagebreak
\subsection{Proteins in P2Y$_{12}$ Pathway}
\subsubsection{P2Y$_{12}$}

The P2Y$_{12}$ model is exactly the same as what Purvis et al. used for P2Y$_1$ \cite{Purvis2008}. 
However, we performed a Quasi-Steady State reduction to reduce the number of equations by four. 
See later in the supplemental for further discussion. 

\begin{figure}[ht!]
\centering\includegraphics[width=\textwidth]{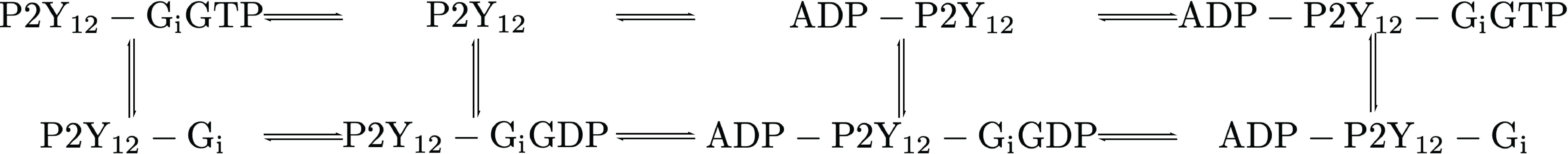}
\caption{P2Y$_{12}$ reaction diagram}
\end{figure}

\begin{table}[htbp]
   \centering
   \begin{tabular}{| c  c  c  c |} 
   \hline
      Parameter & Value & Units & Reference \\ \hline 
  $\bigk{act}{\pTWOyTWELVE}$ & See Table \ref{tab:CaEstimatedParams} & n.d. & \\
  $\alphaS{\pTWOyTWELVE}$ & See Table \ref{tab:CaEstimatedParams} & n.d. & \\
  $\betaS{\pTWOyTWELVE}$ & See Table \ref{tab:CaEstimatedParams} & n.d. & \\
  $\deltaS{\pTWOyTWELVE}$ & See Table \ref{tab:CaEstimatedParams} & n.d. & \\
  $\gammaS{\pTWOyTWELVE}$ & See Table \ref{tab:CaEstimatedParams} & n.d. & \\
  $\smallk{\gi GDP}{\pTWOyTWELVE}$ & See Table \ref{tab:CaEstimatedParams} & dm$^2$/mol-s & \\
  $\smallk{-\gi GDP}{\pTWOyTWELVE}$ & See Table \ref{tab:CaEstimatedParams} & 1/s & \\
  $\smallk{\gi GTP}{\pTWOyTWELVE}$ & See Table \ref{tab:CaEstimatedParams} & dm$^2$/mol-s & \\
  $\smallk{-\gi GTP}{\pTWOyTWELVE}$ & See Table \ref{tab:CaEstimatedParams} & 1/s & \\
  $\smallk{ADP}{\pTWOyTWELVE}$ & See Table \ref{tab:CaEstimatedParams} & 1/M-s & \\
  $\smallk{-ADP}{\pTWOyTWELVE}$ & See Table \ref{tab:CaEstimatedParams} & 1/s & \\
  $\smallk{-GDP}{\pTWOyTWELVE}$ & See Table \ref{tab:CaEstimatedParams} & 1/s & \\
  $\smallk{GDP}{\pTWOyTWELVE}$ & See Table \ref{tab:CaEstimatedParams} & 1/M-s & \\
  $\smallk{-GTP}{\pTWOyTWELVE}$ & See Table \ref{tab:CaEstimatedParams} & 1/s & \\
  $\smallk{GTP}{\pTWOyTWELVE}$ & See Table \ref{tab:CaEstimatedParams} & 1/M-s & \\
  $\conc{GDP}$ & $1.3362\cdot10^{-5}$ & M & \cite{Purvis2008} \\
  $\conc{GTP}$ & $7.117\cdot10^{-4}$ & M & \cite{Purvis2008} \\
  $\NS{\pTWOyTWELVE}$ & 2681 & \# & \cite{Zeiler2014} \\ \hline
   \end{tabular}
   \caption{Parameter values associated with P2Y$_{12}$. At time At time $t=0$ s in the steady-state simulation, $[{\rm P2Y_{12}}]_S = \NS{\pTWOyTWELVE}/S_{\rm PM}$ and all other states are initialized to 0.}
   \label{tab:P2Y12EstimatedParams}
\end{table}

\begin{align}
\begin{split}
S_{\rm PM}\frac{d}{dt}\surf{\pTWOyTWELVE} = {}& \underbrace{-S_{\rm PM}\bigg(\smallk{ADP}{\pTWOyTWELVE}\frac{\alphaS{\pTWOyTWELVE} \bigk{act}{\pTWOyTWELVE}+1}{\bigk{act}{\pTWOyTWELVE}+1}\bigg)\conc{\rm ADP}\surf{\pTWOyTWELVE} + S_{\rm PM}\smallk{-ADP}{\pTWOyTWELVE}\surf{\rm ADP-\pTWOyTWELVE}}_{\text{ADP Binding}} \\[.5em]
& -S_{\rm PM}\bigg(\smallk{G_iGDP}{\pTWOyTWELVE}\frac{\betaS{\pTWOyTWELVE} \bigk{act}{\pTWOyTWELVE}}{\bigk{act}{\pTWOyTWELVE}+1}\bigg)\surf{\gi GDP}\surf{\pTWOyTWELVE} \\
& \underbrace{\qquad + S_{\rm PM}\bigg(\smallk{-\gi GDP}{\pTWOyTWELVE}\frac{\betaS{\pTWOyTWELVE} \bigk{act}{\pTWOyTWELVE}}{\betaS{\pTWOyTWELVE} \bigk{act}{\pTWOyTWELVE}+1}\bigg)\surf{\pTWOyTWELVE-\gi GDP}}_{\text{G$_i$GDP Binding}} \\[.5em]
& -S_{\rm PM}\bigg(\smallk{\gi GTP}{\pTWOyTWELVE}\frac{\bigk{act}{\pTWOyTWELVE}}{\bigk{act}{\pTWOyTWELVE}+1}\bigg)\surf{\gi GTP}\surf{\pTWOyTWELVE} \\
& \underbrace{\qquad + S_{\rm PM}\smallk{-\gi GTP}{\pTWOyTWELVE}\surf{\pTWOyTWELVE-\gi GTP}}_{\text{G$_i$GTP Binding}}
\end{split} \\[1em]
\begin{split}
S_{\rm PM}\frac{d}{dt}\surf{ADP-\pTWOyTWELVE} ={}& \underbrace{S_{\rm PM}\bigg(\smallk{ADP}{\pTWOyTWELVE} \frac{\alphaS{\pTWOyTWELVE} \bigk{act}{\pTWOyTWELVE}+1}{\bigk{act}{\pTWOyTWELVE}+1}\bigg)\conc{ADP}\surf{\pTWOyTWELVE} - S_{\rm PM}\smallk{-ADP}{\pTWOyTWELVE}\surf{ADP-\pTWOyTWELVE}}_{\text{ADP Binding}} \\[.5em]
& - S_{\rm PM}\bigg(\smallk{\gi GDP}{\pTWOyTWELVE}\frac{\alphaS{\pTWOyTWELVE}\betaS{\pTWOyTWELVE} \bigk{act}{\pTWOyTWELVE}}{\alphaS{\pTWOyTWELVE} \bigk{act}{\pTWOyTWELVE}+1}\bigg) \surf{\gi GDP}\surf{ADP-\pTWOyTWELVE} \\
& \underbrace{\qquad + S_{\rm PM}\bigg(\frac{\smallk{-\gi GDP}{\pTWOyTWELVE}}{\gammaS{\pTWOyTWELVE}}\frac{\alphaS{\pTWOyTWELVE}\betaS{\pTWOyTWELVE} \bigk{act}{\pTWOyTWELVE}}{\alphaS{\pTWOyTWELVE}\betaS{\pTWOyTWELVE}\deltaS{\pTWOyTWELVE} \bigk{act}{\pTWOyTWELVE}+1}\bigg)\surf{ADP-\pTWOyTWELVE-\gi GDP}}_{\text{G$_i$GDP Binding}} \\[.5em]
& - S_{\rm PM}\bigg(\smallk{\gi GTP}{\pTWOyTWELVE}\frac{\alphaS{\pTWOyTWELVE} \bigk{act}{\pTWOyTWELVE}}{\alphaS{\pTWOyTWELVE} \bigk{act}{\pTWOyTWELVE}+1}\bigg) \surf{\gi GTP}\surf{ADP-\pTWOyTWELVE} \\
& \underbrace{\qquad + S_{\rm PM}\smallk{-\gi GTP}{\pTWOyTWELVE}\surf{ADP-\pTWOyTWELVE-\gi GTP}}_{\text{G$_i$GTP Binding}}
\end{split}
\end{align}
\begin{align}
\begin{split}
S_{\rm PM}\frac{d}{dt}\surf{\pTWOyTWELVE-\gi GDP} ={}& -S_{\rm PM}\bigg(\smallk{ADP}{\pTWOyTWELVE}\frac{\alphaS{\pTWOyTWELVE}\betaS{\pTWOyTWELVE} \bigk{act}{\pTWOyTWELVE}+1}{\betaS{\pTWOyTWELVE} \bigk{act}{\pTWOyTWELVE}+1}\bigg)\conc{ADP}\surf{\pTWOyTWELVE-\gi GDP} \\
& \underbrace{\qquad +S_{\rm PM}\bigg(\frac{\smallk{-ADP}{\pTWOyTWELVE}}{\gammaS{\pTWOyTWELVE}}\frac{\alphaS{\pTWOyTWELVE}\betaS{\pTWOyTWELVE} \bigk{act}{\pTWOyTWELVE}+1}{\alphaS{\pTWOyTWELVE}\betaS{\pTWOyTWELVE}\deltaS{\pTWOyTWELVE} \bigk{act}{\pTWOyTWELVE}+1}\bigg)\surf{\rm ADP-\pTWOyTWELVE-\gi GDP}}_{\text{ADP Binding}} \\[.5em]
& + S_{\rm PM}\bigg(\smallk{\gi GDP}{\pTWOyTWELVE}\frac{\betaS{\pTWOyTWELVE} \bigk{act}{\pTWOyTWELVE}}{\bigk{act}{\pTWOyTWELVE}+1}\bigg)\surf{\gi GDP}\surf{\pTWOyTWELVE} \\
& \underbrace{\qquad - S_{\rm PM}\bigg(\smallk{-\gi GDP}{\pTWOyTWELVE}\frac{\betaS{\pTWOyTWELVE} \bigk{act}{\pTWOyTWELVE}}{\betaS{\pTWOyTWELVE} \bigk{act}{\pTWOyTWELVE}+1}\bigg)\surf{\pTWOyTWELVE-\gi GDP}}_{\text{G$_i$GDP Binding}} \\[.5em]
& + S_{\rm PM}\smallk{GDP}{\pTWOyTWELVE}\surf{\rm GDP}\surf{\pTWOyTWELVE-\gi } \\
& \underbrace{\qquad - S_{\rm PM}\bigg(\smallk{-GDP}{\pTWOyTWELVE}\frac{\betaS{\pTWOyTWELVE} \bigk{act}{\pTWOyTWELVE}}{\betaS{\pTWOyTWELVE} \bigk{act}{\pTWOyTWELVE}+1}\bigg)\surf{\pTWOyTWELVE-\gi GDP}}_{\text{GDP Binding}}
\end{split}
\end{align}
\begin{align}
\begin{split}
S_{\rm PM}\frac{d}{dt}\surf{ADP-\pTWOyTWELVE-\gi GDP} ={}& 
S_{\rm PM}\bigg(\smallk{ADP}{\pTWOyTWELVE}\frac{\alphaS{\pTWOyTWELVE}\betaS{\pTWOyTWELVE} \bigk{act}{\pTWOyTWELVE}+1}{\betaS{\pTWOyTWELVE} \bigk{act}{\pTWOyTWELVE}+1}\bigg)\conc{ADP}\surf{\pTWOyTWELVE-\gi GDP} \\ 
& \underbrace{\qquad -S_{\rm PM}\bigg(\frac{\smallk{-ADP}{\pTWOyTWELVE}}{\gammaS{\pTWOyTWELVE}}\frac{\alphaS{\pTWOyTWELVE}\betaS{\pTWOyTWELVE} \bigk{act}{\pTWOyTWELVE}+1}{\alphaS{\pTWOyTWELVE}\betaS{\pTWOyTWELVE}\deltaS{\pTWOyTWELVE} \bigk{act}{\pTWOyTWELVE}+1}\bigg)\surf{ADP-\pTWOyTWELVE-\gi GDP}}_{\text{ADP Binding}} \\
& +S_{\rm PM}\bigg(\smallk{\gi GDP}{\pTWOyTWELVE}\frac{\alphaS{\pTWOyTWELVE}\betaS{\pTWOyTWELVE} \bigk{act}{\pTWOyTWELVE}}{\alphaS{\pTWOyTWELVE} \bigk{act}{\pTWOyTWELVE}+1}\bigg) \surf{\gi GDP}\surf{ADP-\pTWOyTWELVE} \\
& \underbrace{\qquad - S_{\rm PM}\bigg(\frac{\smallk{-\gi GDP}{\pTWOyTWELVE}}{\gammaS{\pTWOyTWELVE}}\frac{\alphaS{\pTWOyTWELVE}\betaS{\pTWOyTWELVE} \bigk{act}{\pTWOyTWELVE}}{\alphaS{\pTWOyTWELVE}\betaS{\pTWOyTWELVE}\deltaS{\pTWOyTWELVE} \bigk{act}{\pTWOyTWELVE}+1}\bigg)\surf{ADP-\pTWOyTWELVE-\gi GDP}}_{\text{G$_i$GDP Binding}} \\
& +S_{\rm PM} \smallk{GDP}{\pTWOyTWELVE}\conc{GDP}\surf{ADP-\pTWOyTWELVE-\gi } \\
& \underbrace{\qquad -S_{\rm PM} \bigg(\smallk{-GDP}{\pTWOyTWELVE}\frac{\alphaS{\pTWOyTWELVE}\betaS{\pTWOyTWELVE}\deltaS{\pTWOyTWELVE} \bigk{act}{\pTWOyTWELVE}}{\alphaS{\pTWOyTWELVE}\betaS{\pTWOyTWELVE}\deltaS{\pTWOyTWELVE} \bigk{act}{\pTWOyTWELVE}+1}\bigg)\surf{ADP-\pTWOyTWELVE-\gi GDP}}_{\text{GDP Binding}}
\end{split}  \\[1em]
\begin{split}
S_{\rm PM}\frac{d}{dt}\surf{\pTWOyTWELVE-\gi GTP} ={}& \underbrace{S_{\rm PM}\smallk{GTP}{\pTWOyTWELVE}\surf{\pTWOyTWELVE-\gi }\conc{GTP} - S_{\rm PM}\smallk{-GTP}{\pTWOyTWELVE}\surf{\pTWOyTWELVE-\gi GTP}}_{\text{GTP Binding}} \\[.5em]
 & + S_{\rm PM}\bigg(\smallk{\gi GTP}{\pTWOyTWELVE}\frac{\bigk{act}{\pTWOyTWELVE}}{\bigk{act}{\pTWOyTWELVE}+1}\bigg)\surf{\gi GTP}\surf{\pTWOyTWELVE} \\
 & \underbrace{\qquad - S_{\rm PM}\smallk{-\gi GTP}{\pTWOyTWELVE}\surf{\pTWOyTWELVE-\gi GTP}}_{\text{G$_i$GTP binding}}
\end{split}\\[1em]
\begin{split}
S_{\rm PM}\frac{d}{dt}\surf{ADP-\pTWOyTWELVE-\gi GTP} ={}& S_{\rm PM}\smallk{GTP}{\pTWOyTWELVE}\surf{ADP-\pTWOyTWELVE-\gi }\conc{GTP} \\
& \underbrace{\qquad - S_{\rm PM}\smallk{-GTP}{\pTWOyTWELVE}\surf{ADP-\pTWOyTWELVE-\gi GTP}}_{\text{GTP Binding}} \\[.5em]
 & + S_{\rm PM}\bigg(\smallk{\gi GTP}{\pTWOyTWELVE}\frac{\alphaS{\pTWOyTWELVE} \bigk{act}{\pTWOyTWELVE}}{\alphaS{\pTWOyTWELVE} \bigk{act}{\pTWOyTWELVE}+1}\bigg) \surf{\gi GTP}\surf{ADP-\pTWOyTWELVE} \\
 & \underbrace{\qquad - S_{\rm PM}\smallk{-\gi GTP}{\pTWOyTWELVE}\surf{ADP-\pTWOyTWELVE-\gi GTP}}_{\text{G$_i$GTP binding}}
\end{split}
\end{align}
\begin{align}
\begin{split}
S_{\rm PM}\frac{d}{dt}\surf{\pTWOyTWELVE-\gi } ={}& -S_{\rm PM}\smallk{GTP}{\pTWOyTWELVE}\surf{\pTWOyTWELVE-\gi }\conc{GTP} \\
& \underbrace{\qquad + S_{\rm PM}\smallk{-GTP}{\pTWOyTWELVE}\surf{\pTWOyTWELVE-\gi GTP}}_{\text{GTP Binding}} \\[.5em]
& +S_{\rm PM}\bigg(\smallk{-GDP}{\pTWOyTWELVE}\frac{\betaS{\pTWOyTWELVE} \bigk{act}{\pTWOyTWELVE}}{\betaS{\pTWOyTWELVE} \bigk{act}{\pTWOyTWELVE}+1}\bigg)\surf{\pTWOyTWELVE-\gi GDP} \\
& \underbrace{\qquad - S_{\rm PM}\smallk{GDP}{\pTWOyTWELVE}\surf{\pTWOyTWELVE-\gi }\conc{GDP}}_{\text{GDP binding}}
\end{split} \\[1em]
\begin{split}
S_{\rm PM}\frac{d}{dt}\surf{ADP-\pTWOyTWELVE-\gi } ={}& -S_{\rm PM}\smallk{GTP}{\pTWOyTWELVE}\surf{ADP-\pTWOyTWELVE-\gi }\conc{GTP} \\
&\underbrace{\qquad+ S_{\rm PM}\smallk{-GTP}{\pTWOyTWELVE}\surf{ADP-\pTWOyTWELVE-\gi GTP}}_{\text{GTP Binding}} \\[.5em]
& + S_{\rm PM}\bigg(\smallk{-GDP}{\pTWOyTWELVE}\frac{\alphaS{\pTWOyTWELVE}\betaS{\pTWOyTWELVE}\deltaS{\pTWOyTWELVE} \bigk{act}{\pTWOyTWELVE}}{\alphaS{\pTWOyTWELVE}\betaS{\pTWOyTWELVE}\deltaS{\pTWOyTWELVE} \bigk{act}{\pTWOyTWELVE}+1}\bigg)\surf{ADP-\pTWOyTWELVE-\gi GDP} \\
& \underbrace{\qquad - S_{\rm PM}\smallk{GDP}{\pTWOyTWELVE}\surf{ADP-\pTWOyTWELVE-\gi }\conc{GDP}}_{\text{GDP binding}}
\end{split}
\end{align}

\pagebreak
\subsubsection{G$_i$}

\begin{table}[htbp]
   \centering
   \begin{tabular}{| c  c  c  c |} 
   \hline
      Parameter & Value & Units & Reference \\ \hline 
  $\smallk{GTP}{}$ & See Table \ref{tab:CaEstimatedParams} & 1/s & \\
  $\smallk{\gibg}{\gi}$ & 103584 & 1/M-s & \cite{Kinzer-Ursem2007} \\
  $\smallk{-\gibg}{\gi}$ & 7.7832 & 1/s & \cite{Kinzer-Ursem2007} \\
  $\NS{\gi}$ & 27752 & \# & \cite{Zeiler2014} \\ \hline
   \end{tabular}
   \caption{Parameter values associated with G$_{\rm i}$. At time $t=0$ s in the steady-state simulation, $\surf{\gi GDP} = \NS{\gi}/S_{\rm PM}$ and all other states are initialized to 0.}
   \label{tab:GiParams}
\end{table}

\begin{align}
\begin{split}
S_{\rm PM}\frac{d}{dt}\surf{\gi GDP} ={}& -S_{\rm PM}\bigg(\smallk{\gi GDP}{\pTWOyTWELVE}\frac{\betaS{\pTWOyTWELVE} \bigk{act}{\pTWOyTWELVE}}{\bigk{act}{\pTWOyTWELVE}+1}\bigg)\surf{\gi GDP}\surf{\pTWOyTWELVE} \\
& \underbrace{\qquad + S_{\rm PM}\bigg(\smallk{-\gi GDP}{\pTWOyTWELVE}\frac{\betaS{\pTWOyTWELVE} \bigk{act}{\pTWOyTWELVE}}{\betaS{\pTWOyTWELVE} \bigk{act}{\pTWOyTWELVE}+1}\bigg)\surf{\pTWOyTWELVE-\gi GDP}}_{\text{G$_i$GDP Binding}} \\[.5em]
& - S_{\rm PM}\bigg(\smallk{\gi GDP}{\pTWOyTWELVE}\frac{\alphaS{\pTWOyTWELVE}\betaS{\pTWOyTWELVE} \bigk{act}{\pTWOyTWELVE}}{\alphaS{\pTWOyTWELVE} \bigk{act}{\pTWOyTWELVE}+1}\bigg) \surf{\gi GDP}\surf{ADP-\pTWOyTWELVE} \\
& \underbrace{\qquad + S_{\rm PM}\bigg(\frac{\smallk{-\gi GDP}{\pTWOyTWELVE}}{\gammaS{\pTWOyTWELVE}}\frac{\alphaS{\pTWOyTWELVE}\betaS{\pTWOyTWELVE} \bigk{act}{\pTWOyTWELVE}}{\alphaS{\pTWOyTWELVE}\betaS{\pTWOyTWELVE}\deltaS{\pTWOyTWELVE} \bigk{act}{\pTWOyTWELVE}+1}\bigg)\surf{ADP-\pTWOyTWELVE-\gi GDP}}_{\text{G$_i$GDP Binding}} \\[.5em]
& \underbrace{+S_{\rm PM}\Big(\frac{S_{\rm PM}}{V_{\rm cyt}}\smallk{\gibg}{\rm \gi }\Big)\surf{G_{i\alpha}GDP}\surf{\gibg} - S_{\rm PM}\smallk{-\gibg}{\rm \gi }\surf{\gi GDP}}_{\text{G$_{i\beta\gamma}$ binding G$_{i\alpha}$GDP}}
\end{split} \\[1em]
\begin{split}
S_{\rm PM}\frac{d}{dt}\surf{\gi GTP} ={}& -S_{\rm PM}\bigg(\smallk{\gi GTP}{\pTWOyTWELVE}\frac{\bigk{act}{\pTWOyTWELVE}}{\bigk{act}{\pTWOyTWELVE}+1}\bigg)\surf{\gi GTP}\surf{\pTWOyTWELVE} \\
& \underbrace{\qquad + S_{\rm PM}\smallk{-\gi GTP}{\pTWOyTWELVE}\surf{\pTWOyTWELVE-\gi GTP}}_{\text{G$_i$GTP Binding}} \\[.5em]
& - S_{\rm PM}\bigg(\smallk{\gi GTP}{\pTWOyTWELVE}\frac{\alphaS{\pTWOyTWELVE} \bigk{act}{\pTWOyTWELVE}}{\alphaS{\pTWOyTWELVE} \bigk{act}{\pTWOyTWELVE}+1}\bigg) \surf{\gi GTP}\surf{ADP-\pTWOyTWELVE} \\
& \underbrace{\qquad + S_{\rm PM}\smallk{-\gi GTP}{\pTWOyTWELVE}\surf{ADP-\pTWOyTWELVE-\gi GTP}}_{\text{G$_i$GTP Binding}} \\[.5em]
& \underbrace{-S_{\rm PM}\smallk{autohydrolysis}{\rm \gi }\surf{\gi GTP}}_{\text{Autohydrolysis of G$_i$GTP}} \\[.5em]
& \underbrace{+S_{\rm PM}\Big(\frac{S_{\rm PM}}{V_{\rm cyt}}\smallk{\gibg}{\rm \gi }\Big)\surf{G_{i\alpha}GTP}\surf{\gibg} - S_{\rm PM}\smallk{-\gibg}{\rm \gi }\surf{\gi GTP}}_{\text{G$_{i\beta\gamma}$ binding G$_{i\alpha}$GTP}}
\end{split}
\end{align}
\begin{align}
\begin{split}
S_{\rm PM}\frac{d}{dt}\surf{G_{i\alpha}GTP} ={}& \underbrace{-S_{\rm PM}\Big(\smallk{\gibg}{\rm \gi }\frac{S_{\rm PM}}{V_{\rm cyt}}\Big)\surf{G_{i\alpha}GTP}\surf{\gibg}+ S_{\rm PM}\smallk{-\gibg}{\rm \gi }\surf{\gi GTP}}_{\text{G$_{i\beta\gamma}$ binding G$_{i\alpha}$GTP}} \\[.5em]
& \underbrace{+S_{\rm PM}\smallk{-G_{i\alpha}GTP}{\rm PI3K}\surf{PI3K-G_{i\alpha}GTP} - S_{\rm PM}\smallk{G_{i\alpha}GTP}{\rm PI3K}\surf{PI3K}\surf{G_{i\alpha}GTP}}_{\text{GTP binding (activation)}}
\end{split} \\[1em]
\begin{split}
S_{\rm PM}\frac{d}{dt}\surf{G_{i\alpha}GDP} ={}& \underbrace{S_{\rm PM}\smallk{autohydrolysis}{\rm \gi }\surf{\gi GTP}}_{\text{Autohydrolysis of G$_i$GTP}} \\[.5em]
& \underbrace{-S_{\rm PM}\smallk{\gibg}{\rm \gi }\surf{G_{i\alpha}GDP}\surf{\gibg} + S_{\rm PM}\smallk{-\gibg}{\rm \gi }\surf{\gi GDP}}_{\text{G$_{i\beta\gamma}$ binding G$_{i\alpha}$GDP}} \\[.5em]
& \underbrace{+S_{\rm PM}\smallk{-\gi {i\alpha}GDP}{\rm PI3K}\surf{PI3K-G_{i\alpha}GDP} - S_{\rm PM}\smallk{G_{i\alpha}GDP}{\rm PI3K}\surf{PI3K}\surf{G_{i\alpha}GDP}}_{\text{GDP binding}}
\end{split}
\end{align}
\begin{align}
\begin{split}
S_{\rm PM}\frac{d}{dt}\surf{\gibg} ={}& \underbrace{S_{\rm PM}\smallk{autohydrolysis}{\rm \gi }\surf{\gi GTP}}_{\text{Autohydrolysis of G$_i$GTP}} \\[.5em]
& \underbrace{-S_{\rm PM}\Big(\smallk{\gibg}{\rm \gi }\frac{S_{\rm PM}}{V_{\rm cyt}}\Big)\surf{G_{i\alpha}GTP}\surf{\gibg} + S_{\rm PM}\smallk{-\gibg}{\rm \gi }\surf{\gi GTP}}_{\text{G$_{i\beta\gamma}$ binding G$_{i\alpha}$GTP}} \\[.5em]
& \underbrace{-S_{\rm PM}\Big(\smallk{\gibg}{\rm \gi }\frac{S_{\rm PM}}{V_{\rm cyt}}\Big)\surf{G_{i\alpha}GDP}\surf{\gibg} + S_{\rm PM}\smallk{-\gibg}{\rm \gi }\surf{\gi GDP}}_{\text{G$_{i\beta\gamma}$ binding G$_{i\alpha}$GDP}}
\end{split}
\end{align}

\pagebreak
\subsubsection{PI3K}

\begin{table}[htbp]
   \centering
   \begin{tabular}{| c  c  c  c |} 
   \hline
      Parameter & Value & Units & Reference \\ \hline 
 $\smallk{hydrolyze}{\piTHREEk}$ & See Table \ref{tab:CaEstimatedParams} & 1/s & \\
 $\smallk{-\gia GDP}{\piTHREEk}$ & See Table \ref{tab:CaEstimatedParams} & 1/s & \\
 $\smallk{-\gia GTP}{\piTHREEk}$ & See Table \ref{tab:CaEstimatedParams} & 1/s & \\
 $\smallk{\gia GDP}{\piTHREEk}$ & See Table \ref{tab:CaEstimatedParams} & dm$^2$/mol-s & \\
 $\smallk{\gia GTP}{\piTHREEk}$ & See Table \ref{tab:CaEstimatedParams} & dm$^2$/mol-s & \\
 $\smallk{\pipTHREE}{\piTHREEk}$      & See Table \ref{tab:RAP1EstimatedParams}  & 1/M-s & \\
 $\smallk{-\pipTHREE}{\piTHREEk}$     & See Table \ref{tab:RAP1EstimatedParams} & 1/s &   \\
 $\smallk{\pipTWO,cat}{\piTHREEk}$  & See Table \ref{tab:RAP1EstimatedParams} & 1/s &   \\
 $\NS{\piTHREEk}$ & 1916 & \# & \cite{Zeiler2014} \\ \hline
   \end{tabular}
   \caption{Parameter values associated with PI3K. At time $t=0$ in the steady-state simulation, $\surf{\piTHREEk} = \NS{\piTHREEk}/S_{\rm PM}$ and all other states are initialized to 0.}
   \label{tab:PI3KParams}
\end{table}

\begin{align}
\begin{split}
S_{\rm PM}\frac{d}{dt}\surf{\piTHREEk} ={}& S_{\rm PM}\smallk{-\gia GTP}{\piTHREEk}\surf{\piTHREEk-\gia GTP} \\
& \underbrace{\qquad - S_{\rm PM}\smallk{\gia GTP}{\piTHREEk}\surf{\piTHREEk}\surf{\rm \gia GTP}}_{\text{G$_{i\alpha}$GTP binding (activation)}} \\[.5em]
& +S_{\rm PM}\smallk{-\gia GDP}{\piTHREEk}\surf{\piTHREEk-\gia GDP} \\
& \underbrace{\qquad - S_{\rm PM}\smallk{\gia GDP}{\piTHREEk}\surf{\piTHREEk}\surf{\rm \gia GDP}}_{\text{G$_{i\alpha}$GDP binding}}
\end{split} \\[1em]
\begin{split}
S_{\rm PM}\frac{d}{dt}\surf{\piTHREEk-\gia GTP} ={}& -S_{\rm PM}\smallk{-\gia GTP}{\piTHREEk}\surf{\piTHREEk-\gia GTP} \\
& \underbrace{\qquad + S_{\rm PM}\smallk{\gia GTP}{\piTHREEk}\surf{\piTHREEk}\surf{\rm \gia GTP}}_{\text{GTP binding (activation)}} \\[.5em]
& \underbrace{-S_{\rm PM}\smallk{hydrolyze}{\piTHREEk}\surf{\piTHREEk-\gia GTP}}_{\text{GTP hydrolysis (inactivation)}} \\[.5em]
& - S_{\rm PM}\big(\frac{S_{\rm PM}}{V_{\rm cyt}}\big)\smallk{\pipTWO}{\piTHREEk}\surf{\piTHREEk-\gia GTP}\surf{\rm \pipTWO} \\
& \underbrace{\qquad + S_{\rm PM}\smallk{-\pipTWO}{\piTHREEk}\surf{\piTHREEk-\gia GTP-\pipTWO}}_{\text{PI3K binding PIP$_2$}} \\[.5em]
& \underbrace{ + S_{\rm PM}\smallk{\pipTWO,cat}{\piTHREEk}\surf{\piTHREEk-\gia GTP-\pipTWO}}_{\text{Formation of PIP$_3$}}
\end{split} \\[1em]
\begin{split}
S_{\rm PM}\frac{d}{dt}\surf{\piTHREEk-\gia GDP} ={}& \underbrace{S_{\rm PM}\smallk{hydrolyze}{\piTHREEk}\surf{\piTHREEk-\gia GTP}}_{\text{GTP hydrolysis (inactivation)}} \\[.5em]
& -S_{\rm PM}\smallk{-\gia GDP}{\piTHREEk}\surf{\piTHREEk-\gia GDP} \\
& \underbrace{\qquad + S_{\rm PM}\smallk{\gia GDP}{\piTHREEk}\surf{\piTHREEk}\surf{\rm \gia GDP}}_{\text{G$_{i\alpha}$GDP binding}}
\end{split} 
\end{align}
\begin{align}
\begin{split}
S_{\rm PM}\frac{d}{dt}\surf{\piTHREEk-\gia GTP-\pipTWO} ={}& S_{\rm PM}\big(\frac{S_{\rm PM}}{V_{\rm cyt}}\big)\smallk{\pipTWO}{\piTHREEk}\surf{\piTHREEk-\gia GTP}\surf{\rm \pipTWO} \\
& \underbrace{\qquad - S_{\rm PM}\smallk{-\pipTWO}{\piTHREEk}\surf{\piTHREEk-\gia GTP-\pipTWO}}_{\text{PI3K binding PIP$_2$}} \\[.5em]
& \underbrace{ - S_{\rm PM}\smallk{\pipTWO,cat}{\piTHREEk}\surf{\piTHREEk-\gia GTP-\pipTWO}}_{\text{Formation of PIP$_3$}}
\end{split}
\end{align}

\pagebreak
\subsection{Proteins in P2Y$_1$ pathway}
\subsection{P2Y$_1$}

We continue to use the model from Purvis et al. with a Quasi-Steady State assumption on certain states that rapidly switch between ``active" and ``inactive" states.
We also introduce two new states, ADP-pP2Y$_1$ and ADP-pP2Y$_1$-G$_{\rm q}$GDP, that represent desensitization via PKC phosphorylation.

\begin{figure}[ht!]
\centering\includegraphics[width=.7\textwidth]{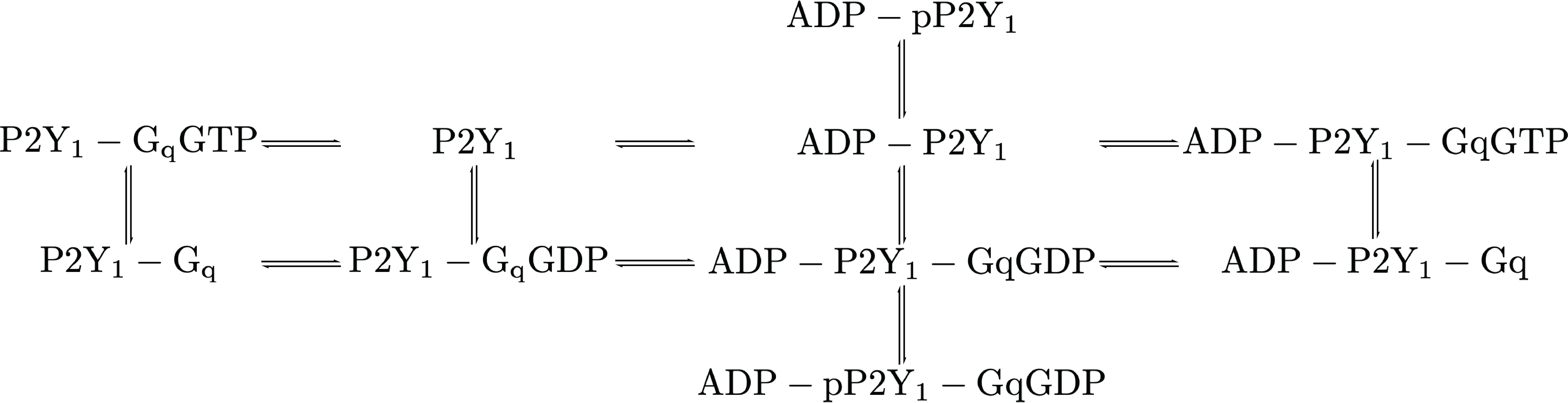}
\caption{P2Y$_1$ reaction diagram}
\end{figure}

\begin{table}[htbp]
   \centering
   \begin{tabular}{| c  c  c  c |} 
   \hline
      Parameter & Value & Units & Reference \\ \hline 
  $\bigk{act}{\pTWOyONE}$ & See Table \ref{tab:CaEstimatedParams} & n.d. & \\
  $\alpha^{\pTWOyONE}$ & See Table \ref{tab:CaEstimatedParams} & n.d. & \\
  $\beta^{\pTWOyONE}$ & See Table \ref{tab:CaEstimatedParams} & n.d. & \\
  $\delta^{\pTWOyONE}$ & See Table \ref{tab:CaEstimatedParams} & n.d. & \\
  $\gamma^{\pTWOyONE}$ & See Table \ref{tab:CaEstimatedParams} & n.d. & \\
  $\smallk{G_qGDP}{\pTWOyONE}$ & See Table \ref{tab:CaEstimatedParams} & dm$^2$/mol-s & \\
  $\smallk{-G_qGDP}{\pTWOyONE}$ & See Table \ref{tab:CaEstimatedParams} & 1/s & \\
  $\smallk{G_qGTP}{\pTWOyONE}$ & See Table \ref{tab:CaEstimatedParams} & dm$^2$/mol-s & \\
  $\smallk{-G_qGTP}{\pTWOyONE}$ & See Table \ref{tab:CaEstimatedParams} & 1/s & \\
  $\smallk{ADP}{\pTWOyONE}$ & See Table \ref{tab:CaEstimatedParams} & 1/M-s & \\
  $\smallk{-ADP}{\pTWOyONE}$ & See Table \ref{tab:CaEstimatedParams} & 1/M-s & \\
  $\smallk{-GDP}{\pTWOyONE}$ & See Table \ref{tab:CaEstimatedParams} & 1/s & \\
  $\smallk{GDP}{\pTWOyONE}$ & See Table \ref{tab:CaEstimatedParams} & 1/M-s & \\
  $\smallk{-GTP}{\pTWOyONE}$ & See Table \ref{tab:CaEstimatedParams} & 1/s & \\
  $\smallk{GTP}{\pTWOyONE}$ & See Table \ref{tab:CaEstimatedParams} & 1/M-s & \\
  $\smallk{phos}{\pTWOyONE}$ & See Table \ref{tab:CaEstimatedParams} & 1/M-s & \\
  $\smallk{dephos}{\pTWOyONE}$ & See Table \ref{tab:CaEstimatedParams} & 1/s & \\
  $\conc{GDP}$ & $1.3362\cdot10^{-5}$ & M & \cite{Purvis2008} \\
  $\conc{GTP}$ & $7.117\cdot10^{-4}$ & M & \cite{Purvis2008} \\
  $\NS{\pTWOyONE}$ & 983 & \# & \cite{Zeiler2014} \\ \hline
   \end{tabular}
   \caption{Parameter values associated with P2Y$_{12}$. At time At time $t=0$ s in the steady-state simulation, $\surf{\pTWOyONE} = \NS{\pTWOyONE}/S_{\rm PM}$ and all other states are initialized to 0.}
   \label{tab:P2Y1Params}
\end{table}

\begin{align}
\begin{split}
S_{\rm PM}\frac{d}{dt}\surf{\pTWOyONE} = {}& \underbrace{-S_{\rm PM}\bigg(\smallk{ADP}{\pTWOyONE}\frac{\alphaS{\pTWOyONE} \bigk{act}{\pTWOyONE}+1}{\bigk{act}{\pTWOyONE}+1}\bigg)\conc{ADP}\surf{\pTWOyONE} + S_{\rm PM}\smallk{-ADP}{\pTWOyONE}\surf{ADP-\pTWOyONE}}_{\text{ADP Binding}} \\[.5em]
& -S_{\rm PM}\bigg(\smallk{\gq GDP}{\pTWOyONE}\frac{\betaS{\pTWOyONE} \bigk{act}{\pTWOyONE} }{\bigk{act}{\pTWOyONE}+1}\bigg)\surf{\gq GDP}\surf{\pTWOyONE} \\
& \underbrace{\qquad + S_{\rm PM}\bigg(\smallk{-\gq GDP}{\pTWOyONE}\frac{\betaS{\pTWOyONE} \bigk{act}{\pTWOyONE}}{\betaS{\pTWOyONE} \bigk{act}{\pTWOyONE}+1}\bigg)\surf{\pTWOyONE-\gq GDP}}_{\text{G$_q$GDP Binding}} \\[.5em]
& -S_{\rm PM}\bigg(\smallk{\gq GTP}{\pTWOyONE}\frac{\bigk{act}{\pTWOyONE}}{\bigk{act}{\pTWOyONE}+1}\bigg)\surf{\gq GTP}\surf{\pTWOyONE} \\
& \underbrace{\qquad + S_{\rm PM}\smallk{-\gq GTP}{\pTWOyONE}\surf{\pTWOyONE-\gq GTP}}_{\text{G$_q$GTP Binding}}
\end{split} \\[1em]
\begin{split}
S_{\rm PM}\frac{d}{dt}\surf{ADP-\pTWOyONE} ={}& \underbrace{S_{\rm PM}\bigg(\smallk{ADP}{\pTWOyONE} \frac{\alphaS{\pTWOyONE} \bigk{act}{\pTWOyONE}+1}{\bigk{act}{\pTWOyONE}+1}\bigg)\conc{ADP}\surf{\pTWOyONE} - S_{\rm PM}\smallk{-ADP}{\pTWOyONE}\surf{ADP-\pTWOyONE}}_{\text{ADP Binding}} \\[.5em]
& - S_{\rm PM}\bigg( \smallk{\gq GDP}{\pTWOyONE}\frac{\alphaS{\pTWOyONE}\betaS{\pTWOyONE} \bigk{act}{\pTWOyONE}}{\alphaS{\pTWOyONE} \bigk{act}{\pTWOyONE}+1}\bigg) \surf{\gq GDP}\surf{ADP-\pTWOyONE} \\
& \underbrace{\qquad + S_{\rm PM}\bigg(\frac{\smallk{-\gq GDP}{\pTWOyONE}}{\gammaS{\pTWOyONE}}\frac{\alphaS{\pTWOyONE}\betaS{\pTWOyONE} \bigk{act}{\pTWOyONE}}{\alphaS{\pTWOyONE}\betaS{\pTWOyONE}\deltaS{\pTWOyONE} \bigk{act}{\pTWOyONE}+1}\bigg)\surf{ADP-\pTWOyONE-\gq GDP}}_{\text{G$_q$GDP Binding}} \\[.5em]
& - S_{\rm PM}\bigg(\smallk{\gq GTP}{\pTWOyONE}\frac{\alphaS{\pTWOyONE} \bigk{act}{\pTWOyONE}}{\alphaS{\pTWOyONE} \bigk{act}{\pTWOyONE}+1}\bigg) \surf{\gq GTP}\surf{ADP-\pTWOyONE} \\
& \underbrace{\qquad + S_{\rm PM}\smallk{-\gq GTP}{\pTWOyONE}\surf{ADP-\pTWOyONE-\gq GTP}}_{\text{G$_q$GTP Binding}} \\[.5em]
& -S_{\rm PM}\smallk{phos}{\pTWOyONE}\surf{ADP-\pTWOyONE}\conc{\pkc a-Ca-DAG} \\
& \underbrace{\qquad + S_{\rm PM}\smallk{dephos}{\pTWOyONE}\surf{ADP-p\pTWOyONE} \qquad \qquad \qquad}_{\text{Inactivation by \pkc}}
\end{split}
\end{align}

\begin{align}
\begin{split}
S_{\rm PM}\frac{d}{dt}\surf{\pTWOyONE-\gq GDP} ={}& -S_{\rm PM}\bigg(\smallk{ADP}{\pTWOyONE}\frac{\alphaS{\pTWOyONE}\betaS{\pTWOyONE} \bigk{act}{\pTWOyONE}+1}{\betaS{\pTWOyONE} \bigk{act}{\pTWOyONE}+1}\bigg)\conc{ADP}\surf{\pTWOyONE-\gq GDP} \\
& \underbrace{\qquad +S_{\rm PM}\bigg(\frac{\smallk{-ADP}{\pTWOyONE}}{\gammaS{\pTWOyONE}}\frac{\alphaS{\pTWOyONE}\betaS{\pTWOyONE} \bigk{act}{\pTWOyONE}+1}{\alphaS{\pTWOyONE}\betaS{\pTWOyONE}\deltaS{\pTWOyONE} \bigk{act}{\pTWOyONE}+1}\bigg)\surf{ADP-\pTWOyONE-\gq GDP}}_{\text{ADP Binding}} \\[.5em]
& + S_{\rm PM}\bigg(\smallk{\gq GDP}{\pTWOyONE}\frac{\betaS{\pTWOyONE} \bigk{act}{\pTWOyONE}}{\bigk{act}{\pTWOyONE}+1}\bigg)\surf{\gq GDP}\surf{\pTWOyONE} \\
& \underbrace{\qquad - S_{\rm PM}\bigg(\smallk{-\gq GDP}{\pTWOyONE}\frac{\betaS{\pTWOyONE} \bigk{act}{\pTWOyONE}}{\betaS{\pTWOyONE} \bigk{act}{\pTWOyONE}+1}\bigg)\surf{\pTWOyONE-\gq GDP}}_{\text{G$_q$GDP Binding}} \\[.5em]
& + S_{\rm PM}\smallk{GDP}{\pTWOyONE}\conc{GDP}\surf{\pTWOyONE-\gq } \\
& \underbrace{\qquad - S_{\rm PM}\bigg(\smallk{-GDP}{\pTWOyONE}\frac{\betaS{\pTWOyONE} \bigk{act}{\pTWOyONE}}{\betaS{\pTWOyONE} \bigk{act}{\pTWOyONE}+1}\bigg)\surf{\pTWOyONE-\gq GDP}}_{\text{GDP Binding}}
\end{split} \\[1em]
\begin{split}
S_{\rm PM}\frac{d}{dt}\surf{ADP-\pTWOyONE-\gq GDP} ={}& S_{\rm PM}\bigg(\smallk{ADP}{\pTWOyONE}\frac{\alphaS{\pTWOyONE}\betaS{\pTWOyONE} \bigk{act}{\pTWOyONE}+1}{\betaS{\pTWOyONE} \bigk{act}{\pTWOyONE}+1}\bigg)\conc{ADP}\surf{\pTWOyONE-\gq GDP} \\
& \underbrace{\qquad -S_{\rm PM}\bigg(\frac{\smallk{-ADP}{\pTWOyONE}}{\gammaS{\pTWOyONE}}\frac{\alphaS{\pTWOyONE}\betaS{\pTWOyONE} \bigk{act}{\pTWOyONE}+1}{\alphaS{\pTWOyONE}\betaS{\pTWOyONE}\deltaS{\pTWOyONE} \bigk{act}{\pTWOyONE}+1}\bigg)\surf{ADP-\pTWOyONE-\gq GDP}}_{\text{ADP Binding}} \\
& +S_{\rm PM}\bigg( \smallk{\gq GDP}{\pTWOyONE}\frac{\alphaS{\pTWOyONE}\betaS{\pTWOyONE} \bigk{act}{\pTWOyONE}}{\alphaS{\pTWOyONE} \bigk{act}{\pTWOyONE}+1}\bigg) \surf{\gq GDP}\surf{ADP-\pTWOyONE} \\
& \underbrace{\qquad - S_{\rm PM}\bigg(\frac{\smallk{-\gq GDP}{\pTWOyONE}}{\gammaS{\pTWOyONE}}\frac{\alphaS{\pTWOyONE}\betaS{\pTWOyONE} \bigk{act}{\pTWOyONE}}{\alphaS{\pTWOyONE}\betaS{\pTWOyONE}\deltaS{\pTWOyONE} \bigk{act}{\pTWOyONE}+1}\bigg)\surf{ADP-\pTWOyONE-\gq GDP}}_{\text{G$_q$GDP Binding}} \\
& + S_{\rm PM}\smallk{GDP}{\pTWOyONE}\conc{GDP}\surf{ADP-\pTWOyONE-\gq } \\
& \underbrace{\qquad - S_{\rm PM}\bigg(\smallk{-GDP}{\pTWOyONE}\frac{\alphaS{\pTWOyONE}\betaS{\pTWOyONE}\deltaS{\pTWOyONE} \bigk{act}{\pTWOyONE}}{\alphaS{\pTWOyONE}\betaS{\pTWOyONE}\deltaS{\pTWOyONE} \bigk{act}{\pTWOyONE}+1}\bigg)\surf{ADP-\pTWOyONE-\gq GDP}}_{\text{GDP Binding}}\\[.5em]
& -S_{\rm PM}\smallk{phos}{\pTWOyONE}\surf{ADP-\pTWOyONE-\gq GDP}\conc{\pkc a-Ca-DAG} \\
& \underbrace{\qquad + S_{\rm PM}\smallk{dephos}{\pTWOyONE}\surf{ADP-p\pTWOyONE-\gq GDP} \qquad \qquad \qquad  }_{\text{Inactivation by \pkc}}
\end{split}
\end{align}

\begin{align}
\begin{split}
S_{\rm PM}\frac{d}{dt}\surf{\pTWOyONE-\gq GTP} ={}& \underbrace{S_{\rm PM}\smallk{GTP}{\pTWOyONE}\surf{\pTWOyONE-\gq }\conc{GTP} - S_{\rm PM}\smallk{-GTP}{\pTWOyONE}\surf{\pTWOyONE-\gq GTP}}_{\text{GTP Binding}} \\[.5em]
 & + S_{\rm PM}\bigg(\smallk{\gq GTP}{\pTWOyONE}\frac{\bigk{act}{\pTWOyONE}}{\bigk{act}{\pTWOyONE}+1}\bigg)\surf{\gq GTP}\surf{\pTWOyONE} \\
 & \underbrace{\qquad - S_{\rm PM}\smallk{-\gq GTP}{\pTWOyONE}\surf{\pTWOyONE-\gq GTP}}_{\text{G$_q$GTP binding}}
\end{split}\\[1em]
\begin{split}
S_{\rm PM}\frac{d}{dt}\surf{ADP-\pTWOyONE-\gq GTP} ={}& S_{\rm PM}\smallk{GTP}{\pTWOyONE}\surf{ADP-\pTWOyONE-\gq }\conc{GTP} \\
& \underbrace{\qquad - S_{\rm PM}\smallk{-GTP}{\pTWOyONE}\surf{ADP-\pTWOyONE-\gq GTP}}_{\text{GTP Binding}} \\[.5em]
 & + S_{\rm PM}\bigg(\smallk{\gq GTP}{\pTWOyONE}\frac{\alphaS{\pTWOyONE} \bigk{act}{\pTWOyONE}}{\alphaS{\pTWOyONE} \bigk{act}{\pTWOyONE}+1}\bigg) \surf{\gq GTP}\surf{ADP-\pTWOyONE} \\
 & \underbrace{\qquad - S_{\rm PM}\smallk{-\gq GTP}{\pTWOyONE}\surf{ADP-\pTWOyONE-\gq GTP}}_{\text{G$_q$GTP binding}}
\end{split} \\[1em]
\begin{split}
S_{\rm PM}\frac{d}{dt}\surf{\pTWOyONE-\gq } ={}& -S_{\rm PM}\smallk{GTP}{\pTWOyONE}\surf{\pTWOyONE-\gq }\conc{GTP} \\
& \underbrace{\qquad + S_{\rm PM}\smallk{-GTP}{\pTWOyONE}\surf{\pTWOyONE-\gq GTP}}_{\text{GTP Binding}} \\[.5em]
& +S_{\rm PM}\bigg(\smallk{-GDP}{\pTWOyONE}\frac{\betaS{\pTWOyONE} \bigk{act}{\pTWOyONE}}{\betaS{\pTWOyONE} \bigk{act}{\pTWOyONE}+1}\bigg)\surf{\pTWOyONE-\gq GDP} \\
& \underbrace{\qquad - S_{\rm PM}\smallk{GDP}{\pTWOyONE}\surf{\pTWOyONE-\gq }\conc{GDP}}_{\text{GDP binding}}
\end{split} \\[1em]
\begin{split}
S_{\rm PM}\frac{d}{dt}\surf{ADP-\pTWOyONE-\gq } ={}& -S_{\rm PM}\smallk{GTP}{\pTWOyONE}\surf{ADP-\pTWOyONE-\gq }\conc{GTP} \\
&\underbrace{\qquad+ S_{\rm PM}\smallk{-GTP}{\pTWOyONE}\surf{ADP-\pTWOyONE-\gq GTP}}_{\text{GTP Binding}} \\[.5em]
& - S_{\rm PM}\smallk{GDP}{\pTWOyONE}\surf{ADP-\pTWOyONE-\gq }\conc{GDP} \\
& \underbrace{\qquad + S_{\rm PM}\bigg(\smallk{-GDP}{\pTWOyONE}\frac{\alphaS{\pTWOyONE}\betaS{\pTWOyONE}\deltaS{\pTWOyONE} \bigk{act}{\pTWOyONE}}{\alphaS{\pTWOyONE}\betaS{\pTWOyONE}\deltaS{\pTWOyONE} \bigk{act}{\pTWOyONE}+1}\bigg)\surf{ADP-\pTWOyONE-\gq GDP}}_{\text{GDP binding}}
\end{split}
\end{align}

\begin{align}
\begin{split}
S_{\rm PM}\frac{d}{dt}\surf{ADP-p\pTWOyONE} ={}& S_{\rm PM}\smallk{phos}{\pTWOyONE}\surf{ADP-P2Y_1}\conc{\pkc a-Ca-DAG} \\
& \underbrace{\qquad - S_{\rm PM}\smallk{dephos}{\pTWOyONE}\surf{ADP-p\pTWOyONE} \qquad \qquad \qquad}_{\text{Inactivation by \pkc}}
\end{split} \\[1em]
\begin{split}
S_{\rm PM}\frac{d}{dt}\surf{ADP-p\pTWOyONE-\gq GDP} ={}& S_{\rm PM}\smallk{phos}{\pTWOyONE}\surf{ADP-P2Y_1-GqGDP}\conc{\pkc a-Ca-DAG} \\
& \underbrace{\qquad - S_{\rm PM}\smallk{dephos}{\pTWOyONE}\surf{ADP-p\pTWOyONE-\gq GDP}\qquad \qquad}_{\text{Inactivation by \pkc}}
\end{split}
\end{align}

\pagebreak
\subsubsection{G$_q$}

\begin{table}[htbp]
   \centering
   \begin{tabular}{| c  c  c  c |} 
   \hline
      Parameter & Value & Units & Reference \\ \hline 
  $\smallk{GTP}{}$ & See Table \ref{tab:CaEstimatedParams} & 1/s &  \\
  $\smallk{\gqbg}{\gq}$ & 103584 & 1/M-s & \cite{Kinzer-Ursem2007} \\
  $\smallk{-\gqbg}{\gq}$ & 7.7832 & 1/s & \cite{Kinzer-Ursem2007} \\
  $\NS{\gq}$ & 18583 & \# & \cite{Zeiler2014} \\ \hline
   \end{tabular}
   \caption{Parameter values associated with G$_{\rm q}$. At time $t=0$ s in the steady-state simulation, $\surf{\gq GDP} = \NS{\gq}/S_{\rm PM}$ and all other states are initialized to 0.}
   \label{tab:GqParams}
\end{table}

\begin{align}
\begin{split}
S_{\rm PM}\frac{d}{dt}\surf{\gqa GTP} ={}& \underbrace{-S_{\rm PM}\Big(\smallk{G_{\beta\gamma}}{\gq}\frac{S_{\rm PM}}{V_{\rm cyt}}\Big)\surf{\gqa GTP}\surf{\gqbg} + S_{\rm PM}\smallk{-\gqbg}{\gq}\surf{\gq GTP}}_{\text{G$_{q\beta\gamma}$ binding G$_{q\alpha}$GTP}} \\[.5em]
& \underbrace{+S_{\rm PM}\smallk{-\gqa GTP}{\plc}\surf{\plc-\gqa GTP} - S_{\rm PM}\smallk{\gqa GTP}{\plc}\surf{PLC}\surf{\gqa GTP}}_{\text{$\rm \plc$ binding $\rm \gqa GTP$}}
\end{split} \\[1em]
\begin{split}
S_{\rm PM}\frac{d}{dt}\surf{\gqa GDP} ={}& \underbrace{S_{\rm PM}\smallk{autohydrolysis}{\gq}\surf{\gq GTP}}_{\text{Autohydrolysis of $\gq$ GTP}} \\[.5em]
& \underbrace{-S_{\rm PM}\Big(\smallk{G_{\beta\gamma}}{\gq}\frac{S_{\rm PM}}{V_{\rm cyt}}\Big)\surf{\gqa GDP}\surf{\gqbg} + V_{\rm cyt}\smallk{-G_{\beta\gamma}}{\gq}\surf{\gq GDP}}_{\text{G$_{q\beta\gamma}$ binding G$_{q\alpha}$GDP}} \\[.5em]
& \underbrace{+S_{\rm PM}\smallk{-\gqa GDP}{\plc}\surf{\plc-\gqa GDP} - S_{\rm PM}\smallk{\gqa GDP}{\plc}\surf{\plc}\surf{\gqa GDP}}_{\text{$\rm \plc$ binding $\rm \gqa GDP$}}
\end{split} \\[1em]
\begin{split}
S_{\rm PM}\frac{d}{dt}\surf{\gqbg} ={}& \underbrace{S_{\rm PM}\smallk{autohydrolysis}{\gq}\surf{\gq GTP}}_{\text{Autohydrolysis of G$_i$GTP}} \\[.5em]
& \underbrace{-S_{\rm PM}\Big(\smallk{\gqbg}{\gq}\frac{S_{\rm PM}}{V_{\rm cyt}}\Big)\surf{\gqa GTP}\surf{\gqbg} + S_{\rm PM}\smallk{-\gqbg}{\gq}\surf{\gq GTP}}_{\text{G$_{q\beta\gamma}$ binding G$_{q\alpha}$GTP}} \\[.5em]
& \underbrace{-S_{\rm PM}\Big(\smallk{\gqbg}{\gq}\frac{S_{\rm PM}}{V_{\rm cyt}}\Big)\surf{\gqa GDP}\surf{\gqbg} + S_{\rm PM}\smallk{-\gqbg}{\gq}\surf{\gq GDP}}_{\text{G$_{q\beta\gamma}$ binding G$_{q\alpha}$GDP}}
\end{split}
\end{align}

\begin{align}
\begin{split}
S_{\rm PM}\frac{d}{dt}\surf{\gq GDP} ={}&  -S_{\rm PM}\bigg(\smallk{\gq GDP}{\pTWOyONE}\frac{\betaS{\pTWOyONE} \bigk{act}{\pTWOyONE} }{\bigk{act}{\pTWOyONE}+1}\bigg)\surf{\gq GDP}\surf{\pTWOyONE} \\
& \underbrace{\qquad + S_{\rm PM}\bigg(\smallk{-\gq GDP}{\pTWOyONE}\frac{\betaS{\pTWOyONE} \bigk{act}{\pTWOyONE}}{\betaS{\pTWOyONE} \bigk{act}{\pTWOyONE}+1}\bigg)\surf{\pTWOyONE-\gq GDP}}_{\text{G$_q$GDP Binding $\rm \pTWOyONE$}} \\[.5em]
& - S_{\rm PM}\bigg( \smallk{\gq GDP}{\pTWOyONE}\frac{\alphaS{\pTWOyONE}\betaS{\pTWOyONE} \bigk{act}{\pTWOyONE}}{\alphaS{\pTWOyONE} \bigk{act}{\pTWOyONE}+1}\bigg) \surf{\gq GDP}\surf{ADP-\pTWOyONE} \\
& \underbrace{\qquad + S_{\rm PM}\bigg(\frac{\smallk{-\gq GDP}{\pTWOyONE}}{\gammaS{\pTWOyONE}}\frac{\alphaS{\pTWOyONE}\betaS{\pTWOyONE} \bigk{act}{\pTWOyONE}}{\alphaS{\pTWOyONE}\betaS{\pTWOyONE}\deltaS{\pTWOyONE} \bigk{act}{\pTWOyONE}+1}\bigg)\surf{ADP-\pTWOyONE-\gq GDP}}_{\text{G$_q$GDP Binding $\rm ADP-\pTWOyONE$}} \\[.5em]
& \underbrace{+S_{\rm PM}\Big(\smallk{\gqbg}{\gq}\frac{S_{\rm PM}}{V_{\rm cyt}}\Big)\surf{\gqa GDP}\surf{\gqbg} - S_{\rm PM}\smallk{-\gqbg}{\gq}\surf{\gq GDP}}_{\text{G$_{q\beta\gamma}$ binding G$_{q\alpha}$GDP}}
\end{split} \\[1em]
\begin{split}
S_{\rm PM}\frac{d}{dt}\surf{\gq GTP} ={}& -S_{\rm PM}\bigg(\smallk{\gq GTP}{\pTWOyONE}\frac{\bigk{act}{\pTWOyONE}}{\bigk{act}{\pTWOyONE}+1}\bigg)\surf{\gq GTP}\surf{\pTWOyONE} \\
& \underbrace{\qquad + S_{\rm PM}\smallk{-\gq GTP}{\pTWOyONE}\surf{\pTWOyONE-\gq GTP}}_{\text{G$_q$GTP Binding}} \\[.5em]
& - S_{\rm PM}\bigg(\smallk{\gq GTP}{\pTWOyONE}\frac{\alphaS{\pTWOyONE} \bigk{act}{\pTWOyONE}}{\alphaS{\pTWOyONE} \bigk{act}{\pTWOyONE}+1}\bigg) \surf{\gq GTP}\surf{ADP-\pTWOyONE} \\
& \underbrace{\qquad + S_{\rm PM}\smallk{-\gq GTP}{\pTWOyONE}\surf{ADP-\pTWOyONE-\gq GTP}}_{\text{G$_q$GTP Binding}} \\[.5em]
& \underbrace{-S_{\rm PM}\smallk{autohydrolysis}{\gq}\surf{\gq GTP}}_{\text{Autohydrolysis of $\gq$ GTP}} \\[.5em]
& \underbrace{+S_{\rm PM}\Big(\smallk{\gqbg}{\gq}\frac{S_{\rm PM}}{V_{\rm cyt}}\Big)\surf{\gqa GTP}\surf{\gqbg} - S_{\rm PM}\smallk{-\gqbg}{\gq}\surf{\gq GTP}}_{\text{G$_{q\beta\gamma}$ binding G$_{q\alpha}$GTP}}
\end{split}
\end{align}

\pagebreak
\subsubsection{PLC}

\begin{table}[htbp]
   \centering
   \begin{tabular}{| c  c  c  c |} 
   \hline
      Parameter & Value & Units & Reference  \\ \hline 
 $\smallk{hydrolyze}{\plc}$ & See Table \ref{tab:CaEstimatedParams} & 1/s &  \\
 $\smallk{-\gqa GDP}{\plc}$ & See Table \ref{tab:CaEstimatedParams} & 1/s & \\
 $\smallk{-\gqa GTP}{\plc}$ & See Table \ref{tab:CaEstimatedParams} & 1/s & \\
 $\smallk{\gqa GDP}{\plc}$ & See Table \ref{tab:CaEstimatedParams} & dm$^2$/mol-s & \\
 $\smallk{\gqa GTP}{\plc}$ & See Table \ref{tab:CaEstimatedParams} & dm$^2$/mol-s & \\
 $\smallk{\pi}{\plc}$        & $1\cdot10^8$  & 1/M-s & \cite{Rittenhouse1983}  \\
 $\smallk{-\pi}{\plc}$       & $7.05\cdot10^4$      & 1/s & \cite{Rittenhouse1983}   \\
 $\smallk{\pi,cat}{\plc}$    & 1.43       & 1/s & \cite{Rittenhouse1983}  \\
 $\smallk{\pip}{\plc}$       & $1\cdot10^8$  & 1/M-s & \cite{Rittenhouse1983} \\
 $\smallk{-\pip}{\plc}$      & $1.9\cdot10^7$   & 1/s & \cite{Rittenhouse1983}   \\
 $\smallk{\pip,cat}{\plc}$   & 0.35       & 1/s & \cite{Rittenhouse1983}  \\
 $\smallk{\pipTWO}{\plc}$      & $1\cdot10^8$  & 1/M-s & \cite{Rittenhouse1983} \\
 $\smallk{-\pipTWO}{\plc}$     & $5\cdot10^4$      & 1/s & \cite{Rittenhouse1983}   \\
 $\smallk{\pipTWO,cat}{\plc}$  & 9.8505     & 1/s & \cite{Rittenhouse1983}   \\
 $\NS{\plc}$ & 1579 & \# & \cite{Zeiler2014} \\ \hline
   \end{tabular}
   \caption{Parameter values associated with PLC. At time $t=0$ s in the steady-state simulation, $\surf{\plc} = \NS{\plc}/S_{\rm PM}$ and all other states are initialized to 0.}
   \label{tab:PLCParams}
\end{table}

\begin{align}
\begin{split}
S_{\rm PM}\frac{d}{dt}\surf{\plc} ={}& \underbrace{S_{\rm PM}\smallk{-\gqa GTP}{\plc}\surf{\plc-\gqa GTP} - S_{\rm PM}\smallk{\gqa GTP}{\plc}\surf{\plc}\surf{\rm \gqa GTP}}_{\text{$\gqa$GTP binding}} \\[.5em]
& \underbrace{+S_{\rm PM}\smallk{-\gqa GDP}{\plc}\surf{\plc-\gqa GDP} - S_{\rm PM}\smallk{\gqa GDP}{\plc}\surf{\plc}\surf{\rm \gqa GDP}}_{\text{$\gqa$GDP binding}} 
\end{split}
\end{align}
\begin{align}
\begin{split}
S_{\rm PM}\frac{d}{dt}\surf{\plc-\gqa GTP} ={}& \underbrace{-S_{\rm PM}\smallk{-\gqa GTP}{\plc}\surf{\plc-\gqa GTP} + S_{\rm PM}\smallk{\gqa GTP}{\plc}\surf{\plc}\surf{\gqa GTP}}_{\text{$\gqa$GTP binding}} \\[.5em]
& \underbrace{-S_{\rm PM}\smallk{hydrolyze}{\plc}\surf{\plc-\gqa GTP}}_{\text{GTP hydrolysis (inactivation)}} \\[.5em]
& -S_{\rm PM}\big(\frac{S_{\rm PM}}{V_{\rm cyt}}\big)\smallk{\pi}{\plc}\surf{\plc-\gqa GTP}\surf{\pi} \\
& \underbrace{\qquad + S_{\rm PM}\smallk{-\pi}{\plc}\surf{\plc-\gqa GTP-\pi}}_{\text{$\rm \plc$ binding $\rm \pi$}} \\[.5em]
& \underbrace{ + S_{\rm PM}\smallk{\pi,cat}{\plc}\surf{\plc-\gqa GTP-\pi}}_{\text{Formation of $\rm \iONEp$}} \\[.5em]
& -S_{\rm PM}\big(\frac{S_{\rm PM}}{V_{\rm cyt}}\big)\smallk{\pip}{\plc}\surf{\plc-\gqa GTP}\surf{\pip} \\
& \underbrace{\qquad + S_{\rm PM}\smallk{-\pip}{\plc}\surf{\plc-\gqa GTP-\pip}}_{\text{PLC binding $\rm \pip$}} \\[.5em]
& \underbrace{ + S_{\rm PM}\smallk{\pip,cat}{\plc}\surf{\plc-\gqa GTP-\pip}}_{\text{Formation of $\rm \ipTWO$}} \\[.5em]
& -S_{\rm PM}\big(\frac{S_{\rm PM}}{V_{\rm cyt}}\big)\smallk{\pipTWO}{\plc}\surf{\plc-\gqa GTP}\surf{\pipTWO} \\
& \underbrace{\qquad + S_{\rm PM}\smallk{-\pipTWO}{\plc}\surf{\plc-\gqa GTP-\pipTWO}}_{\text{PLC binding $\rm \pipTWO$}} \\[.5em]
& \underbrace{ + S_{\rm PM}\smallk{\pipTWO,cat}{\plc}\surf{\plc-\gqa GTP-\pipTWO}}_{\text{Formation of $\rm \ipTHREE$}}
\end{split} \\[1em]
\begin{split}
S_{\rm PM}\frac{d}{dt}\surf{\plc-\gqa GDP} ={}& \underbrace{S_{\rm PM}\smallk{hydrolyze}{\plc}\surf{\plc-\gqa GTP}}_{\text{GTP hydrolysis (inactivation)}} \\[.5em]
& -S_{\rm PM}\smallk{-\gqa GDP}{\plc}\surf{\plc-\gqa GDP} \\
& \underbrace{\qquad + S_{\rm PM}\smallk{\gqa GDP}{\plc}\surf{\plc}\surf{\gqa GDP}}_{\text{$\gqa$ GDP binding}}
\end{split} \\[1em]
\begin{split}
S_{\rm PM}\frac{d}{dt}\surf{\plc-\gqa GTP-\pi} ={}& S_{\rm PM}\big(\frac{S_{\rm PM}}{V_{\rm cyt}}\big)\smallk{\pi}{\plc}\surf{\plc-\gqa GTP}\surf{\pi} \\
& \underbrace{\qquad - S_{\rm PM}\smallk{-\pi}{\plc}\surf{\plc-\gqa GTP-\pi}}_{\text{PLC binding PI}} \\[.5em]
& \underbrace{ - S_{\rm PM}\smallk{\pi,cat}{\plc}\surf{\plc-\gqa GTP-\pi}}_{\text{Formation of I$_1$P}}
\end{split}
\end{align}
\begin{align}
\begin{split}
S_{\rm PM}\frac{d}{dt}\surf{\plc-\gqa GTP-\pip} ={}& S_{\rm PM}\big(\frac{S_{\rm PM}}{V_{\rm cyt}}\big)\smallk{\pip}{\plc}\surf{\plc-\gqa GTP}\surf{\pip} \\
& \underbrace{\qquad - S_{\rm PM}\smallk{-\pip}{\plc}\surf{\plc-\gqa GTP-\pip}}_{\text{PLC binding PI}} \\[.5em]
& \underbrace{ - S_{\rm PM}\smallk{\pip,cat}{\plc}\surf{\plc-\gqa GTP-\pip}}_{\text{Formation of IP$_2$}}
\end{split} \\[1em]
\begin{split}
S_{\rm PM}\frac{d}{dt}\surf{\plc-\gqa GTP-\pipTWO} ={}& S_{\rm PM}\big(\frac{S_{\rm PM}}{V_{\rm cyt}}\big)\smallk{\pipTWO}{\plc}\surf{\plc-\gqa GTP}\surf{\pipTWO} \\
& \underbrace{\qquad - S_{\rm PM}\smallk{-\pipTWO}{\plc}\surf{\plc-\gqa GTP-\pipTWO}}_{\text{PLC binding PIP$_2$}} \\[.5em]
& \underbrace{ - S_{\rm PM}\smallk{\pipTWO,cat}{\plc}\surf{\plc-\gqa GTP-\pipTWO}}_{\text{Formation of IP$_3$}}
\end{split}
\end{align}

\pagebreak
\subsubsection{PKC}

PKC exists in the cytosol at approximately the same copy number as CalDAG; its concentration in this model is controlled by the volume clustering coefficient $\beta_v$.
PKC inactivates P2Y$_1$ through phosphorylation \cite{Paul_Blatt_Schug_Clark_etAl}

\begin{table}[htbp]
   \centering
   \begin{tabular}{| c  c  c  c |} 
   \hline
      Parameter & Value & Units & Reference \\ \hline 
  $\smallk{Ca}{\pkc}$         & $6\cdot10^5$ & 1/M-s & \cite{BallaIyengar1999} \\
  $\smallk{-Ca}{\pkc}$        & 0.5     & 1/s & \cite{BallaIyengar1999}   \\
  $\smallk{\dag}{\pkc-Ca}$      & $8\cdot10^3$   & 1/M-s & \cite{BallaIyengar1999} \\
  $\smallk{-\dag}{\pkc-Ca}$     & 8.6348  & 1/s & \cite{BallaIyengar1999}   \\
  $\smallk{act}{\pkc}$        & 1      & 1/s & \cite{BallaIyengar1999}   \\
  $\smallk{-act}{\pkc}$       & 2      & 1/s & \cite{BallaIyengar1999}   \\
  $\smallk{act}{\pkc-Ca}$     & 1.2706  & 1/M-s & \cite{BallaIyengar1999} \\
  $\smallk{-act}{\pkc_Ca}$    & 3.5026  & 1/s & \cite{BallaIyengar1999}   \\
  $\smallk{act}{\pkc-Ca-\dag}$  & 1      & 1/M-s & \cite{BallaIyengar1999} \\
  $\smallk{-act}{\pkc-Ca-\dag}$ & 0.1     & 1/s & \cite{BallaIyengar1999}   \\
  $\NS{PKC}$ & 36135 & \# & \cite{Zeiler2014} \\ \hline
   \end{tabular}
   \caption{Parameter values associated with PKC. At time $t=0$ s in the steady-state simulation, $\conc{\pkc} = \NS{\pkc}/V_{\rm cyt}$ and all other states are initialized to 0.}
   \label{tab:PKCParam}
\end{table}

\begin{align}
\begin{split}
V_{\rm cyt}\frac{d}{dt}\conc{\pkc} ={}& \underbrace{-V_{\rm cyt}\smallk{act}{\pkc}\conc{\pkc} + V_{\rm cyt}\smallk{-act}{\pkc}\conc{\pkc a}}_{\text{activation of PKC}} \\[.5em]
& \underbrace{-V_{\rm cyt}\smallk{Ca}{\pkc}\conc{\pkc}\conc{\rm Ca_{cyt}} + V_{\rm cyt}\smallk{-Ca}{\pkc}\conc{\pkc-Ca}}_{\text{Ca binding to PKC}}
\end{split} \\[1em]
\begin{split}
V_{\rm cyt}\frac{d}{dt}\conc{\pkc a} ={}& \underbrace{V_{\rm cyt}\smallk{act}{\pkc}\conc{\pkc} - V_{\rm cyt}\smallk{-act}{\pkc}\conc{\pkc a}}_{\text{activation of PKC}} \\[.5em]
& \underbrace{-V_{\rm cyt}\smallk{Ca}{\pkc}\conc{\pkc a}\conc{\rm Ca_{cyt}} + V_{\rm cyt}\smallk{-Ca}{\pkc}\conc{\pkc a-Ca}}_{\text{Ca binding to PKCa}}
\end{split}
\end{align}
\begin{align}
\begin{split}
V_{\rm cyt}\frac{d}{dt}\conc{\pkc-Ca} ={}& \underbrace{-V_{\rm cyt}\smallk{act}{\pkc-Ca}\conc{\pkc-Ca} + V_{\rm cyt}\smallk{-act}{\pkc-Ca}\conc{\pkc a-Ca}}_{\text{activation of PKC-Ca}} \\[.5em]
& \underbrace{+V_{\rm cyt}\smallk{Ca}{\pkc}\conc{\pkc}\conc{\rm Ca_{cyt}} - V_{\rm cyt}\smallk{-Ca}{\pkc}\conc{\pkc-Ca}}_{\text{Ca binding to PKC}} \\[.5em]
& \underbrace{-S_{\rm PM}\smallk{DAG}{\pkc}\conc{\pkc-Ca}\surf{\dag} + V_{\rm cyt}\smallk{-DAG}{\pkc}\conc{\pkc-Ca-DAG}}_{\text{DAG binding to PKC-Ca}}
\end{split} \\[1em]
\begin{split}
V_{\rm cyt}\frac{d}{dt}\conc{\pkc a-Ca} ={}& \underbrace{V_{\rm cyt}\smallk{act}{\pkc-Ca}\conc{\pkc-Ca} - V_{\rm cyt}\smallk{-act}{\pkc-Ca}\conc{\pkc a-Ca}}_{\text{activation of PKC}} \\[.5em]
& \underbrace{V_{\rm cyt}\smallk{Ca}{\pkc}\conc{\pkc a}\conc{\rm Ca_{cyt}} - V_{\rm cyt}\smallk{-Ca}{\pkc}\conc{\pkc a-Ca}}_{\text{Ca binding to PKCa}} \\[.5em]
& \underbrace{-S_{\rm PM}\smallk{DAG}{\pkc}\conc{\pkc a-Ca}\surf{\dag} + V_{\rm cyt}\smallk{-DAG}{\pkc}\conc{\pkc a-Ca-DAG}}_{\text{DAG binding to PKCa-Ca}}
\end{split} \\[1em]
\begin{split}
V_{\rm cyt}\frac{d}{dt}\conc{\pkc-Ca-DAG} ={}& -V_{\rm cyt}\smallk{act}{\pkc-Ca-\dag}\conc{\pkc-Ca-DAG} \\
& \underbrace{\qquad + V_{\rm cyt}\smallk{-act}{\pkc-Ca-\dag}\conc{\pkc a-Ca-DAG}}_{\text{activation of PKC-Ca-DAG}} \\[.5em]
& \underbrace{+ S_{\rm PM}\smallk{DAG}{\pkc}\conc{\pkc-Ca}\surf{\dag} - V_{\rm cyt}\smallk{-DAG}{\pkc}\conc{\pkc-Ca-DAG}}_{\text{DAG binding to PKC-Ca}}
\end{split} \\[1em]
\begin{split}
V_{\rm cyt}\frac{d}{dt}\conc{\pkc a-Ca-DAG} ={}& V_{\rm cyt}\smallk{act}{\pkc-Ca-\dag}\conc{\pkc-Ca-DAG} \\
& \underbrace{\qquad - V_{\rm cyt}\smallk{-act}{\pkc-Ca-\dag}\conc{\pkc a-Ca-DAG}}_{\text{activation of PKC-Ca-DAG}} \\[.5em]
& \underbrace{+S_{\rm PM}\smallk{DAG}{\pkc}\conc{\pkc a-Ca}\surf{\dag} - V_{\rm cyt}\smallk{-DAG}{\pkc}\conc{\pkc a-Ca-DAG}}_{\text{DAG binding to PKCa-Ca}}
\end{split}
\end{align}

\pagebreak
\subsection{Calcium and Inositol Species}
\subsubsection{Inositol}

\begin{figure}[ht!]
\centering\includegraphics[width=.6\textwidth]{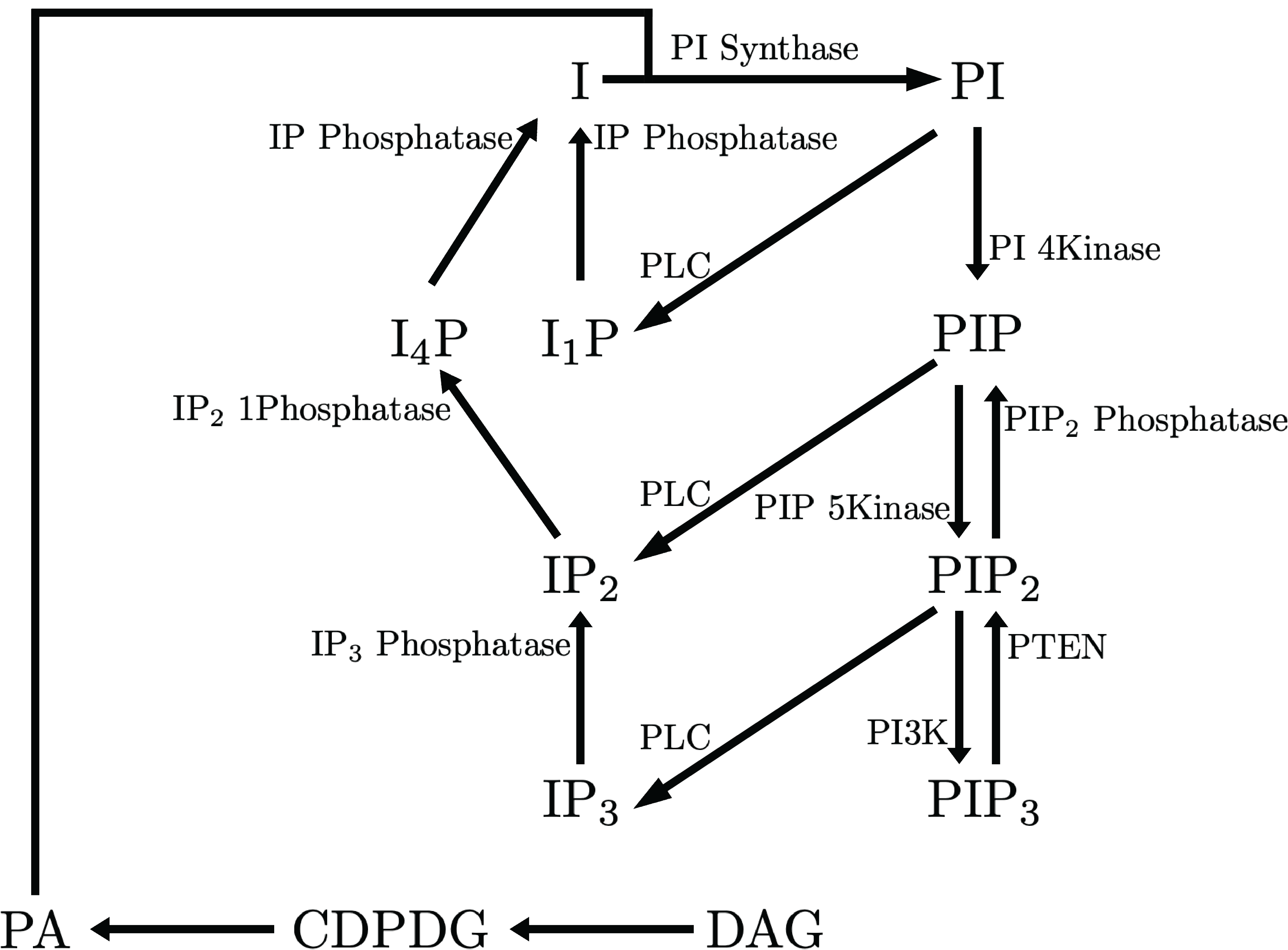}
\caption{Inositol reaction diagram}
\end{figure}

\begin{table}[htbp]
   \centering
   \begin{tabular}{| c  c  c  c |} 
   \hline
      Parameter & Value & Units & Reference \\ \hline 
 $\smallk{I,cat}{\pi Synthase}$            &  13.6    & 1/s & \cite{VARGAS1984123} \\
 $\smallk{\pi,cat}{\pi 4Kinase}$                  &  2.77    & 1/s & \cite{KANOH1990120} \\
 $\bigk{\pi,M}{\pi 4Kinase}$                   &  $1.6\cdot10^{-5}$ & M & \cite{KANOH1990120} \\
 $\smallk{\pip,cat}{\pip 5Kinase}$           &  1.021   & 1/s &  \cite{URUMOW1990152} \\
 $\bigk{\pip,cat}{\pip 5Kinase}$           &  $1.\cdot10^{-5}$  & M & \cite{URUMOW1990152} \\
 $\smallk{-\pipTWO,cat}{\pipTWO Phosphatase}$ &  1      & 1/s & \cite{MATZARIS19943397} \\
 $\bigk{-\pipTWO,M}{\pipTWO Phosphatase}$   &  0.00025 & M & \cite{MATZARIS19943397} \\
 $\smallk{-\iONEp,cat}{\ip Phosphatase}$     &  1      & 1/s & \cite{ATACK1993305} \\
 $\bigk{-\iONEp,M}{\ip Phosphatase}$       &  0.00012 & M & \cite{ATACK1993305} \\
 $\smallk{-\iFOURp,cat}{\ip Phosphatase}$     &  1      & 1/s & \cite{ATACK1993305} \\
 $\bigk{-\iFOURp,M}{\ip Phosphatase}$       &  0.00012 & M & \cite{ATACK1993305} \\
 $\smallk{-\ipTHREE,cat}{\ipTHREE 5Phosphatase}$   &  31.25   & 1/s & \cite{MITCHELL19898873} \\
 $\bigk{-\ipTHREE,M}{\ipTHREE 5Phosphatase}$     &  $2.4\cdot10^{-5}$ & M  & \cite{MITCHELL19898873} \\
 $\smallk{-\ipTWO,cat}{\ipTWO 1Phosphatase}$   &  0.05027 & 1/s  & \cite{MOYER19871018} \\
 $\bigk{\ipTWO,M}{\ipTWO 1Phosphatase}$     &  $9\cdot10^{-7}$   & M  & \cite{MOYER19871018} \\
 $\smallk{-\pipTHREE,cat}{PTEN}$ &  See Table \ref{tab:RAP1EstimatedParams}  & 1/s  & \\
 $\bigk{-\pipTHREE,m}{PTEN}$   &  See Table \ref{tab:RAP1EstimatedParams} & M  & \\
 $\smallk{\dag,cat}{\dag Kinase}$                  &  0.2618  & 1/s  & \cite{Wissing_Heim_Wagner_1989} \\
 $\bigk{\dag,M}{\dag Kinase}$                    &  0.00025 & M  & \cite{Wissing_Heim_Wagner_1989} \\
 $\smallk{PA,cat}{\rm CDPDGSynthase}$             &  8.917   & 1/s  & \cite{KELLEY198714563} \\
 $\bigk{PA,M}{\rm CDPDGSynthase}$                 &  0.0005  & M  & \cite{KELLEY198714563} \\
 $\bigk{CTP,M}{\rm CDPDGSynthase}$                &  0.001   & M  & \cite{KELLEY198714563} \\
 $\bigk{I,M}{\pi Synthase}$              &  $1.3\cdot10^{-5}$ & M  & \cite{VARGAS1984123} \\
 $\bigk{CDPDG,M}{\pi Synthase}$          &  0.00028 & M  & \cite{VARGAS1984123} \\
 $\conc{CTP}$                               &  0.000278 & M & \cite{Purvis2008} \\
  $\conc{\pi Synthase}$ &  $1.482896\cdot10^{-6}$ & M  & \cite{Zeiler2014} \\
  $\conc{\pi 4Kinase}$ &  $491.53105\cdot10^{-9}$ & M & \cite{Zeiler2014} \\
  $\conc{\pip 5Kinase}$ &  $329.3479\cdot10^{-9}$ & M & \cite{Zeiler2014} \\
  $\conc{\pipTWO Phosphatase}$ &  $698.273\cdot10^{-9}$ & M & \cite{Zeiler2014} \\
  $\conc{\ip Phosphatase}$ &  $4.55939\cdot10^{-6}$ & M  & \cite{Zeiler2014} \\
  $\conc{\ipTHREE 5Phosphatase}$ &  $698.273\cdot10^{-9}$ & M & \cite{Zeiler2014} \\
  $\conc{\ipTWO 1Phosphatase}$ &  $2.420569\cdot10^{-6}$ & M & \cite{Zeiler2014} \\
  $\conc{DGKinase}$ &  $692.737739\cdot10^{-9}$ & M & \cite{Zeiler2014} \\
  $\conc{CDPDGSynthase}$ &  $2.159028\cdot10^{-6}$ & M & \cite{Zeiler2014} \\
  $\conc{PTEN}$ &  $100.1882\cdot10^{-9}$ & M & \cite{Zeiler2014} \\
  $\NS{\rm Ins}$ & 291460 & \# & \\ \hline
   \end{tabular}
   \caption{Parameter values associated with inositol. At time $t=0$ s in the steady-state simulation, $[{\rm I}]_S=N_{\rm Ins}/S_{\rm PM}$ and all other states are initialized to 0.}
   \label{tab:InsParams}
\end{table}


\begin{align}
\begin{split}
S_{\rm PM}\frac{d}{dt}\surf{\pi } ={}& S_{\rm PM}\underbrace{\frac{\big(\frac{S_{\rm PM}}{V_{\rm cyt}}\big)\smallk{I,cat}{\pi Synthase}\conc{\pi  Synthase}\surf{\rm CDPDG}}{\bigk{CDPDG,M}{\pi Synthase}\bigk{I,M}{\pi Synthase}+\bigk{CDPDG,M}{\pi Synthase}\conc{\rm I}+ (\bigk{I,M}{\pi Synthase}\surf{\rm CDPDG} + \conc{\rm I}\surf{\rm CDPDG})\big(\frac{S_{\rm PM}}{V_{\rm cyt}}\big)}\conc{\rm I}}_{\text{Synthesis of PI}} \\[.5em]
& \underbrace{-V_{\rm cyt}\frac{\smallk{\pi,cat}{\pi 4Kinase}\conc{\pi 4Kinase}}{\big(\frac{V_{\rm cyt}}{S_{\rm PM}}\big)\bigk{\pi,M}{\pi 4Kinase}+\surf{\pi }}\surf{\pi }}_{\text{Synthesis of $\rm \pip$}} \\[.5em]
& \underbrace{-S_{\rm PM}\big(\frac{S_{\rm PM}}{V_{\rm cyt}}\big)\smallk{\pi}{\plc}\surf{\plc-\gqa GTP}\surf{\pi }
  + S_{\rm PM}\smallk{-\pi}{\plc}\surf{\plc-\gqa GTP-PI}}_{\text{PLC binding PI}}
\end{split} \\[1em]
\begin{split}
S_{\rm PM}\frac{d}{dt}\surf{\pip} ={}& \underbrace{V_{\rm cyt}\frac{\smallk{\pi,cat}{\pi 4Kinase}\conc{\rm \pi 4Kinase}}{\big(\frac{V_{\rm cyt}}{S_{\rm PM}}\big)\bigk{\pi,M}{\pi 4Kinase}+\surf{\pi }}\surf{\pi }}_{\text{Synthesis of $\rm \pip$}} \underbrace{-V_{\rm cyt}\frac{\smallk{\pip,cat}{\pip 5Kinase}\conc{\pip 5Kinase}}{\big(\frac{V_{\rm cyt}}{S_{\rm PM}}\big)\bigk{\pip,M}{\pip 5Kinase}+\surf{\pip}}\surf{\pip}}_{\text{Synthesis of $\rm \pipTWO$}} \\[.5em]
&\underbrace{+ V_{\rm cyt}\frac{\smallk{\pipTWO,cat}{\pipTWO Phosphatase}\conc{\pipTWO Phosphatase}}{\big(\frac{V_{\rm cyt}}{S_{\rm PM}}\big)\bigk{\pipTWO,M}{\pipTWO Phosphatase}+\surf{\rm \pipTWO}}\surf{\rm \pipTWO}}_{\text{Dephosphorylation of $\rm \pipTWO$}} \\[.5em]
& \underbrace{-S_{\rm PM}\big(\frac{S_{\rm PM}}{V_{\rm cyt}}\big)\smallk{\pip}{\plc}\surf{\plc-\gqa GTP}\surf{\pip} + S_{\rm PM}\smallk{-\pip}{\plc}\surf{\plc-\gqa GTP-\pip}}_{\text{PLC binding $\rm \pip$}}
\end{split} \\[1em]
\begin{split}
S_{\rm PM}\frac{d}{dt}\surf{\rm \pipTWO} ={}& \underbrace{V_{\rm cyt}\frac{\smallk{\pip}{\pip 5Kinase}\conc{\pip 5Kinase}}{\big(\frac{V_{\rm cyt}}{S_{\rm PM}}\big)\bigk{\pip,M}{\pip 5Kinase}+\surf{\pip}}\surf{\pip}}_{\text{Synthesis of $\rm \pipTWO$}} \underbrace{ - V_{\rm cyt}\frac{\smallk{\pipTWO}{\pipTWO Phosphatase}\conc{\rm \pipTWO Phosphatase}}{\big(\frac{V_{\rm cyt}}{S_{\rm PM}}\big)\bigk{\pipTWO,M}{\pipTWO Phosphatase}+\surf{\rm \pipTWO}}\surf{\rm \pipTWO}}_{\text{Dephosphorylation of $\rm \pipTWO$}} \\[.5em]
& \underbrace{+V_{\rm cyt}\frac{\smallk{\pipTHREE}{PTEN}\conc{PTEN}}{\big(\frac{V_{\rm cyt}}{S_{\rm PM}}\big)\bigk{\pipTHREE,M}{PTEN}+\surf{\rm \pipTHREE}}\surf{\rm \pipTHREE}}_{\text{$\rm \pipTHREE$ degredation}} \\[.5em]
& -S_{\rm PM}\big(\frac{S_{\rm PM}}{V_{\rm cyt}}\big)\smallk{\pipTWO}{\plc}\surf{\plc-\gqa GTP}\surf{\rm \pipTWO} \\
& \underbrace{\qquad + S_{\rm PM}\smallk{-\pipTWO}{\plc}\surf{\plc-\gqa GTP-\pipTWO}}_{\text{PLC binding $\rm \pipTWO$}} \\[.5em]
& -S_{\rm PM}\big(\frac{S_{\rm PM}}{V_{\rm cyt}}\big)\smallk{\pipTWO}{\piTHREEk}\surf{\piTHREEk-\gia GTP}\surf{\rm \pipTWO} \\
& \underbrace{\qquad + S_{\rm PM}\smallk{-\pipTWO}{\piTHREEk}\surf{\piTHREEk-\gia GTP-\pipTWO}}_{\text{PI3K binding $\rm \pipTWO$}}
\end{split}
\end{align}
\begin{align}
\begin{split}
V_{\rm cyt}\frac{d}{dt}\conc{\iONEp} ={}& \underbrace{S_{\rm PM}\smallk{\pi,cat}{\plc}\surf{\plc-\gqa GTP-PI}}_{\text{Formation of $\rm \iONEp$}} \underbrace{-V_{\rm cyt}\frac{\smallk{\iONEp,cat}{\ip Phosphatase}\conc{\ip Phosphatase}}{\bigk{\iONEp,M}{\ip Phosphatase}+\conc{\iONEp}}\conc{\iONEp}}_{\text{Dephosphorylation of $\rm \iONEp$}}
\end{split} \\[.5em]
\begin{split}
V_{\rm cyt}\frac{d}{dt}\conc{\iFOURp} ={}& \underbrace{V_{\rm cyt}\frac{\smallk{\ipTWO,cat}{\ipTWO Phosphatase}\conc{\ipTWO Phosphatase}}{\bigk{\ipTWO,M}{\ipTWO Phosphatase}+\conc{\ipTWO}}\conc{\ipTWO}}_{\text{Dephosphorylation of $\rm \ipTWO$}} \underbrace{-V_{\rm cyt}\frac{\smallk{\iFOURp,cat}{\ip Phosphatase}\conc{\ip Phosphatase}}{\bigk{\iFOURp,M}{\ip Phosphatase}+\conc{\iFOURp}}\conc{\iFOURp}}_{\text{Dephosphorylation of $\rm \iFOURp$}}
\end{split} \\[1em]
\begin{split}
V_{\rm cyt}\frac{d}{dt}\conc{\ipTWO} ={}& \underbrace{S_{\rm PM}\smallk{\pip,cat}{\plc}\surf{\plc-\gqa GTP-\pip}}_{\text{Formation of $\rm \ipTWO$}} \underbrace{-V_{\rm cyt}\frac{\smallk{\ipTWO,cat}{\ipTWO Phosphatase}\conc{\ipTWO Phosphatase}}{\bigk{\ipTWO,M}{\ipTWO Phosphatase}+\conc{\ipTWO}}\conc{\ipTWO}}_{\text{Dephosphorylation of $\rm \ipTWO$}} \\[.5em]
& \underbrace{+V_{\rm cyt}\frac{\smallk{\ipTHREE,cat}{\ipTHREE 5Phosphatase}\conc{\ipTHREE 5Phosphatase}}{\bigk{\ipTHREE,M}{\ipTHREE 5Phosphatase}+\conc{\ipTHREE}}\conc{\ipTHREE}}_{\text{Dephosphorylation of $\rm \ipTHREE$}}
\end{split} \\[1em]
\begin{split}
V_{\rm cyt}\frac{d}{dt}\conc{I} ={}& \underbrace{V_{\rm cyt}\frac{\smallk{\iONEp,cat}{\ip Phosphatase}\conc{\ip Phosphatase}}{\bigk{\iONEp,M}{\ip Phosphatase}+\conc{\iONEp}}\conc{\iONEp}}_{\text{Dephosphorylation of $\rm \iONEp$}} \underbrace{+V_{\rm cyt}\frac{\smallk{\iFOURp,cat}{\ip Phosphatase}\conc{\rm \ip Phosphatase}}{\bigk{\iFOURp,M}{\ip Phosphatase}+\conc{\iFOURp}}\conc{\iFOURp}}_{\text{Dephosphorylation of $\rm \iFOURp$}} \\[.5em]
& - S_{\rm PM}\underbrace{\frac{\big(\frac{S_{\rm PM}}{V_{\rm cyt}}\big)\smallk{I,cat}{\pi Synthase}\conc{\pi  Synthase}\surf{\rm CDPDG}}{\bigk{CDPDG,M}{\pi Synthase}\bigk{I,M}{\pi Synthase}+\bigk{CDPDG,M}{\pi Synthase}\conc{\rm I}+ (\bigk{I,M}{\pi Synthase}\surf{\rm CDPDG} + \conc{\rm I}\surf{\rm CDPDG})\big(\frac{S_{\rm PM}}{V_{\rm cyt}}\big)}\conc{\rm I}}_{\text{Synthesis of PI}}
\end{split} \\[1em]
\begin{split}
V_{\rm cyt}\frac{d}{dt}\conc{\ipTHREE} ={}& \underbrace{S_{\rm PM}\smallk{\pipTWO,cat}{\plc}\surf{\plc-\gqa GTP-\pipTWO}}_{\text{Formation of $\rm \ipTHREE$}} \underbrace{-V_{\rm cyt}\frac{\smallk{\ipTHREE}{\ipTHREE 5Phosphatase}\conc{\ipTHREE 5Phosphatase}}{\bigk{\ipTHREE,M}{\ipTHREE 5Phosphatase}+\conc{\ipTHREE}}\conc{\ipTHREE}}_{\text{Dephosphorylation of $\rm \ipTHREE$}} \\[.5em]
& \underbrace{-S_{\rm IM}\frac{(k_2L_3+l_4)\conc{Ca_{cyt}^{2+}}}{L_3+\conc{Ca_{cyt}^{2+}}(1+L_1/L_3)}\surf{\rm \ipTHREEr_n}\conc{\ipTHREE} + S_{\rm IM}\frac{k_{-2}+l_{-4}\conc{Ca_{cyt}^{2+}}}{1+\conc{Ca_{cyt}^{2+}}/L_5}\surf{\rm \ipTHREEr_o}}_{\text{$\rm \ipTHREE$ binding to $\rm \ipTHREE$ receptor}}
\end{split} \\[1em]
\begin{split}
S_{\rm PM}\frac{d}{dt}\surf{\rm \pipTHREE} ={}& \underbrace{S_{\rm PM}\smallk{\pipTWO,cat}{\piTHREEk}\surf{\piTHREEk-\gia GTP-\pipTWO}}_{\text{$\rm \pipTHREE$ generation}} \underbrace{-V_{\rm cyt}\frac{\smallk{\pipTHREE,cat}{PTEN}\conc{PTEN}}{\big(\frac{V_{\rm cyt}}{S_{\rm PM}}\big)\bigk{\pipTHREE,M}{PTEN}+\surf{\rm \pipTHREE}}\surf{\rm \pipTHREE}}_{\text{$\rm \pipTHREE$ degradation}} \\[.5em]
& +S_{\rm PM}\smallk{-\pipTHREE}{\rasaTHREE}\surf{\rasaTHREE-\pipTHREE} \\
& \underbrace{\qquad - S_{\rm PM}\big(\frac{S_{\rm PM}}{V_{\rm cyt}}\big)\smallk{\pipTHREE}{\rasaTHREE}\surf{\rm RASA3}\surf{\rm \pipTHREE}}_{\text{$\rm \pipTHREE$ inactivation of $\rm \rasaTHREE$}}
\end{split}
\end{align}

\begin{align}
\begin{split}
S_{\rm PM}\frac{d}{dt}\surf{\dag} ={}& \underbrace{S_{\rm PM}\smallk{\pipTWO,cat}{\plc}\surf{\plc-\gqa GTP-\pipTWO}}_{\text{Conversion to $\rm \ipTHREE$}} \underbrace{+ S_{\rm PM}\smallk{\pip,cat}{\plc}\surf{\plc-\gqa GTP-\pip}}_{\text{Conversion to $\rm \ipTWO$}} \\[.5em]
& \underbrace{+ S_{\rm PM}\smallk{\pi,cat}{\plc}\surf{\plc-\gqa GTP-\pi}}_{\text{Conversion to $\rm \iONEp$}} \underbrace{-V_{\rm cyt}\frac{\smallk{\dag,cat}{\dag Kinase}\conc{\dag Kinase}}{\big(\frac{V_{\rm cyt}}{S_{\rm PM}}\big)\bigk{\dag,M}{\dag Kinase}+\surf{\rm DAG}}\surf{\rm DAG}}_{\text{Phosphorylation of DAG}} \\[.5em]
& \underbrace{-S_{\rm PM}\smallk{\dag}{\pkc-Ca}\conc{\pkc-Ca}\surf{\dag} + V_{\rm cyt}\smallk{-\dag}{\pkc-Ca}\conc{\pkc-Ca-\dag}}_{\text{$\rm \dag$ binding to $\rm \pkc$-Ca}}  \\[.5em]
& \underbrace{-S_{\rm PM}\smallk{\dag}{\pkc-Ca}\conc{\pkc a-Ca}\surf{\dag} + V_{\rm cyt}\smallk{-\dag}{\pkc-Ca}\conc{\pkc a-Ca-\dag}}_{\text{$\rm DAG$ binding to $\rm \pkc$a-Ca}}
\end{split} \\[1em]
\begin{split}
S_{\rm PM}\frac{d}{dt}\surf{PA} ={}& \underbrace{V_{\rm cyt}\frac{\smallk{\dag,cat}{\dag Kinase}\conc{\dag Kinase}}{\big(\frac{V_{\rm cyt}}{S_{\rm PM}}\big)\bigk{\dag,M}{\dag Kinase}+\surf{\dag}}\surf{\dag}}_{\text{Phosphorylation of $\rm \dag$}} \\[.5em]
& \scriptstyle - S_{\rm PM}\underbrace{\scriptstyle \frac{\smallk{PA,cat}{CDPDGSynthase}\conc{CDPDGSynthase}\conc{CTP}}{\bigk{PA,M}{CDPDG Synthase}\bigk{CTP,M}{CDPDG Synthase} + \bigk{PA,M}{CDPDG Synthase}\conc{CTP} + (\bigk{CTP,M}{CDPDG Synthase}\surf{PA} +\surf{PA}\conc{CTP})\big(\frac{S_{\rm PM}}{V_{\rm cyt}}\big)}\surf{PA}}_{\text{Synthesis of CDPDG}}
\end{split} \\[1em]
\begin{split}
S_{\rm PM}\frac{d}{dt}\surf{\rm CDPDG} ={}& \scriptstyle S_{\rm PM}\underbrace{\scriptstyle \frac{\smallk{PA,cat}{CDPDGSynthase}\conc{CDPDGSynthase}\conc{CTP}}{\bigk{PA,M}{CDPDG Synthase}\bigk{CTP,M}{CDPDG Synthase} + \bigk{PA,M}{CDPDG Synthase}\conc{CTP} + (\bigk{CTP,M}{CDPDG Synthase}\surf{PA} +\surf{PA}\conc{CTP})\big(\frac{S_{\rm PM}}{V_{\rm cyt}}\big)}\surf{PA}}_{\text{Synthesis of CDPDG}} \\[.5em]
& - S_{\rm PM}\underbrace{\frac{\big(\frac{S_{\rm PM}}{V_{\rm cyt}}\big)\smallk{I,cat}{\pi Synthase}\conc{\pi  Synthase}\surf{\rm CDPDG}}{\bigk{CDPDG,M}{\pi Synthase}\bigk{I,M}{\pi Synthase}+\bigk{CDPDG,M}{\pi Synthase}\conc{\rm I}+ (\bigk{I,M}{\pi Synthase}\surf{\rm CDPDG} + \conc{\rm I}\surf{\rm CDPDG})\big(\frac{S_{\rm PM}}{V_{\rm cyt}}\big)}\conc{\rm I}}_{\text{Synthesis of PI}}
\end{split}
\end{align}

\pagebreak

\subsubsection{Calcium}
$R$ is the ideal gas constant, $T$ is temperature, $z$ is the charge of a Calcium ion, and $F$ is the Faraday constant.
To calculate the flux of calcium into the cytosol, the Nernst Potentials across the two membranes of our models are needed:
$$ \psi_{\rm IM}=\frac{RT}{zF}\ln\Big(\frac{[{\rm Ca_{dts}}]}{[{\rm Ca_{cyt}}]}\Big) \qquad \psi_{\rm PM}=\frac{RT}{zF}\ln\Big(\frac{[{\rm Ca_{prp}}]}{[{\rm Ca_{cyt}}]}\Big)$$

\begin{table}[htbp]
   \centering
   \begin{tabular}{| c  c  c  c |} 
   \hline
      Parameter & Value & Units & Reference \\ \hline 
  $\gammaS{\ipTHREEr}$ & See Table \ref{tab:CaEstimatedParams} & S & \\
  $\gammaS{leak}$  & See Table \ref{tab:CaEstimatedParams} & S/dm$^2$ & \\
  $\smallk{\ca}{Fura}$ &  $6\cdot10^8$ & 1/M-s & \\
  $\smallk{-\ca}{Fura}$ & 180 & 1/s & \\ \hline
   \end{tabular}
   \caption{Parameter values associated with calcium. At time $t=0$ s in the steady-state simulation, $\conc{\caprp} = 1\cdot10^{-3}$ M, $\conc{\cacyt} = 40\cdot10^{-9}$ M, $\conc{\cadts} = 100\cdot10^{-6}$ M, $\conc{Fura-\ca} = 0$ M. At each time step, we compute $\conc{Fura} = 5\cdot10^{-6} - \conc{Fura-\ca}$. Fura (the fluorescent calcium probe) is only considered during the parameter estimation of calcium data. For all other cases, $\smallk{\ca}{Fura}{\rm Fura}=0$.}
   \label{tab:CaParams}
\end{table}

\begin{align}
\begin{split}
V_{\rm cyt}\frac{d}{dt}\conc{Fura-\ca} ={}& \underbrace{V_{\rm cyt}\smallk{\ca}{\rm Fura}\conc{Fura}\conc{\cacyt} - V_{\rm cyt}\smallk{-\ca}{\rm Fura}\conc{Fura-\ca}}_{\text{Calcium binding to calcium probe}}
\end{split} \\[1em]
\begin{split}
V_{\rm DTS}\frac{d}{dt}\conc{\cadts} ={}& \underbrace{2S_{\rm IM}\smallk{5f}{\rm SERCA}\surf{\serca_{ E2}P\ca_2} - 2S_{\rm IM}\smallk{5r}{\serca}\surf{\serca_{E2}P}\conc{\cadts}}_{\text{$2\rm \cadts$ ions binding to SERCA}} \\[.5em]
& \underbrace{-\frac{\NS{\ipTHREEr}}{4}\frac{\gammaS{\rm IP_3R}}{zF}\Big(\frac{0.1\surf{IP_3R_o}+0.9\surf{IP_3R_a}}{[\rm \ipTHREEr]_{\rm S,tot}}\Big)^4\psi_{\rm IM}}_{\text{Calcium release from $\rm \ipTHREEr$}}
\end{split} \\[1em]
\begin{split}
V_{\rm PRP}\frac{d}{dt}\conc{\caprp} ={}& \underbrace{2S_{\rm PM}\smallk{5f}{\rm PMCA}\surf{\pmca_{E2}P\ca_2} - 2S_{\rm PM}\smallk{5r}{\pmca}\surf{\pmca_{E2}P}\conc{\caprp}}_{\text{2$\rm \caprp$ ions binding to PMCA}} \\
{}& \qquad \underbrace{-S_{\rm PM}\frac{\gammaS{\rm leak}}{zF}\psi_{\rm PM}}_{\text{Calcium leak into cytoplasm}}
\end{split}
\end{align}
\begin{align}
\begin{split}
V_{\rm cyt}\frac{d}{dt}\conc{\cacyt} ={}& \underbrace{\frac{N_{\rm IP_3R}}{4}\frac{\gammaS{\rm IP_3R}}{zF}\Big(\frac{0.1\surf{\ipTHREEr_o}+0.9\surf{\ipTHREEr_a}}{[{\rm \ipTHREEr}]_{\rm S,tot}}\Big)^4\psi_{\rm IM}}_{\text{Calcium release from IP$_3$R}} \underbrace{+ S_{\rm PM}\frac{\gammaS{\rm leak}}{zF}\psi_{\rm PM}}_{\text{Calcium leak into cytoplasm}} \\[.5em]
& \underbrace{-S_{\rm IM}\frac{(\smallk{1}{\ipTHREEr}\bigl{1}{\ipTHREEr}+\smalll{2}{\ipTHREEr})\conc{\cacyt}}{\bigl{1}{\ipTHREEr}+\conc{\cacyt}(1+\bigl{1}{\ipTHREEr}/\bigl{3}{\ipTHREEr})}\surf{\ipTHREEr_n} + S_{\rm IM}(\smallk{-1}{\ipTHREEr}+\smalll{-2}{\ipTHREEr})\surf{\ipTHREEr_{i1}}}_{\text{Calcium inactivating resting IP$_3$R}} \\[.5em]
& \underbrace{-S_{\rm IM}\frac{(\smallk{4}{\ipTHREEr}\bigl{5}{\ipTHREEr}+\smalll{6}{\ipTHREEr})\conc{\cacyt}}{\bigl{5}{\ipTHREEr}+\conc{\cacyt}}\surf{\ipTHREEr_o} + S_{\rm IM}\frac{\bigl{1}{\ipTHREEr}(\smallk{-4}{\ipTHREEr}+\smalll{-6}{\ipTHREEr})}{\bigl{1}{\ipTHREEr}+\conc{\cacyt}}\surf{\ipTHREEr_a}}_{\text{Calcium activating receptor}} \\[.5em]
& \underbrace{-S_{\rm IM}\frac{\smallk{1}{\ipTHREEr}\bigl{1}{\ipTHREEr}+\smalll{2}{\ipTHREEr}}{\bigl{1}{\ipTHREEr}+\conc{\cacyt}}\surf{\ipTHREEr_a}\conc{\cacyt} + S_{\rm IM}(\smallk{-1}{\ipTHREEr}+\smalll{-2}{\ipTHREEr})\surf{\ipTHREEr_{i2}}}_{\text{Calcium inactivating active IP$_3$R}} \\[.5em]
& \underbrace{-2S_{\rm IM}\smallk{2f}{\rm SERCA}\surf{\serca_{E1}}\conc{\cacyt}^2 + 2S_{\rm IM}\smallk{2r}{\rm SERCA}\surf{\serca_{ E1}-2\ca}}_{\text{2Ca$_{\rm cyt}$ ions binding to SERCA}} \\[.5em]
& \underbrace{-2S_{\rm IM}\smallk{2f}{\rm PMCA}\surf{\pmca_{E1}}\conc{\cacyt}^2 + 2S_{\rm IM}\smallk{2r}{\rm PMCA}\surf{\pmca_{ E1}-2\ca}}_{\text{2Ca$_{\rm cyt}$ ions binding to PMCA}} \\[.5em]
& \underbrace{-V_{\rm cyt}\smallk{\ca}{\rm PKC}\conc{\pkc }\conc{\cacyt} + V_{\rm cyt}\smallk{-\ca}{\rm PKC}\conc{\pkc -Ca}}_{\text{Ca binding to PKC}} \\[.5em]
& \underbrace{-V_{\rm cyt}\smallk{\ca}{\rm PKC}\conc{\pkc a}\conc{\cacyt} + V_{\rm cyt}\smallk{-\ca}{\rm PKC}\conc{\pkc a-Ca}}_{\text{Ca binding to PKCa}} \\[.5em]
& \underbrace{-V_{\rm cyt}\smallk{\ca}{\cdgi}\conc{\cacyt}\conc{\cdgi} + V_{\rm cyt}\smallk{-\ca}{\cdgi}\conc{\cdgi-Ca}}_{\text{CalDAG-GEFI binding calcium}} \\[.5em]
& \underbrace{-V_{\rm cyt}\smallk{\ca}{\cdgi}\conc{\cacyt}\conc{\cdgi-Ca} + V_{\rm cyt}\smallk{-\ca}{\cdgi}\conc{\cdgi-2Ca}}_{\text{CalDAG-GEFI-Ca binding calcium}} \\[.5em]
& -\underbrace{V_{\rm cyt}\smallk{\ca}{Fura}[{\rm Ca}{2+}][{\rm Fura}] + V_{\rm cyt}\smallk{-\ca}{\rm Fura}[{\rm Fura-Ca}]}_{\text{Fluorescent Probe binding calcium}}
\end{split}
\end{align}

\pagebreak
\subsection{Calcium channels/pumps}
\subsubsection{IP$_3$R}

 \begin{figure}[ht!]
\centering\includegraphics[width=.5\textwidth]{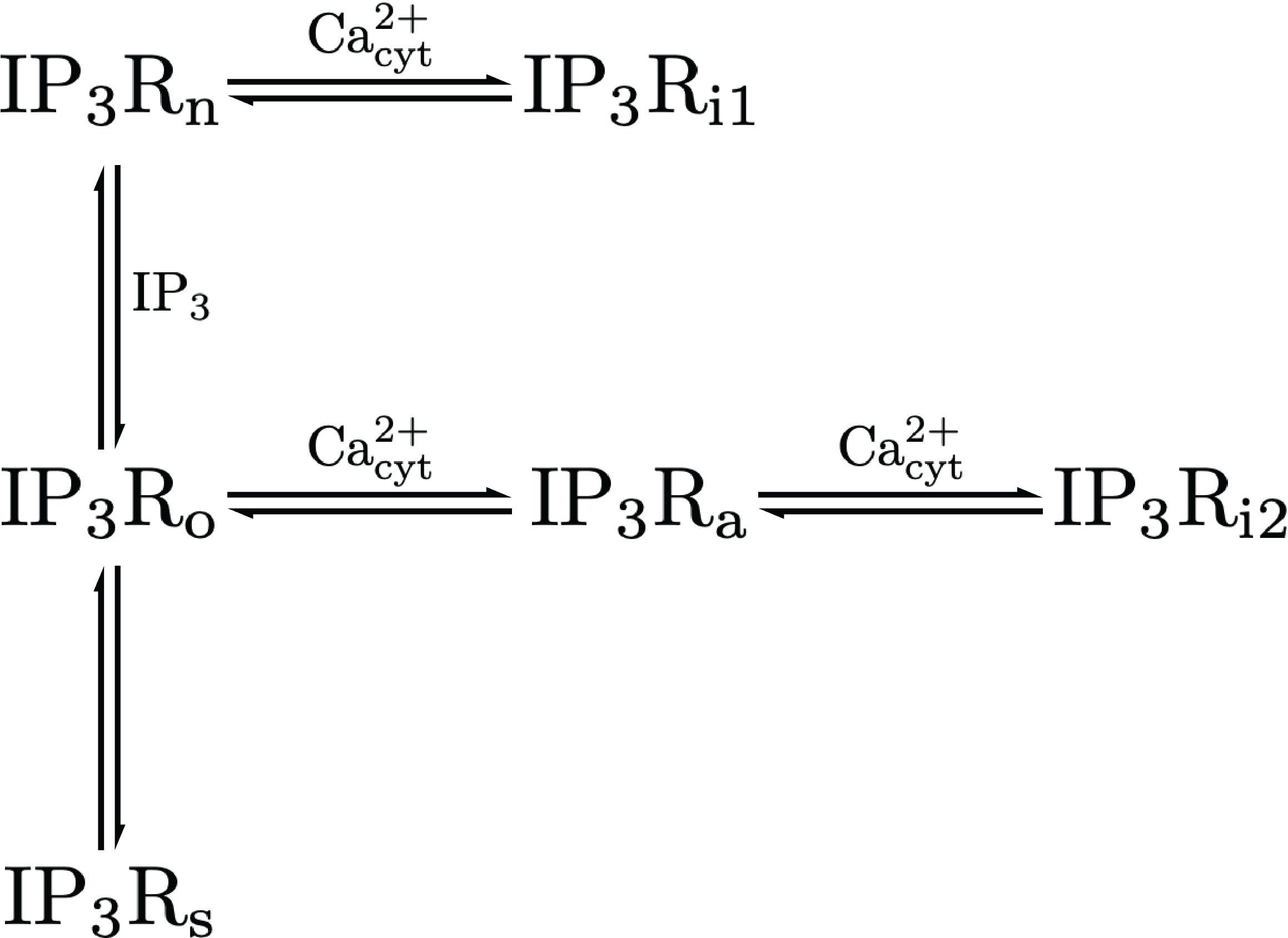}
\caption{IP$_3$R reaction diagram. Model setup and states are exactly the same as in Sneyd and Dufour \cite{SneydDufour2002}}
\end{figure}

\begin{table}[htbp]
   \centering
   \begin{tabular}{| c  c  c  c |} 
   \hline
      Parameter & Value & Units & Reference \\ \hline 
  $\smallk{1}{\ipTHREEr}$ &  640000 & 1/M-s & \cite{SneydDufour2002} \\
  $\smallk{-1}{\ipTHREEr}$ &  0.04 & 1/s & \cite{SneydDufour2002} \\
  $\smallk{2}{\ipTHREEr}$ &  37400000 & 1/M-s & \cite{SneydDufour2002} \\
  $\smallk{-2}{\ipTHREEr}$ &  1.4 & 1/s & \cite{SneydDufour2002} \\
  $\smallk{3}{\ipTHREEr}$ &  0.11 & 1/M-s & \cite{SneydDufour2002} \\
  $\smallk{-3}{\ipTHREEr}$ &  29.8 & 1/s & \cite{SneydDufour2002} \\
  $\smallk{4}{\ipTHREEr}$ &  4000000 & 1/M-s & \cite{SneydDufour2002} \\
  $\smallk{-4}{\ipTHREEr}$ &  0.54 & 1/s & \cite{SneydDufour2002} \\
  $\bigl{1}{\ipTHREEr}$ &  1.2e-07 & M & \cite{SneydDufour2002} \\
  $\bigl{3}{\ipTHREEr}$ &  2.5e-08 & M & \cite{SneydDufour2002} \\
  $\bigl{5}{\ipTHREEr}$ &  5.47e-05 & M & \cite{SneydDufour2002} \\
  $\smalll{2}{\ipTHREEr}$ &  1.7 & 1/s & \cite{SneydDufour2002} \\
  $\smalll{4}{\ipTHREEr}$ &  1700000 & 1/M-s & \cite{SneydDufour2002} \\
  $\smalll{6}{\ipTHREEr}$ &  4707 & 1/s & \cite{SneydDufour2002} \\
  $\smalll{-2}{\ipTHREEr}$ &  0.8 & 1/s & \cite{SneydDufour2002} \\
  $\smalll{-4}{\ipTHREEr}$ &  2500000 & 1/M-s & \cite{SneydDufour2002} \\
  $\smalll{-6}{\ipTHREEr}$ &  11.4 & 1/s & \cite{SneydDufour2002} \\
  $\NS{\ipTHREEr}$ & 3361 & \# & \cite{Zeiler2014} \\ \hline
   \end{tabular}
   \caption{Parameter values associated with IP$_3$R. At time $t=0$ s in the steady-state simulation, $\surf{\ipTHREEr_n} = \NS{\ipTHREEr}/S_{\rm IM}$ and all other states are initialized to 0.}
   \label{tab:IP3RParams}
\end{table}

\begin{align}
\begin{split}
S_{\rm IM}\frac{d}{dt}\surf{\ipTHREEr_n} ={}& \underbrace{-S_{\rm IM}\frac{(\smallk{1}{\ipTHREEr}\bigl{1}{\ipTHREEr}+\smalll{2}{\ipTHREEr})\conc{\cacyt}}{\bigl{1}{\ipTHREEr}+\conc{\cacyt}(1+\bigl{1}{\ipTHREEr}/\bigl{3}{\ipTHREEr})}\surf{\ipTHREEr_n} + S_{\rm IM}(\smallk{-1}{\ipTHREEr}+\smalll{-2}{\ipTHREEr})\surf{\ipTHREEr_{i1}}}_{\text{Calcium inactivating resting IP$_3$R}} \\[.5em]
& -S_{\rm IM}\frac{\smallk{2}{\ipTHREEr}\bigl{3}{\ipTHREEr}+\smalll{4}{\ipTHREEr}\conc{\cacyt}}{\bigl{3}{\ipTHREEr}+\conc{\cacyt}(1+\bigl{3}{\ipTHREEr}/\bigl{1}{\ipTHREEr})}\surf{\ipTHREEr_n}\conc{\ipTHREE} \\
& \underbrace{\qquad + S_{\rm IM}\frac{\smallk{-2}{\ipTHREEr}+\smalll{-4}{\ipTHREEr}\conc{\cacyt}}{1+\conc{\cacyt}/\bigl{5}{\ipTHREEr}}\surf{\ipTHREEr_o}\qquad\qquad\qquad}_{\text{IP$_3$ binding to receptor}}
\end{split} \\[1em]
\begin{split}
S_{\rm IM}\frac{d}{dt}\surf{\ipTHREEr_{i1}} ={}& \underbrace{S_{\rm IM}\frac{(\smallk{1}{\ipTHREEr}\bigl{1}{\ipTHREEr}+\smalll{2}{\ipTHREEr})\conc{\cacyt}}{\bigl{1}{\ipTHREEr}+\conc{\cacyt}(1+\bigl{1}{\ipTHREEr}/\bigl{3}{\ipTHREEr})}\surf{\ipTHREEr_n} - S_{\rm IM}(\smallk{-1}{\ipTHREEr}+\smalll{-2}{\ipTHREEr})\surf{\ipTHREEr_{i1}}}_{\text{Calcium inactivating resting IP$_3$R}}
\end{split}
\end{align}
\begin{align}
\begin{split}
S_{\rm IM}\frac{d}{dt}\surf{\ipTHREEr_o} ={}& S_{\rm IM}\frac{\smallk{2}{\ipTHREEr}\bigl{3}{\ipTHREEr}+\smalll{4}{\ipTHREEr}\conc{\cacyt}}{\bigl{3}{\ipTHREEr}+\conc{\cacyt}(1+\bigl{3}{\ipTHREEr}/\bigl{1}{\ipTHREEr})}\surf{\ipTHREEr_n}\conc{\ipTHREE} \\
& \underbrace{\qquad - S_{\rm IM}\frac{\smallk{-2}{\ipTHREEr}+\smalll{-4}{\ipTHREEr}\conc{\cacyt}}{1+\conc{\cacyt}/\bigl{5}{\ipTHREEr}}\surf{\ipTHREEr_o}\qquad\qquad\qquad}_{\text{IP$_3$ binding to receptor}} \\[.5em]
& \underbrace{- S_{\rm IM}\frac{(\smallk{4}{\ipTHREEr}\bigl{5}{\ipTHREEr}+\smalll{6}{\ipTHREEr})}{\bigl{5}{\ipTHREEr}+\conc{\cacyt}}\surf{\ipTHREEr_o}\conc{\cacyt} + S_{\rm IM}\frac{\bigl{1}{\ipTHREEr}(\smallk{-4}{\ipTHREEr}+\smalll{-6}{\ipTHREEr})}{\bigl{1}{\ipTHREEr}+\conc{\cacyt}}\surf{\ipTHREEr_a}}_{\text{Calcium activating receptor}} \\[.5em]
& \underbrace{-S_{\rm IM}\frac{\smallk{3}{\ipTHREEr}\bigl{5}{\ipTHREEr}}{\bigl{5}{\ipTHREEr}+\conc{\cacyt}}\surf{\ipTHREEr_o}+S_{\rm IM}\smallk{-3}{\ipTHREEr}\surf{\ipTHREEr_s}}_{\text{receptor closing}}
\end{split} \\[1em]
\begin{split}
S_{\rm IM}\frac{d}{dt}\surf{\ipTHREEr_a} ={}& \underbrace{S_{\rm IM}\frac{(\smallk{4}{\ipTHREEr}\bigl{5}{\ipTHREEr}+\smalll{6}{\ipTHREEr})}{\bigl{5}{\ipTHREEr}+\conc{\cacyt}}\surf{\ipTHREEr_o}\conc{\cacyt} - S_{\rm IM}\frac{\bigl{1}{\ipTHREEr}(\smallk{-4}{\ipTHREEr}+\smalll{-6}{\ipTHREEr})}{\bigl{1}{\ipTHREEr}+\conc{\cacyt}}\surf{\ipTHREEr_a}}_{\text{Calcium activating receptor}} \\[.5em]
& \underbrace{-S_{\rm IM}\frac{\smallk{1}{\ipTHREEr}\bigl{1}{\ipTHREEr}+\smalll{2}{\ipTHREEr}}{\bigl{1}{\ipTHREEr}+\conc{\cacyt}}\surf{\ipTHREEr_a}\conc{\cacyt} + S_{\rm IM}(\smallk{-1}{\ipTHREEr}+\smalll{-2}{\ipTHREEr})\surf{\ipTHREEr_{i2}}}_{\text{Calcium inactivating active IP$_3$R}}
\end{split}
\end{align}
\begin{align}
\begin{split}
S_{\rm IM}\frac{d}{dt}\surf{\ipTHREEr_s} ={}& \underbrace{-S_{\rm IM}\frac{\smallk{3}{\ipTHREEr}\bigl{5}{\ipTHREEr}}{\bigl{5}{\ipTHREEr}+\conc{\cacyt}}\surf{\ipTHREEr_o} + S_{\rm IM}\smallk{-3}{\ipTHREEr}\surf{\ipTHREEr_s}}_{\text{receptor closing}}
\end{split} \\[1em]
\begin{split}
S_{\rm IM}\frac{d}{dt}\surf{\ipTHREEr_{i2}} ={}& \underbrace{S_{\rm IM}\frac{\smallk{1}{\ipTHREEr}\bigl{1}{\ipTHREEr}+\smalll{2}{\ipTHREEr}}{\bigl{1}{\ipTHREEr}+\conc{\cacyt}}\surf{\ipTHREEr_a}\conc{\cacyt} - S_{\rm IM}(\smallk{-1}{\ipTHREEr}+\smalll{-2}{\ipTHREEr})\surf{\ipTHREEr_{i2}}}_{\text{Calcium inactivating active IP$_3$R}}
\end{split}
\end{align}

\pagebreak
\subsubsection{SERCA}

The full SERCA pump model is a 6 state model that involves shuttling two calcium ions across the DTS membrane at a time.
The three states without a calcium bound are assumed to be in a quasi-steady state equilibrium with each other.
Let $\surf{\serca} = \surf{\serca_{E1}} + \surf{\serca_{E2}} + \surf{\serca_{E2}P}$.
Then the ratios as follows:
\begin{align*}
\surf{\serca_{E1}} ={}& (0.24996)\surf{\serca} \\
\surf{\serca_{E2}} ={}& (0.74989)\surf{\serca} \\
\surf{\serca_{E2}P} ={}& (1.4998\cdot10^{-4})\surf{\serca}
\end{align*}

 \begin{figure}[ht!]
\centering\includegraphics[width=.7\textwidth]{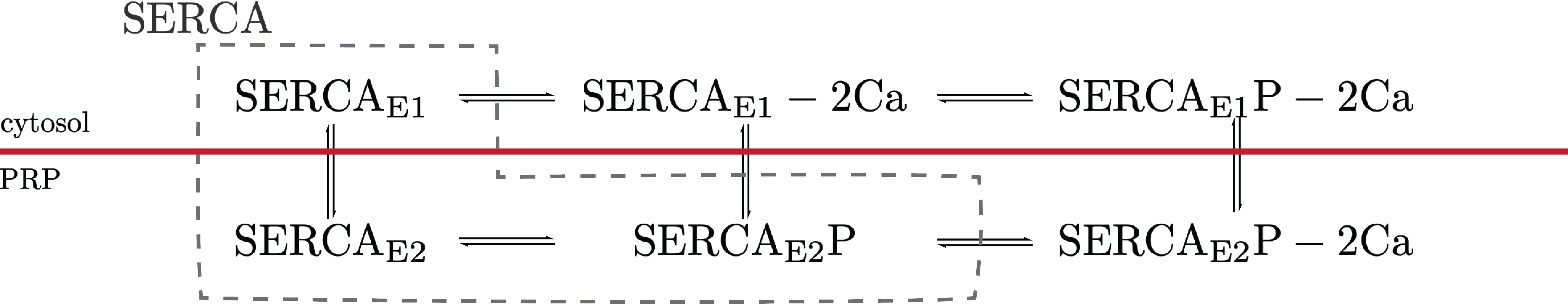}
\caption{SERCA reaction diagram. The three states without a calcium bound are all assumed to be in equilibrium with each other.}
\end{figure}

\begin{table}[h]
   \centering
   \begin{tabular}{| c  c  c  c |} 
   \hline
      Parameter & Value & Units & Reference \\ \hline 
  $\smallk{2f}{\serca}$ & See Table \ref{tab:CaEstimatedParams} & 1/M$^2$-s & \\
  $\smallk{2r}{\serca}$ & See Table \ref{tab:CaEstimatedParams} & 1/s & \\
  $\smallk{3f}{\serca}$ & See Table \ref{tab:CaEstimatedParams} & 1/s & \\
  $\smallk{3r}{\serca}$ & See Table \ref{tab:CaEstimatedParams} & 1/s & \\
  $\smallk{4f}{\serca}$ & See Table \ref{tab:CaEstimatedParams} & 1/s & \\
  $\smallk{4r}{\serca}$ & See Table \ref{tab:CaEstimatedParams} & 1/s & \\
  $\smallk{5f}{\serca}$ & See Table \ref{tab:CaEstimatedParams} & 1/s & \\
  $\smallk{5r}{\serca}$ & See Table \ref{tab:CaEstimatedParams} & 1/M$^2$-s & \\
  $\NS{\serca}$ & 57083 & \# & \cite{Zeiler2014} \\ \hline
   \end{tabular}
   \caption{Parameter values associated with SERCA. At time $t=0$ s in the steady-state simulation, $\surf{\serca} = \NS{\serca}/S_{\rm IM}$ and all other states are initialized to 0.}
   \label{tab:SERCAParams}
\end{table}

\begin{align}
\begin{split}
S_{\rm IM}\frac{d}{dt}\surf{\serca} ={}& \underbrace{-S_{\rm IM}\smallk{2f}{\rm SERCA}\surf{\serca_{E1}}\conc{\cacyt}^2 + S_{\rm IM}\smallk{2r}{\rm SERCA}\surf{\serca_{E1}-2\ca}}_{\text{2Ca$_{\rm cyt}$ ions binding to SERCA}} \\[.5em]
& \underbrace{+S_{\rm IM}\smallk{5f}{\rm SERCA}\surf{\serca_{ E2}P-2\ca} - S_{\rm IM}\smallk{5r}{\rm SERCA}\surf{\serca_{E2}P}\conc{\cadts}^2}_{\text{2Ca$_{\rm dts}$ ions binding to SERCA}}
\end{split} \\[1em]
\begin{split}
S_{\rm IM}\frac{d}{dt}\surf{\serca_{E1}-2\ca} ={}& \underbrace{S_{\rm IM}\smallk{2f}{\rm SERCA}\surf{\serca_{E1}}\conc{\cacyt}^2 - S_{\rm IM}\smallk{2r}{\rm SERCA}\surf{\serca_{E1}-2\ca}}_{\text{2Ca$_{\rm cyt}$ ions binding to SERCA}} \\[.5em]
& \underbrace{-S_{\rm IM}\smallk{3f}{\rm SERCA}\surf{\serca_{E1}-Ca_2} + S_{\rm IM}\smallk{3r}{\rm SERCA}\surf{\serca_{E1}P-2\ca}}_{\text{SERCA phosphorylation}}
\end{split} \\[1em]
\begin{split}
S_{\rm IM}\frac{d}{dt}\surf{\serca_{E1}P-2\ca} ={}& \underbrace{S_{\rm IM}\smallk{3f}{\rm SERCA}\surf{\serca_{E1}-2\ca} + S_{\rm IM}\smallk{3r}{\rm SERCA}\surf{\serca_{E1}P-2\ca}}_{\text{SERCA phosphorylation}} \\[.5em]
& \underbrace{-S_{\rm IM}\smallk{4f}{\rm SERCA}\surf{\serca_{E1}P-2\ca} + S_{\rm IM}\smallk{4r}{\rm SERCA}\surf{\serca_{E2}P-2\ca}}_{\text{SERCA shuttling into DTS}}
\end{split} \\[1em]
\begin{split}
S_{\rm IM}\frac{d}{dt}\surf{\serca_{E2}P-2\ca} ={}& \underbrace{S_{\rm IM}\smallk{4f}{\rm SERCA}\surf{\serca_{E1}P-2\ca} - S_{\rm IM}\smallk{4r}{\rm SERCA}\surf{\serca_{E2}P-2\ca}}_{\text{SERCA shuttling into DTS}} \\[.5em]
& \underbrace{-S_{\rm IM}\smallk{5f}{\rm SERCA}\surf{\serca_{ E2}P-2\ca} + S_{\rm IM}\smallk{5r}{\rm SERCA}\surf{\serca_{E2}P}\conc{\cadts}^2}_{\text{2Ca$_{\rm dts}$ ions binding to SERCA}}
\end{split}
\end{align}

\pagebreak
\subsubsection{PMCA}

The full PMCA pump model is a 6 state model that involves shuttling two calcium ions across the plasma membrane at a time.
The three states without a calcium bound are assumed to be in a quasi-steady state equilibrium with each other.
Let $\surf{\pmca} = \surf{\pmca_{E1}} + \surf{\pmca_{E2}} + \surf{\pmca_{E2}P}$.
Then the ratios as follows:

\begin{align*}
\surf{\pmca_{E1}} ={}& (0.24996)\surf{\pmca} \\
\surf{\pmca_{E2}} ={}& (0.74989)\surf{\pmca} \\
\surf{\pmca_{E2}P} ={}& (1.4998\cdot10^{-4})\surf{\pmca}
\end{align*}

 \begin{figure}[ht!]
\centering\includegraphics[width=.7\textwidth]{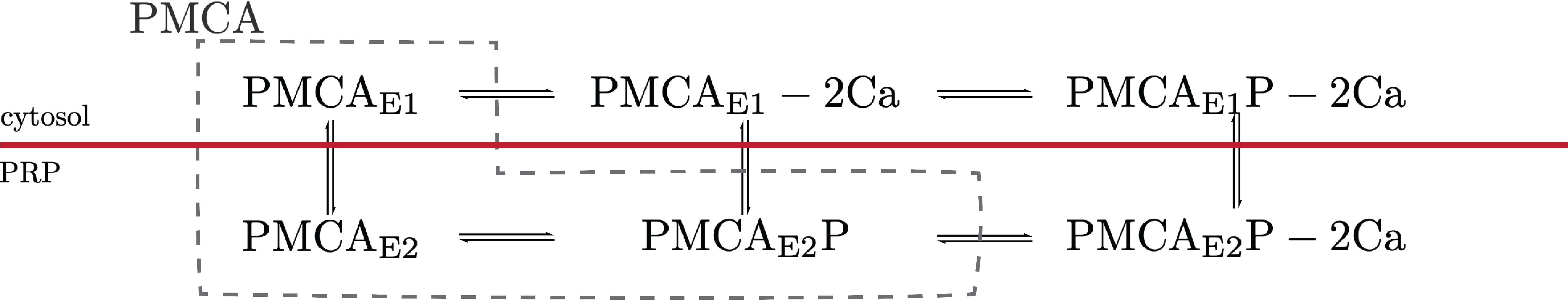}
\caption{PMCA reaction diagram. The three states without a calcium bound are all assumed to be in equilibrium with each other.}
\end{figure}

\begin{table}[htbp]
   \centering
   \begin{tabular}{| c  c  c  c |} 
   \hline
      Parameter & Value & Units & Reference \\ \hline 
  $\smallk{2f}{\pmca}$ & See Table \ref{tab:CaEstimatedParams} & 1/M$^2$-s & \\
  $\smallk{2r}{\pmca}$ & See Table \ref{tab:CaEstimatedParams} & 1/s & \\
  $\smallk{3f}{\pmca}$ & See Table \ref{tab:CaEstimatedParams} & 1/s & \\
  $\smallk{3r}{\pmca}$ & See Table \ref{tab:CaEstimatedParams} & 1/s & \\
  $\smallk{4f}{\pmca}$ & See Table \ref{tab:CaEstimatedParams} & 1/s & \\
  $\smallk{4r}{\pmca}$ & See Table \ref{tab:CaEstimatedParams} & 1/s & \\
  $\smallk{5f}{\pmca}$ & See Table \ref{tab:CaEstimatedParams} & 1/s & \\
  $\smallk{5r}{\pmca}$ & See Table \ref{tab:CaEstimatedParams} & 1/M$^2$-s & \\
  $\NS{\pmca}$ & 611 & \# & \cite{Zeiler2014} \\ \hline
   \end{tabular}
   \caption{Parameter values associated with PMCA. At time $t=0$ s in the steady-state simulation, $\surf{\pmca} = \NS{\pmca}/S_{\rm PM}$ and all other states are initialized to 0.}
   \label{tab:PMCAParams}
\end{table}

\begin{align}
\begin{split}
S_{\rm PM}\frac{d}{dt}\surf{\pmca} ={}& \underbrace{-S_{\rm PM}\smallk{2f}{\rm PMCA}\surf{\pmca_{E1}}\conc{\cacyt}^2 + S_{\rm PM}\smallk{2r}{\rm PMCA}\surf{\pmca_{ E1}-2\ca}}_{\text{2Ca$_{\rm cyt}$ ions binding to PMCA}} \\[.5em]
& \underbrace{+S_{\rm PM}\smallk{5f}{\rm PMCA}\surf{\pmca_{ E2}P-2\ca} - S_{\rm PM}\smallk{5r}{\rm PMCA}\surf{\pmca_{E2}P}\conc{\caprp}^2}_{\text{2Ca$_{\rm dts}$ ions binding to PMCA}}
\end{split} \\[1em]
\begin{split}
S_{\rm PM}\frac{d}{dt}\surf{\pmca_{E1}-2\ca} ={}& \underbrace{S_{\rm PM}\smallk{2f}{\rm PMCA}\surf{\pmca_{E1}}\conc{\cacyt}^2 - S_{\rm PM}\smallk{2r}{\rm PMCA}\surf{\pmca_{ E1}-2\ca}}_{\text{2Ca$_{\rm cyt}$ ions binding to PMCA}} \\[.5em]
& \underbrace{-S_{\rm PM}\smallk{3f}{\rm PMCA}\surf{\pmca_{E1}-2\ca} + S_{\rm PM}\smallk{3r}{\rm PMCA}\surf{\pmca_{E1}P-2\ca}}_{\text{PMCA phosphorylation}}
\end{split} \\[1em]
\begin{split}
S_{\rm PM}\frac{d}{dt}\surf{\pmca_{E1}P-2\ca} ={}& \underbrace{S_{\rm PM}\smallk{3f}{\rm PMCA}\surf{\pmca_{E1}-2\ca} + S_{\rm PM}\smallk{3r}{\rm PMCA}\surf{\pmca_{E1}P-2\ca}}_{\text{PMCA phosphorylation}} \\[.5em]
& \underbrace{-S_{\rm PM}\smallk{4f}{\rm PMCA}\surf{\pmca_{E1}P-2\ca} + S_{\rm PM}\smallk{4r}{\rm PMCA}\surf{\pmca_{E2}P-2\ca}}_{\text{PMCA shuttling into PRP}}
\end{split} \\[1em]
\begin{split}
S_{\rm PM}\frac{d}{dt}\surf{\pmca_{E2}P-2\ca} ={}& \underbrace{S_{\rm PM}\smallk{4f}{\rm PMCA}\surf{\pmca_{E1}P-2\ca} - S_{\rm PM}\smallk{4r}{\rm PMCA}\surf{\pmca_{E2}P-2\ca}}_{\text{PMCA shuttling into PRP}} \\[.5em]
& \underbrace{-S_{\rm PM}\smallk{5f}{\rm PMCA}\surf{\pmca_{ E2}P-2\ca} + S_{\rm PM}\smallk{5r}{\rm PMCA}\surf{\pmca_{E2}P}\conc{\caprp}^2}_{\text{2Ca$_{\rm dts}$ ions binding to PMCA}}
\end{split}
\end{align}

\pagebreak
\section{Quasi Steady State Approximation }

 \begin{figure}[ht!]
\centering\includegraphics[width=.7\textwidth]{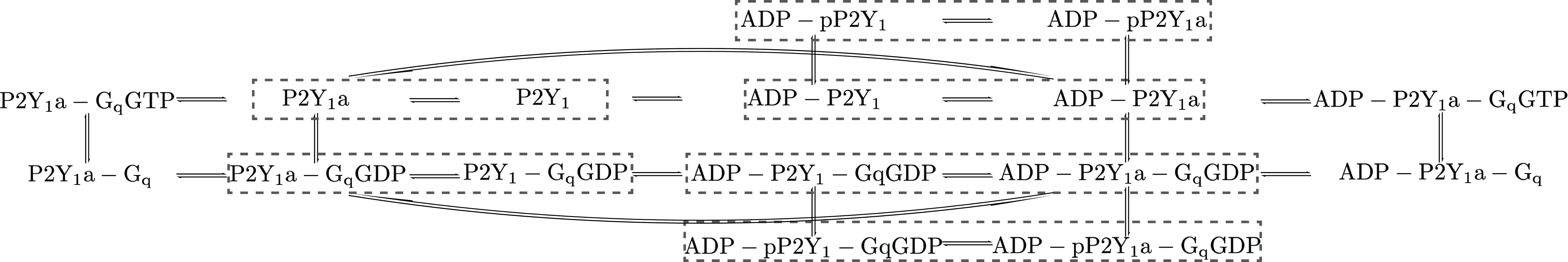}
\caption{$\rm\pTWOyONE$ reaction diagram without a quasi steady state approximation.
Four states are added on top of the model used in Purvis et al. \cite{Purvis2008}, $\rm p-\pTWOyONE-\gq GDP$, $\rm ADP-p\pTWOyONE-\gq GDP$, $\rm p-\pTWOyONE a-\gq GDP$, and $\rm ADP-p\pTWOyONE a-\gq GDP$
These states correspond to a desensitized receptor mediated by PKC.}
\label{fig:p2y1QSS}
\end{figure}

We briefly outline the process of reducing the number of equations in Purvis et al. \cite{Purvis2008} by utilizing a quasi steady state (QSS) assumption.
The work here is for the P2Y$_1$ receptor only; the exact same computation can be done to the P2Y$_{12}$ receptor to yield analogous equations.
Figure~\ref{fig:p2y1QSS} shows the entire reaction diagram for the $\rm \pTWOyONE$ receptor. 
It is generally assumed for G-protein coupled receptors that the receptors must be in an ``active'' conformation in order to process G protein \cite{Kinzer-Ursem2007}.
Our assumption is that the receptor in the ``active'' conformation is in a quasi-steady state equilibrium with the corresponding ``inactive'' conformation, showed by the grey boxes in the figure.
Under this assumption (to be made more precise below), we can reduce the number of equations by six

Below, we write out each of the 12 equations that are contained in the grey boxes in Figure~\ref{fig:p2y1QSS}.
For concision, we divide all terms by $S_{\rm PM}$.
The full equations are as follows:

\begin{align}
\begin{split}
\frac{d}{dt}[{\rm \pTWOyONE }]_S ={}& \underbrace{-k_{\rm act}^{\rm \pTWOyONE }[{\rm \pTWOyONE }]_S + k_{\rm -act}^{\rm \pTWOyONE }[{\rm \pTWOyONE a}]_S}_{\text{activation}} - \underbrace{ k_{\rm ADP}^{\rm \pTWOyONE }[{\rm \pTWOyONE }]_S[{\rm ADP}] + k_{\rm -ADP}^{\rm \pTWOyONE }[{\rm ADP-\pTWOyONE }]_S}_{\text{ADP Binding}} 
\label{eqn:p2y1}
\end{split} \\[1em]
\begin{split}
\frac{d}{dt}[{\rm \pTWOyONE a}]_S ={}& \underbrace{k_{\rm act}^{\rm \pTWOyONE }[{\rm \pTWOyONE }]_S - k_{\rm -act}^{\rm \pTWOyONE }[{\rm \pTWOyONE a}]_S}_{\text{activation}} - \underbrace{ \alpha k_{\rm ADP}^{\rm \pTWOyONE }[{\rm \pTWOyONE a}]_S[{\rm ADP}] + k_{\rm -ADP}^{\rm \pTWOyONE }[{\rm ADP-\pTWOyONE a}]_S}_{\text{ADP Binding}} \\[.5em]
& \underbrace{-\beta k_{\rm G_qGDP}^{\rm \pTWOyONE }[{\rm \pTWOyONE a}]_S[{\rm G_qGDP}]_S + k_{\rm -G_qGDP}^{\rm \pTWOyONE }[{\rm \pTWOyONE a-G_qGDP}]_S}_{\text{G$_q$GDP binding}} \\[.5em]
& \underbrace{- k_{\rm G_qGTP}^{\rm \pTWOyONE }[{\rm \pTWOyONE a}]_S[{\rm G_qGTP}]_S + k_{\rm -G_qGTP}^{\rm \pTWOyONE }[{\rm \pTWOyONE a-G_qGTP}]_S}_{\text{G$_q$GTP binding}}
\end{split} \\[1em]
\begin{split}
\frac{d}{dt}[{\rm ADP-\pTWOyONE }]_S ={}& \underbrace{-\alpha k_{\rm act}^{\rm \pTWOyONE }[{\rm ADP-\pTWOyONE }]_S + k_{\rm -act}^{\rm \pTWOyONE }[{\rm ADP-\pTWOyONE a}]_S}_{\text{activation}} + \underbrace{ k_{\rm ADP}^{\rm \pTWOyONE }[{\rm \pTWOyONE }]_S[{\rm ADP}] - k_{\rm -ADP}^{\rm \pTWOyONE }[{\rm ADP-\pTWOyONE }]_S}_{\text{ADP Binding}} \\[.5em]
& \underbrace{-k_{\rm phos}^{\rm \pTWOyONE }[{\rm ADP-\pTWOyONE }]_S[{\rm PKCa-Ca-DAG}] + k_{\rm dephos}^{\rm \pTWOyONE }[{\rm ADP-p\pTWOyONE }]_S   }_{\text{Desensitization by PKC}}
\label{eqn:adp-p2y1}
\end{split} \\[1em]
\begin{split}
\frac{d}{dt}[{\rm ADP-\pTWOyONE a}]_S ={}& \underbrace{\alpha k_{\rm act}^{\rm \pTWOyONE }[{\rm ADP-\pTWOyONE }]_S - k_{\rm -act}^{\rm \pTWOyONE }[{\rm ADP-\pTWOyONE a}]_S}_{\text{activation}} +\underbrace{ \alpha k_{\rm ADP}^{\rm \pTWOyONE }[{\rm \pTWOyONE a}]_S[{\rm ADP}] - k_{\rm -ADP}^{\rm \pTWOyONE }[{\rm ADP-\pTWOyONE a}]_S}_{\text{ADP Binding}} \\[.5em]
& \underbrace{-k_{\rm phos}^{\rm \pTWOyONE }[{\rm ADP-\pTWOyONE a}]_S[{\rm PKCa-Ca-DAG}] + k_{\rm dephos}^{\rm \pTWOyONE }[{\rm ADP-p\pTWOyONE a}]_S   }_{\text{Desensitization by PKC}} \\[.5em]
& -\underbrace{\beta k_{\rm G_qGDP}^{\rm \pTWOyONE }[{\rm ADP-\pTWOyONE a}]_S[{\rm G_qGDP}]_S + \frac{k_{\rm -G_qGDP}^{\rm \pTWOyONE }}{\delta\gamma}[{\rm ADP-\pTWOyONE a-G_qGDP}]_S}_{\text{G$_q$GDP binding}} \\[.5em]
& - \underbrace{k_{\rm G_qGTP}^{\rm \pTWOyONE }[{\rm ADP-\pTWOyONE a}]_S[{\rm G_qGTP}]_S + k_{\rm -G_qGTP}^{\rm \pTWOyONE }[{\rm ADP-\pTWOyONE a-G_qGTP}]_S}_{\text{G$_q$GTP binding}}
\end{split}
\end{align}

\begin{align}
\begin{split}
\frac{d}{dt}[{\rm P2Y_1-G_qGDP}]_S ={}& \underbrace{-\beta k_{\rm act}^{\rm P2Y_1}[{\rm P2Y_1-G_qGDP}]_S + k_{\rm -act}^{\rm P2Y_1}[{\rm P2Y_1a-G_qGDP}]_S}_{\text{activation}} \\[.5em]
& - \underbrace{k_{\rm ADP}^{\rm P2Y_1}[{\rm P2Y_1-G_qGDP}]_S[{\rm ADP}] + \frac{k_{\rm -ADP}^{\rm P2Y_1}}{\gamma}[{\rm ADP-P2Y_1-G_qGDP}]_S}_{\text{ADP Binding}} 
\label{eqn:p2y1-gqgdp}
\end{split} \\[1em]
\begin{split}
\frac{d}{dt}[{\rm P2Y_1a-G_qGDP}]_S ={}& \underbrace{\beta k_{\rm act}^{\rm P2Y_1}[{\rm P2Y_1-G_qGDP}]_S - k_{\rm -act}^{\rm P2Y_1}[{\rm P2Y_1a-G_qGDP}]_S}_{\text{activation}} \\[.5em]
& - \underbrace{\alpha k_{\rm ADP}^{\rm P2Y_1}[{\rm P2Y_1a-G_qGDP}]_S[{\rm ADP}] + \frac{k_{\rm -ADP}^{\rm P2Y_1}}{\delta\gamma}[{\rm ADP-P2Y_1a-G_qGDP}]_S}_{\text{ADP Binding}} \\[.5em]
& +\underbrace{\beta k_{\rm G_qGDP}^{\rm P2Y_1}[{\rm P2Y_1a}]_S[{\rm G_qGDP}]_S - k_{\rm -G_qGDP}^{\rm P2Y_1}[{\rm P2Y_1a-G_qGDP}]_S}_{\text{G$_q$GDP binding}} \\[.5em]
& \underbrace{-k_{\rm GDP}^{\rm P2Y_1}[{\rm P2Y_1a-G_qGDP}]_S + k_{\rm -GDP}^{\rm P2Y_1}[{\rm P2Y_1a-G_q}]_S[{\rm GDP}]}_{\text{GDP binding}}
\end{split} \\[1em]
\begin{split}
\frac{d}{dt}[{\rm ADP-P2Y_1-G_qGDP}]_S ={}& -\underbrace{\alpha\beta\delta k_{\rm act}^{\rm P2Y_1}[{\rm ADP-P2Y_1-G_qGDP}]_S + k_{\rm -act}^{\rm P2Y_1}[{\rm ADP-P2Y_1a-G_qGDP}]_S}_{\text{activation}} \\[.5em]
& + \underbrace{k_{\rm ADP}^{\rm P2Y_1}[{\rm P2Y_1-G_qGDP}]_S[{\rm ADP}] - \frac{k_{\rm -ADP}^{\rm P2Y_1}}{\gamma}[{\rm ADP-P2Y_1-G_qGDP}]_S}_{\text{ADP Binding}} \\[.5em]
& \underbrace{-k_{\rm phos}^{\rm P2Y_1}[{\rm ADP-P2Y_1-G_qGDP}]_S[{\rm PKCa-Ca-DAG}] + k_{\rm dephos}^{\rm P2Y_1}[{\rm ADP-pP2Y_1-G_qGDP}]_S}_{\text{Desensitization by PKC}}
\label{eqn:adp-p2y1-gqgdp}
\end{split} \\[1em]
\begin{split}
\frac{d}{dt}[{\rm ADP-P2Y_1a-G_qGDP}]_S ={}& \underbrace{\alpha\beta\delta k_{\rm act}^{\rm P2Y_1}[{\rm ADP-P2Y_1-G_qGDP}]_S - k_{\rm -act}^{\rm P2Y_1}[{\rm ADP-P2Y_1a-G_qGDP}]_S}_{\text{activation}} \\[.5em]
& + \underbrace{\alpha k_{\rm ADP}^{\rm P2Y_1}[{\rm P2Y_1a-G_qGDP}]_S[{\rm ADP}] - \frac{k_{\rm -ADP}^{\rm P2Y_1}}{\delta\gamma}[{\rm ADP-P2Y_1a-G_qGDP}]_S}_{\text{ADP Binding}} \\[.5em]
& \underbrace{-k_{\rm phos}^{\rm P2Y_1}[{\rm ADP-P2Y_1a-G_qGDP}]_S[{\rm PKCa-Ca-DAG}] + k_{\rm dephos}^{\rm P2Y_1}[{\rm ADP-pP2Y_1a-G_qGDP}]_S   }_{\text{Desensitization by PKC}} \\[.5em]
& \underbrace{+\beta k_{\rm G_qGDP}^{\rm P2Y_1}[{\rm ADP-P2Y_1a}]_S[{\rm G_qGDP}]_S - \frac{k_{\rm -G_qGDP}^{\rm P2Y_1}}{\delta\gamma}[{\rm ADP-P2Y_1a-G_qGDP}]_S}_{\text{G$_q$GDP binding}} \\[.5em]
& \underbrace{-k_{\rm G_qGDP}^{\rm P2Y_1}[{\rm ADP-P2Y_1a-G_qGDP}]_S  + k_{\rm -GDP}^{\rm P2Y_1}[{\rm ADP-P2Y_1-G_q}]_S[{\rm GDP}]}_{\text{GDP binding}}
\end{split}
\end{align}

\begin{align}
\begin{split}
\frac{d}{dt}[{\rm ADP-p\pTWOyONE }]_S ={}& \underbrace{-\alpha k_{\rm act}^{\rm \pTWOyONE }[{\rm ADP-p\pTWOyONE }]_S + k_{\rm -act}^{\rm \pTWOyONE }[{\rm ADP-p\pTWOyONE a}]_S}_{\text{activation}} \\[.5em]
& \underbrace{-k_{\rm phos}^{\rm \pTWOyONE }[{\rm ADP-\pTWOyONE }]_S[{\rm PKCa-Ca-DAG}] + k_{\rm dephos}^{\rm \pTWOyONE }[{\rm ADP-p\pTWOyONE }]_S   }_{\text{Desensitization by PKC}}
\label{eqn:adp-p2y1}
\end{split} \\[1em]
\begin{split}
\frac{d}{dt}[{\rm ADP-p\pTWOyONE a}]_S ={}& \underbrace{\alpha k_{\rm act}^{\rm \pTWOyONE }[{\rm ADP-p\pTWOyONE }]_S - k_{\rm -act}^{\rm \pTWOyONE }[{\rm ADP-p\pTWOyONE a}]_S}_{\text{activation}} \\[.5em]
& \underbrace{-k_{\rm phos}^{\rm \pTWOyONE }[{\rm ADP-\pTWOyONE a}]_S[{\rm PKCa-Ca-DAG}] + k_{\rm dephos}^{\rm \pTWOyONE }[{\rm ADP-p\pTWOyONE a}]_S   }_{\text{Desensitization by PKC}}
\end{split} \\[1em]
\begin{split}
\frac{d}{dt}[{\rm ADP-p\pTWOyONE -G_qGDP}]_S ={}& \underbrace{-\alpha\beta\delta k_{\rm act}^{\rm \pTWOyONE }[{\rm ADP-p\pTWOyONE -G_qGDP}]_S + k_{\rm -act}^{\rm \pTWOyONE }[{\rm ADP-p\pTWOyONE a-G_qGDP}]_S}_{\text{activation}} \\[.5em]
& \underbrace{+k_{\rm phos}^{\rm \pTWOyONE }[{\rm ADP-\pTWOyONE -G_qGDP}]_S[{\rm PKCa-Ca-DAG}] - k_{\rm dephos}^{\rm \pTWOyONE }[{\rm ADP-p\pTWOyONE -G_qGDP}]_S}_{\text{Desensitization by PKC}}
\label{eqn:adp-p2y1-gqgdp}
\end{split} \\[1em]
\begin{split}
\frac{d}{dt}[{\rm ADP-p\pTWOyONE a-G_qGDP}]_S ={}& \underbrace{\alpha\beta\delta k_{\rm act}^{\rm \pTWOyONE }[{\rm ADP-p\pTWOyONE -G_qGDP}]_S - k_{\rm -act}^{\rm \pTWOyONE }[{\rm ADP-p\pTWOyONE a-G_qGDP}]_S}_{\text{activation}} \\[.5em]
& \underbrace{+k_{\rm phos}^{\rm \pTWOyONE }[{\rm ADP-\pTWOyONE a-G_qGDP}]_S[{\rm PKCa-Ca-DAG}] - k_{\rm dephos}^{\rm \pTWOyONE }[{\rm ADP-p\pTWOyONE a-G_qGDP}]_S}_{\text{Desensitization by PKC}}
\end{split}
\end{align}

If we make the assumption that $k_{\rm act}^{\rm P2Y_1}$ is large compared to other reaction rates (i.e. $k_{\rm -ADP}^{\rm P2Y_1}$ and $k_{\rm -G_qGDP}^{\rm P2Y_1}$) under the condition that $k_{\rm -act}^{\rm P2Y_1}/k_{\rm act}^{\rm P2Y_1}=K_{\rm act}^{\rm P2Y_1}$ is constant, then Equations~\ref{eqn:p2y1},~\ref{eqn:adp-p2y1},~\ref{eqn:p2y1-gqgdp}, and~\ref{eqn:adp-p2y1-gqgdp} reduce to the following algebraic relationships:
\begin{subequations}
\begin{align}
[{\rm P2Y_1a}]_S={}& K_{\rm act}[{\rm P2Y_1}]_S \\
[{\rm ADP-P2Y_1a}]_S={}& \alpha K_{\rm act}[{\rm ADP-P2Y_1}]_S  \\
[{\rm P2Y_1a-G_qGDP}]_S={}& \beta K_{\rm act}[{\rm P2Y_1-G_qGDP}]_S \\
[{\rm ADP-P2Y_1a-G_qGDP}]_S={}& \alpha\beta\delta K_{\rm act}[{\rm ADP-P2Y_1-G_qGDP}]_S \\
[{\rm ADP-pP2Y_1a}]_S={}& \alpha K_{\rm act}[{\rm ADP-pP2Y_1}]_S  \\
[{\rm ADP-pP2Y_1a-G_qGDP}]_S={}& \alpha\beta\delta K_{\rm act}[{\rm ADP-pP2Y_1-G_qGDP}]_S
\end{align}
\end{subequations}
We then define the concentrations $\overline{[{\rm P2Y_1}]}_S=[{\rm P2Y_1a}]_S+[{\rm P2Y_1}]_S$, and similar for the other three equations above.
Plugging in the above equations into these definitions yield the following relations:

\begin{subequations}
\begin{align}
[{\rm P2Y_1a}]_S={}& \frac{K_{\rm act}}{K_{\rm act}+1}\overline{[{\rm P2Y_1}]}_S \\
[{\rm P2Y_1}]_S={}& \frac{1}{K_{\rm act}+1}\overline{[{\rm P2Y_1}]}_S \\
[{\rm ADP-P2Y_1a}]_S={}& \frac{\alpha K_{\rm act}}{\alpha K_{\rm act}+1}\overline{[{\rm ADP-P2Y_1}]}_S \\
[{\rm ADP-P2Y_1}]_S={}& \frac{1}{\alpha K_{\rm act}+1}\overline{[{\rm ADP-P2Y_1}]}_S \\
[{\rm P2Y_1a-GDP}]_S={}& \frac{\beta K_{\rm act}}{\beta K_{\rm act}+1}\overline{[{\rm P2Y_1-GDP}]}_S \\
[{\rm P2Y_1-GDP}]_S={}& \frac{1}{\beta K_{\rm act}+1}\overline{[{\rm P2Y_1-GDP}]}_S \\
[{\rm ADP-P2Y_1a-G_qGDP}]_S={}& \frac{\alpha\beta\delta K_{\rm act}}{\alpha\beta\delta K_{\rm act}+1}\overline{[{\rm P2Y_1-G_qGDP}]}_S \\
[{\rm ADP-P2Y_1-G_qGDP}]_S={}& \frac{1}{\alpha\beta\delta K_{\rm act}+1}\overline{[{\rm P2Y_1-G_qGDP}]}_S \\
[{\rm ADP-p\pTWOyONE a}]_S={}& \frac{\alpha K_{\rm act}}{\alpha K_{\rm act}+1}\overline{[{\rm ADP-p\pTWOyONE }]}_S \\
[{\rm ADP-p\pTWOyONE }]_S={}& \frac{1}{\alpha K_{\rm act}+1}\overline{[{\rm ADP-p\pTWOyONE }]}_S \\
[{\rm ADP-p\pTWOyONE a-G_qGDP}]_S={}& \frac{\alpha\beta\delta K_{\rm act}}{\alpha\beta\delta K_{\rm act}+1}\overline{[{\rm p\pTWOyONE -G_qGDP}]}_S \\
[{\rm ADP-p\pTWOyONE -G_qGDP}]_S={}& \frac{1}{\alpha\beta\delta K_{\rm act}+1}\overline{[{\rm p\pTWOyONE -G_qGDP}]}_S
\end{align}
\end{subequations}

Finally, we compute $\frac{d}{dt}\overline{[{\rm P2Y_1}]}_S$ by adding the derivatives of $[{\rm P2Y_1a}]_S$ and $[{\rm P2Y_1}]_S$ and substituting for $\overline{[{\rm P2Y_1}]}_S$ (and equivalently for the other reduced states):

\begin{align}
\begin{split}
\frac{d}{dt}\overline{[{\rm P2Y_1}]}_S = {}& -\bigg(k_{\rm ADP}^{\rm P2Y_1}\frac{1}{K_{\rm act}+1}+ k_{\rm ADP}^{\rm P2Y_1}\frac{\alpha K_{\rm act}}{K_{\rm act}+1}\bigg)[{\rm ADP}]\overline{[{\rm P2Y_1}]}_S \\
& + \bigg(k_{\rm -ADP}^{\rm P2Y_1}\frac{1}{\alpha K_{\rm act}+1}+k_{\rm -ADP}^{\rm P2Y_1}\frac{\alpha K_{\rm act}}{\alpha K_{\rm act}+1}\bigg)\overline{[{\rm ADP-P2Y_1}]}_S \\
& -\bigg(\beta k_{\rm G_qGDP}^{\rm P2Y_1}\frac{1}{K_{\rm act}+1}\bigg)[{\rm G_qGDP}]\overline{[{\rm P2Y_1}]}_S \\
& + \bigg(k_{\rm -G_qGDP}^{\rm P2Y_1}\frac{\beta K_{\rm act}}{\beta K_{\rm act}+1}\bigg)\overline{[{\rm P2Y_1-G_qGDP}]}_S \\
& -\bigg(k_{\rm G_qGTP}^{\rm P2Y_1}\frac{K_{\rm act}}{K_{\rm act}+1}\bigg)[{\rm G_qGTP}]\overline{[{\rm P2Y_1}]}_S \\
& + k_{\rm -G_qGTP}^{\rm P2Y_1}[{\rm P2Y_1a-G_qGTP}]_S
\end{split} \\[1em]
\begin{split}
\frac{d}{dt}\overline{[{\rm ADP-P2Y_1}]}_S ={}& \bigg(k_{\rm ADP}^{\rm P2Y_1} \frac{1}{K_{\rm act}+1}+\alpha k_{\rm ADP}^{\rm P2Y_1}\frac{K_{\rm act}}{K_{\rm act}+1}\bigg)[{\rm ADP}]\overline{[{\rm P2Y_1}]}_S \\
& - \bigg(k_{\rm -ADP}^{\rm P2Y_1}\frac{1}{\alpha K_{act}+1}+k_{\rm -ADP}^{\rm P2Y_1}\frac{\alpha K_{\rm act}}{\alpha K_{\rm act}+1}\bigg)\overline{[{\rm ADP-P2Y_1}]}_S \\
& - \bigg(\beta k_{\rm G_qGDP}^{\rm P2Y_1}\frac{\alpha K_{\rm act}}{\alpha K_{\rm act}+1}\bigg) [{\rm G_qGDP}]\overline{[{\rm ADP-P2Y_1}]}_S \\
& + \bigg(\frac{k_{\rm -G_qGDP}^{\rm P2Y_1}}{\delta\gamma}\frac{\alpha\beta\delta K_{\rm act}}{\alpha\beta\delta k_{\rm act}+1}\bigg)\overline{[{\rm ADP-P2Y1-G_qGDP}]}_S \\
& - \bigg(k_{\rm G_qGTP}^{\rm P2Y_1}\frac{\alpha K_{\rm act}}{\alpha K_{\rm act}+1}\bigg) [{\rm G_qGTP}]\overline{[{\rm ADP-P2Y_1}]}_S \\
& + k_{\rm -G_qGTP}^{\rm P2Y_1}[{\rm ADP-P2Y_1a-G_qGTP}]_S \\
& -k_{\rm phos}^{\rm \pTWOyONE }\overline{[{\rm ADP-\pTWOyONE}]}_S[{\rm PKCa-Ca-DAG}] + k_{\rm dephos}^{\rm \pTWOyONE }\overline{[{\rm ADP-p\pTWOyONE}]}_S
\end{split} \\[1em]
\begin{split}
\frac{d}{dt}\overline{[{\rm P2Y_1-G_qGDP}]}_S ={}& -\bigg(k_{\rm ADP}^{\rm P2Y_1}\frac{1}{\beta K_{\rm act}+1} + \alpha k_{\rm ADP}^{\rm P2Y_1}\frac{\beta K_{\rm act}}{\beta K_{\rm act}+1}\bigg)[{\rm ADP}]\overline{[{\rm P2Y_1-G_qGDP}]}_S \\
& +\bigg(\frac{k_{\rm -ADP}^{\rm P2Y_1}}{\gamma}\frac{1}{\alpha\beta\delta K_{\rm act}+1} + \frac{k_{\rm -ADP}^{\rm P2Y_1}}{\delta\gamma}\frac{\alpha\beta\delta K_{\rm act}}{\alpha\beta\delta K_{\rm act}+1}\bigg)\overline{[{\rm ADP-P2Y_1-G_qGDP}]}_S \\
& + \bigg(\beta k_{\rm G_qGDP}^{\rm P2Y_1}\frac{K_{\rm act}}{K_{\rm act}+1}\bigg)[{\rm G_qGDP}]\overline{[{\rm P2Y_1}]}_S \\
& - \bigg(k_{\rm -G_qGDP}^{\rm P2Y_1}\frac{\beta K_{\rm act}}{\beta K_{\rm act}+1}\bigg)\overline{[{\rm P2Y_1-G_qGDP}]}_S \\
& +k_{\rm GDP}^{\rm P2Y_1}[{\rm GDP}][{\rm P2Y_1a-G_q}]_S - \bigg(k_{\rm -GDP}^{\rm P2Y_1}\frac{\beta K_{\rm act}}{\beta K_{\rm act}+1}\bigg)\overline{[{\rm P2Y_1-G_qGDP}]}_S
\end{split}
\end{align}
\begin{align}
\begin{split}
\frac{d}{dt}\overline{[{\rm ADP-P2Y_1-G_qGDP}]}_S = {} & \bigg(k_{\rm ADP}^{\rm P2Y_1}\frac{1}{\beta K_{\rm act}+1} + k_{\rm -ADP}^{\rm P2Y_1}\frac{\alpha\beta K_{\rm act}}{\beta K_{\rm act}+1}\bigg)[{\rm ADP}]\overline{[{\rm P2Y_1-G_qGDP}]}_S \\
& -\bigg(\frac{k_{\rm -ADP}^{\rm P2Y_1}}{\gamma}\frac{1}{\alpha\beta\delta K_{\rm act}+1} + \frac{k_{\rm -ADP}^{\rm P2Y_1}}{\delta\gamma}\frac{\alpha\beta\delta K_{\rm act}}{\alpha\beta\delta K_{\rm act}+1}\bigg)\overline{[{\rm ADP-P2Y_1-G_qGDP}]}_S \\
& \bigg(k_{\rm G_qGDP}^{\rm P2Y_1}\frac{\alpha\beta K_{\rm act}}{\alpha K_{\rm act}+1}\bigg) [{\rm G_qGDP}]\overline{[{\rm ADP-P2Y_1}]}_S \\
& - \bigg(\frac{k_{\rm -G_qGDP}^{\rm P2Y_1}}{\delta\gamma}\frac{\alpha\beta\delta K_{\rm act}}{\alpha\beta\delta k_{\rm act}+1}\bigg)\overline{[{\rm ADP-P2Y1-G_qGDP}]}_S \\
& + k_{\rm GDP}^{\rm P2Y_1}[{\rm GDP}][{\rm ADP-P2Y_1a-G_q}]_S \\
& - \bigg(k_{\rm -GDP}^{\rm P2Y_1}\frac{\alpha\beta\delta K_{\rm act}}{\alpha\beta\delta K_{\rm act}+1}\bigg)\overline{[{\rm ADP-P2Y1-G_qGDP}]}_S \\
& -k_{\rm phos}^{\rm \pTWOyONE }\overline{[{\rm ADP-\pTWOyONE -\gq GDP}]}_S[{\rm PKCa-Ca-DAG}] \\
& +k_{\rm dephos}^{\rm \pTWOyONE }\overline{[{\rm ADP-p\pTWOyONE -\gq GDP}]}_S
\end{split} \\[1em]
\begin{split}
\frac{d}{dt}\overline{[{\rm ADP-p\pTWOyONE }]}_S ={}& k_{\rm phos}^{\rm \pTWOyONE }\overline{[{\rm ADP-\pTWOyONE}]}_S[{\rm PKCa-Ca-DAG}] - k_{\rm dephos}^{\rm \pTWOyONE }\overline{[{\rm ADP-p\pTWOyONE}]}_S
\end{split} \\[1em]
\begin{split}
\frac{d}{dt}[{\rm ADP-p\pTWOyONE -G_qGDP}]_S ={}& k_{\rm phos}^{\rm \pTWOyONE }\overline{[{\rm ADP-\pTWOyONE -G_qGDP}]}_S[{\rm PKCa-Ca-DAG}] - k_{\rm dephos}^{\rm \pTWOyONE }\overline{[{\rm ADP-p\pTWOyONE -G_qGDP}]}_S
\end{split}
\end{align}
Which, after simplifying and dropping the bars and ``a'' from activated states, yields the equations in the earlier sections.

\bibliographystyle{abbrv}
\bibliography{main}